\pdfoutput=1
\documentclass[11pt]{article}
\usepackage{dsfont}

\usepackage{amsmath,amssymb,graphicx}
\usepackage{diagbox}
\usepackage{hyperref}
\usepackage[nosort]{cite}

\usepackage{tikz}
\usetikzlibrary{calc} \usetikzlibrary{patterns} \usetikzlibrary{decorations.pathreplacing} \usetikzlibrary{decorations.markings} \usetikzlibrary{decorations.pathmorphing} \usetikzlibrary{positioning}
\usepackage{color}
\definecolor{darkred}{rgb}{0.65,0.15,0}
\definecolor{newgreen}{rgb}{0.2,0.62,0.14}
\hypersetup{pdfborder={0 0 0},colorlinks=true,urlcolor=blue,citecolor=blue,linkcolor=darkred,linktocpage=true}

\DeclareFontFamily{U}{mathx}{}
\DeclareFontShape{U}{mathx}{m}{n}{ <-> mathx10 }{}
\DeclareSymbolFont{mathx}{U}{mathx}{m}{n}
\DeclareFontSubstitution{U}{mathx}{m}{n}

\DeclareMathAccent{\widecheck}{0}{mathx}{"71}

\numberwithin{equation}{section}

\def\nn{\nonumber}

\def\spa#1.#2{\left\langle#1\,#2\right\rangle}
\def\spb#1.#2{\left[#1\,#2\right]}

\def\ep{\epsilon}

\newcommand{\te}{\textrm}

\newcommand{\GBR}[3]{{\cal G}\! \left[\begin{smallmatrix}#1\\#2\end{smallmatrix};#3\right]}
\newcommand{\GBRno}[2]{{\cal G}\! \left[\begin{smallmatrix}#1\\#2\end{smallmatrix}\right]}
\newcommand{\bsv}{\beta^{\rm sv}}
\newcommand{\betasv}[1]{\beta^{\rm sv}\! \left[\begin{smallmatrix}#1\end{smallmatrix}\right]}

\newcommand{\betasvtaualpha}[1]{\beta_{(\alpha)}^{\rm sv}\! \left[\begin{smallmatrix}#1\end{smallmatrix}; \tau\right]}

\newcommand{\betasvStaualpha}[1]{\beta_{(\alpha)}^{\rm sv}\! \left[\begin{smallmatrix}#1\end{smallmatrix};-\tfrac{1}{\tau}\right]}
\newcommand{\bsvBR}[3]{\beta^{\rm sv} \! \left[\begin{smallmatrix}#1\\#2\end{smallmatrix};#3\right]}

\newcommand{\alphaBR}[3]{\alpha\! \left[\begin{smallmatrix}#1\\#2\end{smallmatrix};#3\right]}
\newcommand{\alphaBRno}[2]{\alpha\! \left[\begin{smallmatrix}#1\\#2\end{smallmatrix}\right]}

\newcommand{\cc}{\text{c.c.}}

\font\tenshuffle=shuffle10 \font\sevenshuffle=shuffle7 \font\fiveshuffle=shuffle7 at 5pt
\def\shuffle{{%
\def\Dshuffle{\mathbin{\hbox{\tenshuffle\char'001}}}%
\def\Sshuffle{\mathbin{\hbox{\sevenshuffle\char'001}}}%
\def\SSshuffle{\mathbin{\hbox{\fiveshuffle\char'001}}}%
\mathchoice{\Dshuffle}{\Dshuffle}{\Sshuffle}{\SSshuffle}}}

\newcommand{\series}{\Phi}

\newcommand{\ourh}{t}

\def\beq{\begin{equation}}
\def\eeq{\end{equation}}
\let\Re\relax
\let\Im\relax
\DeclareMathOperator{\Re}{Re}
\DeclareMathOperator{\Im}{Im}

\newcommand{\eq}{\begin{equation}}
\newcommand{\eqe}{\end{equation}}
\newcommand{\eqa}{\begin{eqnarray}}
\newcommand{\eqae}{\end{eqnarray}}

\newcommand{\bea}{\begin{eqnarray}}
\newcommand{\eea}{\end{eqnarray}}
\newcommand{\dd}{\mathrm{d}}

\newcommand{\RR}{\mathbb R}

\newcommand{\NN}{\mathbb N}
\newcommand{\ZZ}{\mathbb Z}
\newcommand{\QQ}{\mathbb Q}

\newcommand{\FFpm}[3]{{\rm F}^{\pm(#1)}_{#2,#3}}
\newcommand{\FFp}[3]{{\rm F}^{+(#1)}_{#2,#3}}
\newcommand{\FFm}[3]{{\rm F}^{-(#1)}_{#2,#3}}
\newcommand{\seedpm}[3]{f^{\pm(#1)}_{#2,#3}}
\newcommand{\seedp}[3]{f^{+(#1)}_{#2,#3}}
\newcommand{\seedm}[3]{f^{-(#1)}_{#2,#3}}

\newcommand{\cFFpm}[3]{\widecheck{\rm F}^{\pm(#1)}_{#2,#3}}
\newcommand{\cFFp}[3]{\widecheck{\rm F}^{+(#1)}_{#2,#3}}
\newcommand{\cFFm}[3]{\widecheck{\rm F}^{-(#1)}_{#2,#3}}
\newcommand{\Hpm}{{\rm H}^{\pm}}
\newcommand{\Hp}{{\rm H}^{+}}
\newcommand{\Hm}{{\rm H}^{-}}

\newcommand{\MM}[2]{{\mathfrak{m}}\! \left[\begin{smallmatrix}#1\\#2\end{smallmatrix}\right]}
\newcommand{\MMsv}[2]{{\cal M}\! \left[\begin{smallmatrix}#1\\#2\end{smallmatrix}\right]}
\newcommand{\textMM}[2]{{\mathfrak{m}}[\begin{smallmatrix}#1\\#2\end{smallmatrix}]}
\newcommand{\textMMsv}[2]{{\cal M}[\begin{smallmatrix}#1\\#2\end{smallmatrix}]}

\newcommand{\SLtwoZ}{{\rm SL}(2,\mathbb{Z})}
\newbox\charbox
\newbox\slabox
\def\s#1{{      % Feynman slash
        \setbox\charbox=\hbox{$#1$}
        \setbox\slabox=\hbox{$/$}
        \dimen\charbox=\ht\slabox
        \advance\dimen\charbox by -\dp\slabox
        \advance\dimen\charbox by -\ht\charbox
        \advance\dimen\charbox by \dp\charbox
        \divide\dimen\charbox by 2
        \raise-\dimen\charbox\hbox to \wd\charbox{\hss/\hss}
        \llap{$#1$}
}}

\reversemarginpar
\newcounter{todocounter}

\colorlet{ddcolor}{green!40!white}

\newcommand{\ddinline}[2][]{
  \ifthenelse { \equal {#1} {} }
    { \def\temp {#2} }  % if #1 == blank
    { \def\temp {#1} }   % else (not blank)
  \refstepcounter{todocounter}\todo[color=ddcolor,inline,caption={\textbf{\thetodocounter. DD} \temp}]{\textbf{\thetodocounter. DD:} #2}{}}

\colorlet{oscolor}{blue!20!white}

\newcommand{\osinline}[2][]{
  \ifthenelse { \equal {#1} {} }
    { \def\temp {#2} }  % if #1 == blank
    { \def\temp {#1} }   % else (not blank)
  \refstepcounter{todocounter}\todo[color=oscolor,inline,caption={\textbf{\thetodocounter. OS} \temp}]{\textbf{\thetodocounter. OS:} #2}{}}

\colorlet{akcolor}{yellow!40!white}

\newcommand{\akinline}[2][]{
  \ifthenelse { \equal {#1} {} }
    { \def\temp {#2} }  % if #1 == blank
    { \def\temp {#1} }   % else (not blank)
  \refstepcounter{todocounter}\todo[color=akcolor,inline,caption={\textbf{\thetodocounter. AK} \temp}]{\textbf{\thetodocounter. AK:} #2}{}}

\newcommand{\GG}{ {\rm G} }
\newcommand{\EE}{ {\rm E} }

\newcommand{\PS}{\sum_{\gamma \in B(\ZZ)\backslash {\rm SL}(2,\ZZ)}}

\begin{document}

 {\flushright  
 UUITP-42/21\\
 DCPT-21/15\\[15mm]}

\begin{center}

{\bf {\LARGE \sc  Poincar\'e series for \\[2mm] modular graph forms at depth two}\\[4mm]
{\Large II. Iterated integrals of cusp forms}}\\[5mm]

\vspace{6mm}
\normalsize
{\large  Daniele Dorigoni${}^{1}$, Axel Kleinschmidt${}^{2,3}$ and Oliver Schlotterer${}^4$}

\vspace{10mm}
${}^1${\it Centre for Particle Theory \& Department of Mathematical Sciences\\
Durham University, Lower Mountjoy, Stockton Road, Durham DH1 3LE, UK}
\vskip 1 em
${}^2${\it Max-Planck-Institut f\"{u}r Gravitationsphysik (Albert-Einstein-Institut)\\
Am M\"{u}hlenberg 1, DE-14476 Potsdam, Germany}
\vskip 1 em
${}^3${\it International Solvay Institutes\\
ULB-Campus Plaine CP231, BE-1050 Brussels, Belgium}
\vskip 1 em
${}^4${\it  Department of Physics and Astronomy, Uppsala University, 75108 Uppsala, Sweden}

\vspace{10mm}

\hrule

\vspace{5mm}

\begin{tabular}{p{14cm}}

We continue the analysis of modular invariant functions, subject to inhomogeneous Laplace eigenvalue equations, that were determined in terms of Poincar\'e series in a companion paper. The source term of the Laplace equation is a product of (derivatives of) two non-holomorphic Eisenstein series whence the modular invariants are assigned depth two. These modular invariant functions can sometimes be expressed in terms of single-valued iterated integrals of holomorphic Eisenstein series as they appear in generating series of modular graph forms. We show that the set of iterated integrals of Eisenstein series has to be extended to include also iterated integrals of holomorphic cusp forms to find expressions for all modular invariant functions of depth two. The coefficients of these cusp forms are identified as ratios of their L-values inside and outside the critical strip.
\end{tabular}

\vspace{6mm}
\hrule
\end{center}

\thispagestyle{empty}

\newpage
\setcounter{page}{1}

\setcounter{tocdepth}{2}
\tableofcontents

\bigskip

%%%%%%%%%%%%%%%%%%%%%%%%%%%%%%%%%%%%%%%%%%%%%%%%%%%%%%%%%%%
\section{Introduction}
\label{sec:1}
%%%%%%%%%%%%%%%%%%%%%%%%%%%%%%%%%%%%%%%%%%%%%%%%%%%%%%%%%%%

In the companion Part~I~\cite{PartI} to this paper we introduced the Laplace equations
\begin{subequations}
\label{eq:Laps}
\begin{align}
(\Delta-s(s-1)) \FFp{s}{m}{k}& = \EE_m \EE_k\,,\\
(\Delta-s(s-1)) \FFm{s}{m}{k}& = \frac{ ( \nabla \EE_m) ( \overline{\nabla}\EE_k) -  (\nabla \EE_k) ( \overline{\nabla}\EE_m)}{2(\Im \tau)^2}\,,
\label{eq:Laps:odd}
\end{align}
\end{subequations}
with integers $s\geq 2$ and $2\leq m\leq k$, and where $\EE_k$ are non-holomorphic Eisenstein series
\begin{align}
\label{eq:Ekdef}
\EE_k = \frac{(\Im \tau)^k}{\pi^k} \sum_{(m,n)\neq (0,0)} \frac{1}{|m\tau+n|^{2k}}\, .
\end{align}
The modular parameter $\tau$ is in the upper half-plane, and $\EE_k$ is invariant under the modular transformations
\begin{align}
 \tau\to \gamma \cdot \tau = \frac{a\tau+b}{c\tau+d}
 \quad \text{ for }
\gamma = \begin{pmatrix}a&b\\c&d\end{pmatrix}\in \SLtwoZ\, .
 \end{align}
The Cauchy--Riemann derivatives $\nabla= 2i (\Im \tau)^2 \partial_\tau$ of $\EE_k$ are modular forms
of weight $(0,-2)$ and the Laplacian $\Delta = 4(\Im\tau)^2 \partial_\tau\partial_{\bar\tau}=\overline{\nabla}\big((\Im \tau)^{-2} \nabla\big)$ is modular invariant.
The superscripts $\pm$ on $\FFpm{s}{m}{k}$ indicate that these functions are required to be even/odd under the involution $\tau\to-\bar\tau$ of the upper half-plane, in line with the respective right-hand sides of~\eqref{eq:Laps}.
The spectrum of eigenvalues appearing in~\eqref{eq:Laps} is
\begin{subequations}
\label{eq:Spectrum}
\begin{align}
\FFp{s}{m}{k}\,&: \quad s\in \left\lbrace k{-}m{+}2,k{-}m{+}4,\ldots,k{+}m{-}4,k{+}m{-}2 \right\rbrace\,,\\
\FFm{s}{m}{k}\,&: \quad s\in \left\lbrace k{-}m{+}1,k{-}m{+}3,\ldots,k{+}m{-}3,k{+}m{-}1 \right\rbrace\,.
\end{align}
\end{subequations}
As all the objects in~\eqref{eq:Laps} are modular invariant, we focus on modular invariant solutions $\FFpm{s}{m}{k}$ to the Laplace problem. The transcendental weight of $\FFpm{s}{m}{k}$ is $m{+}k$ from~\eqref{eq:Laps} given that $\EE_m$ and $\EE_k$ have transcendental weight  $m$ and $k$, respectively.

In Part I, we constructed solutions to~\eqref{eq:Laps} in terms of absolutely convergent\footnote{Absolute convergence is guaranteed for $m<k$ and for $m=k$ a suitable regularisation was described in Part~I.} Poincar\'e series
\begin{align}
\label{eq:PSsol}
\FFpm{s}{m}{k}(\tau) = \PS  \seedpm{s}{m}{k}(\gamma \cdot \tau)\,,
\end{align}
where the seed functions $\seedpm{s}{m}{k}$ are invariant under shifts $\tau\to\tau{+}n$ for $n\in\mathbb{Z}$ which form the stabiliser of the cusp $\tau \rightarrow i \infty$
\begin{align}
B(\mathbb{Z}) = \left\{ \begin{pmatrix} \pm 1 & n \\0 & \pm 1\end{pmatrix} \,\middle|\, n\in \mathbb{Z}\right\}\subset\SLtwoZ\,.
\end{align}
For the convenience of the reader, appendix \ref{app:seeds} recaps the explicit form for our choice of representatives of these seeds $\seedpm{s}{m}{k}$.

Even though the solution~\eqref{eq:PSsol} is fully explicit and has interesting structures analysed in Part~I, extracting the complete Fourier expansion of $\FFpm{s}{m}{k}$ from the Poincar\'e-series representations is fairly involved. For the Fourier zero mode one can use the methods of~\cite{Ahlen:2018wng,Dorigoni:2019yoq,Dorigoni:2020oon,Basu:2020kka} but the non-zero modes with respect to $\tau\to\tau{+}1$ are hard to obtain. For this reason it is desirable to find alternative expressions for the modular invariants $\FFpm{s}{m}{k}$. 

A family of functions with well-defined modular transformation properties is provided by modular graph forms (MGFs)~\cite{DHoker:2015gmr, DHoker:2015wxz, DHoker:2016mwo}. These arise in the $\alpha'$-expansion of configuration-space integrals of genus-one closed-string amplitudes and have been studied from a physical perspective in~\cite{Green:1999pv, Green:2008uj, Green:2013bza, DHoker:2015gmr, DHoker:2015sve, Basu:2015ayg, DHoker:2015wxz, DHoker:2016mwo, Basu:2016xrt, Basu:2016kli, Basu:2016mmk, DHoker:2016quv, Kleinschmidt:2017ege, Basu:2017nhs, Broedel:2018izr, Ahlen:2018wng,Gerken:2018zcy, Gerken:2018jrq, DHoker:2019txf, Dorigoni:2019yoq, DHoker:2019xef, DHoker:2019mib, DHoker:2019blr, Basu:2019idd,  Gerken:2019cxz, Hohenegger:2019tii, Gerken:2020yii, Basu:2020kka, Vanhove:2020qtt, Basu:2020pey, Basu:2020iok , Hohenegger:2020slq} and a mathematical perspective in~\cite{Brown:mmv, Zerbini:2015rss, Brown:I, Brown:II, DHoker:2017zhq,Zerbini:2018sox, Zerbini:2018hgs, Zagier:2019eus, Berg:2019jhh}. 
As they arise from string amplitudes, MGFs possess a lattice-sum description over discrete momenta of Feynman graphs drawn on the genus-one string world-sheet. 

In particular, generating functions of closed-string integrals and their associated differential equations \cite{Gerken:2019cxz,Gerken:2020yii} lead to expressions for MGFs in terms of real-analytic objects denoted by 
\begin{align}
\label{intro:bsv}
 \bsvBR{j_1&j_2&\ldots & j_\ell}{k_1& k_2 & \ldots & k_\ell}{\tau}\,\,\,  \mbox{with}\, \, \, k_i \in \{4,6,8,\ldots\} \, \,\, \mbox{and}\, \,\,  0\leq j_i\leq k_i{-}2  \, ,
\end{align} 
where $\ell \in \mathbb N$ is called the depth of $\bsv$.\footnote{A more detailed review of the construction and 
properties of the $\bsv$ can be found in section~\ref{sec:2}.}  Depth serves as a filtration, and the 
highest-depth terms in the complex-conjugation and modular properties of the $\bsv$ 
take a simple form. The $\bsv$ are constructed from (single-valued) iterated integrals over holomorphic Eisenstein series and should be closely related to Brown's non-holomorphic modular forms~\cite{Brown:I,Brown:II}, although a precise dictionary between the two formalisms is still missing beyond depth one. Together with certain antiholomorphic integration constants determined in Part~I, the complete Fourier expansion of the $\bsv$ at depths one and two is known. Therefore it seems desirable to express the $\FFpm{s}{m}{k}$ in terms of the $\bsv$. A further advantage of using such a representation of a modular-invariant function in terms of iterated integrals is that it is unique~\cite{Nilsnewarticle}, unlike lattice-sum representations that are more frequent for MGFs.

In many cases, the Poincar\'e-series representations in this work may be viewed as interpolating between
double sums over lattice momenta and double integrals over holomorphic Eisenstein series: The seeds $\seedpm{s}{m}{k}$ in (\ref{eq:PSsol}) are constructed from depth-one integrals, and the sum over $\SLtwoZ$ transformations is comparable to a single lattice momentum. However, the Poincar\'e sums in (\ref{eq:PSsol}) often produce MGFs that require three and more lattice momenta (see Part I for details) or modular invariant functions without any known lattice-sum representation.

In Part~I, we have presented a procedure for obtaining linear combinations $\cFFpm{s}{m}{k}$ of $\bsv$ of depths two and one, together with Laurent polynomial terms in $y= \pi \Im \tau$ (that can be thought of as depth zero). These linear combinations were constructed by starting from depth-two terms that solve the Laplace equation~\eqref{eq:Laps} modulo terms of lower depth.
The latter were fixed from certain requirements on the desired solutions concerning their Cauchy--Riemann derivatives and asymptotics at the cusp, see Part I for further details.
However, this procedure was tailored towards solving the Laplace system in terms of 
the building blocks $\bsv$ of MGFs and does not guarantee that the resulting expression is modular invariant.

By comparing the dimensions of the space of solutions to~\eqref{eq:Laps} and the space of MGFs at depth two we have seen in Part~I that the MGFs do not suffice to span the space of $\FFpm{s}{m}{k}$. This is reflected in the fact that certain $\cFFpm{s}{m}{k}$ fail to be
modular invariant exactly in those cases when the dimensions of the function spaces differ. In the present paper, we shall discuss how to augment the $\cFFpm{s}{m}{k}$ so that they become modular invariant and therefore equal the corresponding Poincar\'e series $\FFpm{s}{m}{k}$ in (\ref{eq:PSsol}).  
In other words, we illustrate through a variety of examples that MGFs do not exhaust the modular-invariant combinations of iterated integrals of holomorphic modular forms and their complex conjugates.

As we shall see, the missing ingredients beyond the $\bsv$ are (real and imaginary parts of) iterated integrals of holomorphic cusp forms. From the Eichler--Shimura theorem~\cite{Eichler:1957,Shimura:1959} and the work of Brown~\cite{Brown:mmv, Brown:I,Brown:II,Brown:III} on iterated integrals of general holomorphic modular forms, it is not surprising that restricting to the $\bsv$, that only involve iterated integrals of holomorphic Eisenstein series, is insufficient to describe the full space of modular-invariant solutions to~\eqref{eq:Laps}.
The first discrepancy in the dimensions of the function spaces $\FFpm{s}{m}{k}$ and MGFs appears for eigenvalues $s=6,8,9,10,\ldots$ which coincide exactly with half the modular weight of the first holomorphic cusp forms of $\SLtwoZ$. This can be seen as a hint that cusp forms are the missing piece of the puzzle. 

A further indication for the relevance of holomorphic cusp forms stems from the appearance of
conjectural matrix representations of Tsunogai's derivation algebra~\cite{Tsunogai} in the
generating series of MGFs~\cite{Gerken:2019cxz,Gerken:2020yii}. Relations in the derivation algebra
are also tied to holomorphic cusp forms~\cite{Pollack} and imply that, 
starting from depth two, there are combinations of the $\bsv$
that are not contained in the generating series of MGFs.
Completing these to modular invariants requires holomorphic cusp forms as we shall see. This follows from the S-modular transformations of the various $\bsv$ which contain interesting so-called multiple modular values~\cite{Brown:mmv} that involve the values of completed L-functions of cusp forms at integers~\cite{Brown2019} and extend the set of single-valued multiple zeta values. In order to cancel these L-values from S-modular transformations in general one has to combine the $\bsv$ with iterated integrals of cusp forms. We shall work out these ideas in detail in this paper and spell out a variety of examples.

As a byproduct of our analysis we derive that the series expansions in $q=\exp(2\pi i \tau)$ and
$\bar q = \exp(-2\pi i \bar \tau)$ of these non-holomorphic modular objects $\FFpm{s}{m}{k}$ display 
very interesting structures. Firstly, 
the leading terms in the expansion of the even functions $\FFp{s}{m}{k}$ around the cusp $\Im \tau \gg 1$
are Laurent polynomials in $y= \pi \Im \tau$ that will also be referred to as ``perturbative''.
These Laurent polynomials have a single term with a rational coefficient, 
a single term with a $\mathbb Q$-multiple of the product $\zeta_{2m-1}\zeta_{2k-1}$, while all other 
coefficients are $\mathbb Q$-multiples of odd zeta values $\zeta_{2m-1},\zeta_{2k-1},\zeta_{m+k+s-1}$,
see Part~I for further details.
Secondly, the infinite tower of exponentially suppressed, non-perturbative terms of the form $q^n \bar{q}^m$, with both of $n,m>0$, have Laurent polynomials in $y$ with rational coefficients for both the even and odd $\FFpm{s}{m}{k}$. Finally, and perhaps more interestingly, due to the presence of iterated integrals of  holomorphic cusp forms we find that the exponentially suppressed terms of the form $q^n \bar{q}^0$ (and their complex conjugates $q^0 \bar{q}^n$) with $n>0$ are multiplied by particularly rich Laurent polynomials in $y$: Their coefficients are either rationals, or $\mathbb Q$-multiples of single odd zeta values or surprisingly rationals (or more general number-field extensions of $\mathbb Q$) times special ratios of completed L-values associated to whichever cusp form is at play.

These results allow us to make novel predictions regarding the non-zero Fourier-mode decomposition of the Poincar\'e series \eqref{eq:PSsol}. In particular in Part~I, we have thoroughly explained how, for all the constructed seed functions $\seedpm{s}{m}{k}( \tau)$, one can exploit the results of \cite{Dorigoni:2019yoq} to obtain the purely perturbative Laurent polynomials in $y$. To pass from the seed function $\seedpm{s}{m}{k}( \tau)$ to the actual associated Poincar\'e series \eqref{eq:PSsol} one needs to use a particular integral transform detailed for instance in appendix A of Part I. Such a mapping between seed and modular function can also be used to formally obtain the non-zero modes for the modular invariant Poincar\'e series. However, the computation of non-zero modes from this integral transform of the seed involves very 
complicated Kloosterman sums and the analogue of the analysis in \cite{Dorigoni:2019yoq} to this case 
is currently unknown.
Despite this lack of full control over Kloosterman sums, our results imply that these Kloosterman sums must
contain completed L-values of holomorphic cusp forms. It would be extremely interesting to extend the results of \cite{Dorigoni:2019yoq} to the non-zero Fourier mode sectors, thus deriving directly from the seed functions the 
exponentially suppressed terms $q^n \bar{q}^0$ and $q^0 \bar{q}^n$ with $n>0$ including their Laurent polynomials.

\subsubsection*{Outlook}

The results of Part I and this work raise a variety of follow-up questions of relevance to 
string perturbation theory, algebraic geometry and number theory. Most obviously, the Fourier expansion of 
depth-two MGFs and their extension by iterated integrals of holomorphic cusp forms call for generalisations
to higher depth. Among other things, (single-valued) multiple zeta values beyond depth one, iterated 
integrals that mix holomorphic Eisenstein series with cusp forms and generalisations of L-values~\cite{Brown:Elliptic20} are expected to play a key role starting from 
depth  three. The respective  seed should have one unit of depth less than its modular  invariant Poincar\'e sum, and it will be rewarding to study this
kind of recursive structure at general depth. 

Furthermore, a detailed connection with the recent mathematics literature promises powerful synergies.
Various important properties of the $\beta^{\rm sv}$ at general depth will follow once their precise relation to 
Brown's non-holomorphic modular forms is established. Moreover, we note that iterated integrals of cusp forms and their Poincar\'e sums have featured prominently in recent work~\cite{Diamantis:2020} that also relates to so-called higher-order modular forms. Certain Laplace systems similar to~\eqref{eq:Laps} but at depth three have also been studied recently in~\cite{Drewitt:2021}. These references can provide useful guidance when generalising our work.

\subsubsection*{Outline}

In section~\ref{sec:2}, we review the basic properties of iterated integrals of holomorphic modular forms, with particular emphasis on their modular properties and certain $\SLtwoZ$ group cocycles that arise.  In section~\ref{sec:3}, we then use these results in the analysis of the modular invariant solutions to the Laplace equations. We further show how to combine the $\bsv$ with iterated integrals of cusp forms based on the vanishing of the cocycles thus restoring modularity. We explain the relation between Tsunogai's derivation algebra and the modular invariant Laplace eigenfunctions in section~\ref{sec:select}. Further properties of the solutions to~\eqref{eq:Laps}, such as connections to Kloosterman sums, are discussed in section~\ref{sec:5}. An ancillary file that accompanies the arXiv submission and journal publication of this work contains many examples and explicit expressions related to the functions $\FFpm{s}{m}{k}$.

%%%%%%%%%%%%%%%%%%%%%%%%%%%%%%%%%%%%%%%%%%%%%%%%%%%%%%%%%%%
\section{Basics of iterated integrals}
\label{sec:2}
%%%%%%%%%%%%%%%%%%%%%%%%%%%%%%%%%%%%%%%%%%%%%%%%%%%%%%%%%%%

This section is dedicated to the central aspects of iterated integrals as well as their differential 
and modular properties as they enter our analysis. Frequent use will be 
made of the Cauchy--Riemann derivatives
\begin{align}
\nabla = 2i (\Im\tau)^2 \partial_\tau\,,\quad \overline{\nabla} = -2i (\Im \tau)^2\partial_{\bar\tau}
\end{align}
and the Laplace operator
\begin{align}
\Delta = 4 (\Im\tau)^2 \partial_\tau \partial_{\bar\tau} = (\pi\overline{\nabla}) ( y^{-2} \pi \nabla)\,.
\end{align}
As in the equation above, we often use the symbol $y=\pi \Im \tau$, and powers of $y$ satisfy
\begin{align}
\label{eq:diffy}
(\pi \nabla)^k y^a = (\pi \overline{\nabla})^k y^a = \frac{\Gamma(a{+}k)}{\Gamma(a)} y^{a+k}\,,\quad \Delta y^a = a(a{-}1) y^a\, .
\end{align}

\subsection{Iterated integrals of Eisenstein series}
\label{sec:2.1}

In the present work, we shall only require the depth-one and depth-two versions of the single-valued iterated Eisenstein integrals~\eqref{intro:bsv}. These are defined by the integrals \cite{Gerken:2020yii}\footnote{We shall often suppress the argument $\tau$ of various functions to simplify the notation.}
\begin{subequations}
\label{eq:beta}
\begin{align}
\label{eq:bsv1}
\betasv{j\\k} =  \frac{(2\pi i)^{-1}}{(4y)^{k-2-j}} \bigg\{ \int\limits_{\tau}^{i \infty} \dd \tau_1 (\tau{-}\tau_1)^{k-2-j} (\bar\tau{-}\tau_1)^{j} \GG_k(\tau_1) - \int\limits_{\bar\tau}^{-i\infty}  \dd\bar\tau_1 (\tau{-}\bar\tau_1)^{k-2-j} (\bar\tau{-}\bar\tau_1)^{j} \overline{\GG_k(\tau_1)} \bigg\}
\end{align}
with $0 \leq j \leq k{-}2$ and
\begin{align}
\label{eq:bsv2}
\betasv{j_1 &j_2\\k_1 &k_2} &=
\sum_{p_1=0}^{k_1{-}2{-}j_1} \sum_{p_2=0}^{k_2{-}2{-}j_2} \frac{\binom{k_1{-}2{-}j_1}{p_1}\binom{k_2{-}2{-}j_2}{p_2}}{(4y)^{p_1+p_2}} \overline{\alphaBRno{j_1 +p_1&j_2+p_2}{k_1 &k_2}}
 + \frac{(2\pi i)^{-2}}{(4y)^{k_1+k_2-j_1-j_2-4}} \\
&\quad \times \bigg\{
\int\limits^{i\infty}_\tau \dd\tau_2  (\tau{-}\tau_2)^{k_2-j_2-2} (\bar\tau{-}\tau_2)^{j_2} \GG_{k_2}(\tau_2) \int\limits^{i\infty}_{\tau_2} \dd\tau_1 (\tau{-}\tau_1)^{k_1-j_1-2}(\bar\tau{-}\tau_1)^{j_1}  \GG_{k_1}(\tau_1) \nn\\
&\quad\quad - \int\limits^{i\infty}_\tau \dd\tau_2(\tau{-}\tau_2)^{k_2-j_2-2} (\bar\tau{-}\tau_2)^{j_2}  \GG_{k_2}(\tau_2) \int\limits^{-i\infty}_{\bar\tau} \dd\bar\tau_1  (\tau{-}\bar\tau_1)^{k_1-j_1-2}(\bar\tau{-}\bar\tau_1)^{j_1}\overline{\GG_{k_1}(\tau_1)}\nn\\
&\quad\quad + \int\limits^{-i\infty}_{\bar\tau}\dd\bar\tau_1(\tau{-}\bar\tau_1)^{k_1-j_1-2}(\bar\tau{-}\bar\tau_1)^{j_1} \overline{\GG_{k_1}(\tau_1)} \int\limits^{-i\infty}_{\bar\tau_1}  \dd\bar\tau_2  (\tau{-}\bar\tau_2)^{k_2-j_2-2} (\bar\tau{-}\bar\tau_2)^{j_2} \overline{\GG_{k_2}(\tau_2)}\bigg\}\nn
\end{align}
\end{subequations}
with $0 \leq j_i \leq k_i{-}2$. The holomorphic Eisenstein series are normalised as
\begin{align}
\label{eq:GG}
\GG_k(\tau) = \sum_{(m,n)\neq (0,0)} \frac{1}{(m\tau+n)^{k}} = 2\zeta_k + \frac{2 (2\pi i)^k}{(k{-}1)!} \sum_{n=1}^{\infty} \sigma_{k-1}(n) q^n
\end{align}
with divisor sum $\sigma_s(n) = \sum_{d|n} d^s$.
The integrals~\eqref{eq:beta} have to be understood with tangential base-point regularisation~\cite{Brown:mmv} and satisfy the shuffle relations
\begin{align}
\label{eq:bsvshuffle}
\betasv{j_1\\k_1}\betasv{j_2\\k_2} = \betasv{j_1&j_2\\k_1&k_2} + \betasv{j_2&j_1\\k_2&k_1}\,,
\end{align}
as well as the differential equations \cite{Gerken:2020yii}
\begin{subequations}
\label{eq:betaCR}
\begin{align}
-4\pi \nabla 
\bsvBR{j}{k}{\tau} &= (k-2-j) \bsvBR{j+1}{k}{\tau}  -\delta_{j,k-2} (\tau{-}\bar\tau)^k \GG_k(\tau) \,,\\
-4\pi \nabla 
\bsvBR{j_1 &j_2}{k_1 &k_2}{\tau} &=
 (k_1{-}j_1{-}2) \bsvBR{j_1+1 &j_2}{k_1 &k_2}{\tau} +  (k_2{-}j_2{-}2) \bsvBR{j_1 &j_2+1}{k_1 &k_2}{\tau}   \notag \\
 &\quad - \delta_{j_2,k_2-2} (\tau{-}\bar \tau)^{k_2} {\rm G}_{k_2}(\tau) \bsvBR{j_1}{k_1}{\tau} \,.
\end{align}
\end{subequations}
The objects $\overline{\alpha[\begin{smallmatrix} j_1 &j_2\\ k_1 &k_2 \end{smallmatrix}]}$ appearing in~\eqref{eq:bsv2} are purely antiholomorphic functions and constrained by the shuffle relation~\eqref{eq:bsvshuffle}. They are not fixed by the differential equation (\ref{eq:betaCR}) and therefore referred to as integration constants --- see \cite{Gerken:2020yii} for
a detailed discussion. A method to determine them from the reality properties and 
Laplace equations of $\FFpm{s}{m}{k}$ 
is discussed in Part~I, and a large number of examples can be found in an ancillary file.

%%%%%%
\subsubsection{Fourier expansions of iterated Eisenstein integrals}
%%%%%%

The compact definition~\eqref{eq:beta} of the $\bsv$ can be unpackaged to yield expressions in terms of other iterated integrals of the form~\cite{Broedel:2015hia, Broedel:2018izr}
\begin{subequations}
\label{eq:defE0}
\begin{align}
\label{eq:E0depth1}
\mathcal{E}_0(k,0^p;\tau) &= \frac{(2\pi i)^{p+1-k}}{p!}  \int_\tau^{i \infty} \dd\tau_1 (\tau{-}\tau_1)^p \GG_{k}^0(\tau_1)\,,\\
\label{eq:E0depth2}
\mathcal{E}_0(k_1,0^{p_1}, k_2,0^{p_2};\tau)  &= \frac{(2\pi i)^{p_1+p_2+2-k_1-k_2}}{p_1! p_2!} \int_\tau^{i \infty} \dd\tau_2 (\tau{-}\tau_2)^{p_2} \GG_{k_2}^0(\tau_2)\int_{\tau_2}^{i \infty} \dd\tau_1 (\tau_2{-}\tau_1)^{p_1} \GG_{k_1}^0(\tau_1)
\end{align}
\end{subequations}
and their complex conjugates with integers $k,k_1,k_2 = 4,6,8,\ldots$ and $p,p_1,p_2 \geq 0$. In the above expressions, $0^p$ is a placeholder for $p$ successive zeroes (reminiscent of integration kernels ${\rm G}_0^0=-1$ \cite{Broedel:2015hia}), and the $\GG_k^0$ are obtained from the holomorphic Eisenstein series $\GG_k$ by removing the zero mode
\begin{align}
\label{eq:G0G}
{\rm G}_k^0(\tau) &= \left\{ \begin{array}{cl}
{\rm G}_k(\tau) - 2 \zeta_k &: \ k>0 \ {\rm even} \, ,\\[1mm]
0&: \ k>0 \ {\rm odd} \, .
\end{array} \right.
\end{align}
The removal of the zero mode destroys the good modular transformations of $\GG_k$ but renders the integrals convergent without regularisation. Moreover, the integrals ${\cal E}_0(\ldots)$ have fully explicit $q$-expansions, e.g.
\begin{subequations}
\begin{align}
\mathcal{E}_0(k,0^p;\tau)&= -\frac{2}{(k{-}1)!}\sum_{m=1}^{\infty} \frac{\sigma_{k-1}(m)}{m^{p+1}} q^m\,,\\
\mathcal{E}_0(k_1,0^{p_1}, k_2,0^{p_2};\tau)&= \frac{4}{(k_1{-}1)!(k_2{-}1)!}\sum_{m,n=1}^{\infty} 
\frac{\sigma_{k_1-1}(m)\sigma_{k_2-1}(n)}{m^{p_1+1} (m{+}n)^{p_2+1}  } q^{m+n}
 \, ,
\end{align}
\end{subequations}
see~\cite[Eq.~(2.21)]{Broedel:2018izr} for arbitrary depth. This can be used to obtain the full Fourier expansions of the $\bsv$. The rewriting of the $\bsv$ in terms of the ${\cal E}_0(\ldots)$ requires a number of steps that are well-understood and whose precise form can be found in section 3.3 and appendix D of~\cite{Broedel:2018izr} as well as appendix G of \cite{Gerken:2020yii}.

\subsubsection{Differential equations and non-holomorphic Eisenstein series}

The definition (\ref{eq:E0depth1}) readily implies the Cauchy--Riemann derivative
\begin{align}
\pi \nabla \mathcal{E}_0(k,0^p) = -4 y^2  \mathcal{E}_0(k,0^{p-1})
\end{align}
for $p\geq0$, where we define $\mathcal{E}_0(k,0^{-1}) = -(2\pi i)^{-k} \GG_k^0$, and the Laplace equation $\Delta \mathcal{E}_0(k,0^p)=0$.

The non-holomorphic Eisenstein series $\EE_k$ defined in~\eqref{eq:Ekdef} can be decomposed in terms of iterated Eisenstein integrals as follows~\cite{Ganglzagier,DHoker:2015wxz}:
\begin{align}
\EE_k (\tau) &= (-1)^{k-1} \frac{ {\rm B}_{2k} }{(2k)!} (4y)^k + 
\frac{4(2k{-}3)!  \zeta_{2k-1}}{(k{-}2)!(k{-}1)!} (4y)^{1-k}
\nn\\
&\quad\quad
- 2 \frac{\Gamma(2k)}{\Gamma(k)} \sum_{\ell=0}^{k-1}  (4y)^{-\ell} \frac{\Gamma(k{+}\ell)}{\ell!\Gamma(k{-}\ell)} \Re \mathcal{E}_0\big(2k, 0^{k-1+\ell};\tau\big)\nn\\
&= (-1)^{k-1} \frac{ {\rm B}_{2k} }{(2k)!} (4y)^k + 
\frac{4(2k{-}3)!  \zeta_{2k-1}}{(k{-}2)!(k{-}1)!} (4y)^{1-k}
\label{eq:EII}\\
&\quad\quad  +\left[- \frac{1}{2\pi i } \frac{\Gamma(2k)}{[\Gamma(k)]^2} (4y)^{1-k}  \int_\tau^{i \infty} \dd\tau_1 (\tau{-}\tau_1)^{k-1}(\bar\tau{-}\tau_1)^{k-1} \GG_{2k}^0(\tau_1) + \cc \right]  \nn \\
&= \frac{ (2k{-}1)! }{ [ (k{-}1)! ]^2} \bigg\{ {-} \betasv{k-1\\2k} + \frac{ 2 \zeta_{2k-1} }{(2k{-}1) (4y)^{k-1} } \bigg\}\,,
\nn
\end{align}
where $+ \cc$ instructs us to add the complex conjugate, making $\EE_k$ real-analytic and even under $\tau\to -\bar{\tau}$. We also used $y=\pi \Im \tau$ as we shall do frequently.
This relation between $\EE_k$ and the depth-one $\bsv$, already present in \cite{Gerken:2020yii}, comes directly from~\eqref{eq:bsv1} when using the relation between $\GG_k$ and $\GG_k^0$ as well as tangential base-point regularisation.

From both the lattice-sum representation (\ref{eq:Ekdef}) and the final form of (\ref{eq:EII}),
 one can show the well-known formula for the $k$-th Cauchy--Riemann derivative of $\EE_k$ \cite{DHoker:2016mwo}:
\begin{align}
\label{eq:CREk}
(\pi \nabla)^k \EE_k (\tau) &= 
\frac{\Gamma(2k)}{\Gamma(k)}   (\Im \tau)^{2k} \left[ 2\zeta_{2k} + \GG_{2k}^0(\tau)\right] = \frac{\Gamma(2k)}{\Gamma(k)}   (\Im \tau)^{2k}  \GG_{2k}(\tau)\,,
\end{align}
where we have used Euler's formula relating the even Bernoulli numbers to the even Riemann zeta values
\begin{align}
2\zeta_{2n} = (-1)^{n+1} \frac{4^n \pi^{2n}}{(2n)!} {\rm B}_{2n}\,, \quad\quad n=1,2,3,\ldots\,.
\end{align}
We also record the following general formula
\begin{align}
\label{eq:CRII}
(\pi \nabla)^s \left[ i y^{1-s} \int_{\tau}^{i \infty} \dd\tau_1 (\tau{-}\tau_1)^{s-1} (\bar\tau{-}\tau_1)^{s-1} f(\tau_1)\right]  =  2^{2s-1}\pi \Gamma(s)  (\Im\tau)^{2s} f(\tau)
\end{align}
for any integer $s>0$ and holomorphic function $f(\tau)$ irrespective of its modular properties.
If $f(\tau)$ has a $q$-expansion in terms of positive powers of $q$ only, the integral in (\ref{eq:CRII}) 
is well-defined without tangential base-point regularisation.
With~\eqref{eq:CRII} and~\eqref{eq:diffy} it is easy to demonstrate~\eqref{eq:CREk}.

For the Laplacian there is a similar lemma given by
\begin{align}
\label{eq:LapII}
\big(\Delta-s(s{-}1) \big) \left[ i y^{1-s} \int_{\tau}^{i \infty} \dd\tau_1 (\tau{-}\tau_1)^{s-1} (\bar\tau{-}\tau_1)^{s-1} f(\tau_1)\right]   =0\,.
\end{align}

Besides direct evaluation of the Laplacian on the integral, we can also consider~\eqref{eq:LapII} by Fourier expanding the integrand $f(\tau)$. Specialising to the case of a single Fourier mode $f(\tau) = e^{2\pi i n\tau}$ with $n> 0$, the integral can be evaluated in terms of Bessel functions $K_{s-1/2}$ giving
\begin{equation}\label{eq:BesselLap}
 i y^{1-s} \int_{\tau}^{i \infty} \dd\tau_1 (\tau{-}\tau_1)^{s-1} (\bar\tau{-}\tau_1)^{s-1} e^{2\pi i n \tau_1}= \frac{(-1)^s \Gamma(s)}{\pi^{2s-1} n^s} \sqrt{\frac{n y}{\pi}} K_{s-1/2}(2ny) e^{2\pi i n \Re\tau}\,,
\end{equation}
which is the well-known solution to the Laplace equation \eqref{eq:LapII} in the $n^{\rm th}$ Fourier mode sector.

From this and~\eqref{eq:diffy} one can also prove the classic Laplace equation
\begin{align}
\label{eq:LapEk}
\big(\Delta - k(k{-}1)\big) \EE_k(\tau) = 0 \,,
\end{align}
since the Laurent monomials $y^k$ and $y^{1-k}$
in~\eqref{eq:EII} are in the kernel of $(\Delta-k(k{-}1))$.

\subsubsection{Multiple modular values}
\label{sec:trfe0}

Besides the version of iterated Eisenstein integrals in~\eqref{eq:defE0}, we shall also make use of
\begin{subequations}
\label{eq:defEE}
\begin{align}
\GBR{j}{k}{\tau} &= \int_{\tau}^{i\infty} \dd \tau_1\, \tau_1^j \,\GG_k(\tau_1)\,,\\
\GBR{j_1&j_2}{k_1&k_2}{\tau} &= \int_\tau^{i\infty} \dd \tau_2\, \tau_2^{j_2}\, \GG_{k_2}(\tau_2)
 \int_{\tau_2}^{i\infty} \dd \tau_1 \, \tau_1^{j_1} \, \GG_{k_1}(\tau_1)\,,
\end{align}
\end{subequations}
that are, up to normalisation conventions, Brown's holomorphic iterated Eisenstein integrals and require tangential base-point regularisation~\cite{Brown:mmv}. 
At depth one, this regularisation means
\begin{align}
\GBR{j}{k}{\tau} &= \int_{\tau}^{i\infty} \dd \tau_1\, \tau_1^j \,\GG_k^0(\tau_1) - 2\zeta_k \int_0^\tau \dd \tau_1  \tau_1^j \,,
\end{align}
treating the zero mode of (\ref{eq:GG}) differently, while the depth-two generalisation can be found in~\cite[Eq.~(4.13)]{Brown:mmv}.
A more general translation of~\eqref{eq:defEE} into the integrals~\eqref{eq:defE0} can be found in~\cite{Broedel:2018izr}, and the depth-one instance of the dictionary is
\begin{align}
\mathcal{E}_0(k,0^p) = (2\pi i)^{p+1-k} \sum_{a=0}^p \frac{(-1)^a \tau^{p-a}}{a! (p{-}a)!} \GBRno{a}{k} + 2\zeta_k (2\pi i)^{p+1-k} \frac{\tau^{p+1}}{(p{+}1)!}\,.
\label{calEvscalG}
\end{align}
The extra term proportional to $\zeta_k$ is due to $\mathcal{E}_0(k,0^p)$ being defined in terms of $\GG_k^0$, thus lacking the zero mode $\zeta_k$ when compared to $\GG_k$ appearing in~\eqref{eq:defEE}, see~\eqref{eq:G0G}.

The virtue of the definition~\eqref{eq:defEE} is that it is easier to describe the behaviour under S-modular transformations \cite{Brown:mmv, Broedel:2018izr}:
\begin{subequations}
\label{eq:EESmod}
\begin{align}
\GBR{j}{k}{-\tfrac1{\tau}} &= (-1)^j \GBR{k-2-j}{k}{\tau} + \MM{j}{k}\,,\\
\GBR{j_1&j_2}{k_1&k_2}{-\tfrac1{\tau}} &= (-1)^{j_1+j_2} \GBR{k_1-2-j_1&k_2-2-j_2}{k_1&k_2}{\tau} + (-1)^{j_2} \GBR{k_2-2-j_2}{k_2}{\tau} \MM{j_1}{k_1} + \MM{j_1&j_2}{k_1&k_2}\,.
\end{align}
\end{subequations}
The objects $\MM{\cdots}{\cdots}$ appearing in this equation do not depend on $\tau$ -- they are 
examples of multiple modular values~\cite{Brown:mmv} and correspond to period integrals
\begin{subequations}
\label{eq:defmmvs}
\begin{align}
\MM{j}{k} &= \int_{0}^{i\infty} \dd \tau_1\, \tau_1^j \,\GG_k(\tau_1)\,,\\
\MM{j_1&j_2}{k_1&k_2} &= \int_0^{i\infty} \dd \tau_2\, \tau_2^{j_2}\, \GG_{k_2}(\tau_2)
 \int_{\tau_2}^{i\infty} \dd \tau_1 \, \tau_1^{j_1} \, \GG_{k_1}(\tau_1)\,,
\end{align}
\end{subequations}
which are obtained formally as limits $\tau \to 0$ of~\eqref{eq:defEE}.
This limit is divergent and has to be treated again with tangential base-point regularisation. One way of doing this is to consider at depths one and two
\begin{subequations}
\label{eq:MMVnum}
\begin{align}
\MM{j}{k} &= \GBR{j}{k}{i} - (-1)^j \GBR{k-2-j}{k}{i}\,,\\
\MM{j_1&j_2}{k_1&k_2} &= \GBR{j_1&j_2}{k_1&k_2}{i} - (-1)^{j_1+j_2} \GBR{k_1-2-j_1&k_2-2-j_2}{k_1&k_2}{i}-(-1)^{j_2} \GBR{k_2-2-j_2}{k_2}{i} \MM{j_1}{k_1}\,,
\end{align}
\end{subequations}
where we rewrote the S-modular behaviour~\eqref{eq:EESmod} and evaluated this expression at the self-dual point $\tau=i$. We recall that the integrals~\eqref{eq:defEE} are well-defined for any finite $\tau$, using tangential base-point regularisation at the upper integration boundary $\tau\to i\infty$. The choice of the self-dual point $\tau=i$ in~\eqref{eq:MMVnum} is arbitrary (any pair of S-dual points would do) but convenient for numerical evaluations.

For depth one we can work out the multiple modular values explicitly as
\begin{align}
\label{eq:MMV1}
\MM{j}{k} = 
\left\{\begin{array}{cl}
\displaystyle \Bigg. \frac{2(-2\pi i)^{k-j-1} j!}{(k{-}1)!}  \zeta_{j+1} \zeta_{j+2-k} \Bigg.& \text{for $j>0\, $,}\\[2mm]
\displaystyle -\frac{2\pi i \zeta_{k-1}}{k{-}1} &  \text{for $j=0\, $,}
\end{array}\right.
\end{align}
and these correspond to periods of the holomorphic Eisenstein series~\cite{ZagierF}. The $j=0$ case can also be obtained as a limit after using the functional relations of the zeta function. Since $k\geq 4$ is an even integer, the multiple modular values of depth one vanish for even $0<j<k{-}2$ as they involve the zeta function evaluated at a negative even integer.

For depth two, numbers beyond (multiple) zeta values can occur~\cite{Brown:mmv, Brown2019}. We will discuss further properties of multiple modular values and how they arise directly in the S-modular transformation of the $\bsv$ in section~\ref{sec:mmvS}.

\subsection{Iterated integrals of cusp forms}

We now let $\Delta_{2s}(\tau)$ denote a holomorphic cusp form of weight $2s\in \{12,16,18,\ldots\}$. Then we define the analogue of~\eqref{eq:E0depth1} as
\begin{align}
\label{eq:EDelta}
{\cal E}_0(\Delta_{2s},0^p;\tau) & = (-1)^p(2\pi i)^{p+1} \int_\tau^{i\infty} \dd \tau_1\int_{\tau_1}^{i\infty}\dd \tau_2\ldots\int_{\tau_p}^{i\infty} \dd \tau_{p+1}\, \Delta_{2s}(\tau_{p+1})\nn\\
&= \frac{(2\pi i)^{p+1}}{p!} \int_\tau^{i\infty} \dd \tau_1\, (\tau{-}\tau_1)^p \Delta_{2s}(\tau_1)\,.
\end{align}
Since $\Delta_{2s}$ is a cusp form, this integral is well-defined for any $p\geq0$; however, in everything that follows we shall only encounter the usual range of values $0\leq p\leq 2s{-}2$. The cusp forms in this definition are Hecke normalised with $\Delta_{2s}(\tau)= q + O(q^2)$ such that the transcendentality of the iterated integral~\eqref{eq:EDelta} is given by $p{+}1$, just like for~\eqref{eq:E0depth1}.\footnote{The factor of $(-1)^p$ in the first line of~\eqref{eq:EDelta} is due to the insertion of $p$ copies of $\GG_0^0=-1$ in the iterated integral which is the meaning of the notation $0^p$~\cite{Broedel:2015hia, Broedel:2018izr}.}
The objects that are on a similar footing are $\Delta_{2s}$ and $\frac{\GG_k}{(2\pi i)^k}$ since both have algebraic Fourier coefficients, for instance
\begin{align}
\Delta_{12}(\tau) &= \frac{1}{1728} \bigg\{ \left( \frac{ {\rm G}_4(\tau) }{2 \zeta_4} \right)^3
-  \left( \frac{ {\rm G}_6(\tau) }{2 \zeta_6} \right)^2 \bigg\} =
q - 24 q^2 + 252 q^3 - 1472 q^4 +O(q^5)   \label{excuspforms} \\
\Delta_{16}(\tau) &= \frac{1}{1728} \frac{ {\rm G}_4(\tau) }{2 \zeta_4} \bigg\{ \left( \frac{ {\rm G}_4(\tau) }{2 \zeta_4} \right)^3
-  \left( \frac{ {\rm G}_6(\tau) }{2 \zeta_6} \right)^2 \bigg\} =
q + 216 q^2 - 3348 q^3 + 13888 q^4 + O(q^5) \notag \\
\Delta_{18}(\tau) &= \frac{1}{1728} \frac{ {\rm G}_6(\tau) }{2 \zeta_6} \bigg\{ \left( \frac{ {\rm G}_4(\tau) }{2 \zeta_4} \right)^3
-  \left( \frac{ {\rm G}_6(\tau) }{2 \zeta_6} \right)^2 \bigg\} =
q - 528 q^2 - 4284 q^3 + 147712 q^4 + O(q^5) \, .\notag 
\end{align}
If the cusp form has Fourier expansion $\Delta_{2s}(\tau) = \sum_{n=1}^\infty a(n) q^n$ then
\begin{align} 
\label{eq:IteratedDelta}
{\cal E}_0(\Delta_{2s},0^p;\tau)  = -\sum_{n=1}^\infty n^{-p-1} a(n) q^n\,.
\end{align}
While the seed functions of $\FFp{s}{m}{k}$ and $\FFm{s}{m}{k}$ in (\ref{eq:PSsol}) determined in Part I are essentially\footnote{The seed functions of $\FFp{s}{m}{k}$ 
constructed in Part I additionally involve $\QQ$-multiples of $y^{m+k}$ and $\zeta_{2m-1}y^{1-m+k}$.} 
constructed from real and imaginary parts of ${\cal E}_0(2m,0^p;\tau)$ in (\ref{eq:E0depth1}), respectively, 
Poincar\'e sums of ${\cal E}_0(\Delta_{2s},0^p;\tau) $ have been discussed in \cite{Diamantis:2020}.

\subsubsection{Real-analytic integrals of holomorphic cusp forms}

From the fourth line of~\eqref{eq:EII}, we see that one can define even and odd analogues of the 
non-holomorphic Eisenstein series by trading ${\rm G}_k$ in the integration kernel for
holomorphic cusp forms~$\Delta_{2s}$,
\begin{align}
\label{eq:defH}
\Hpm_{\Delta_{2s}} (\tau)  = (-1)^s \frac{\pi^{2s-1}i}{\Gamma(s)}   y^{1-s} \int_\tau^{i\infty} \dd\tau_1\, (\tau{-}\tau_1)^{s-1} (\bar\tau{-}\tau_1)^{s-1} {\Delta_{2s}}(\tau_1) \pm \cc\,,
\end{align}
where we have fixed a convenient normalisation.
This function satisfies from~\eqref{eq:CRII} and~\eqref{eq:LapII}
\begin{subequations}
\begin{align}
(\pi \nabla)^s \Hpm_{\Delta_{2s}} &= \frac12 (2\pi i)^{2s} (\Im\tau)^{2s} {\Delta_{2s}}(\tau)\,,\\
\big(\Delta -s(s{-}1)\big) \Hpm_{\Delta_{2s}} &=0\,.
\end{align}
\end{subequations}
Clearly, the even function $\Hp_{\Delta_{2s}}$ is obtained from the cusp form ${\Delta_{2s}}$ in the
same way as $\EE_s$ is obtained from $\GG_{2s}$. Moreover $\Hm_{\Delta_{2s}}$ is its odd cousin,
and the appearance of an odd analogue of $\EE_k$ (denoted by $\EE_k^{(-)}$) in the lower-depth 
terms of $\FFm{s}{m}{k}$ is discussed in section 5.5 of Part~I. Variants of (\ref{eq:defH}) with more
general exponents $(\tau{-}\tau_1)^{j} (\bar\tau{-}\tau_1)^{2s-2-j}, \ j=0,1,\ldots,2s{-}2$ arise
from Cauchy--Riemann derivatives of $\Hpm_{\Delta_{2s}}$ and have been studied in \cite{Diamantis:2020}.

Following our discussion around \eqref{eq:BesselLap} we expose the $q$-expansion of
 $\Hpm_{\Delta_{2s}}$ by rewriting them as a finite sum over the ${\cal E}_0({\Delta_{2s}},0^p;\tau)$ in
 (\ref{eq:EDelta}) and (\ref{eq:IteratedDelta}),
\begin{align}
\Hpm_{\Delta_{2s}} (\tau)  &= \frac12\sum_{\ell=0}^{s-1} (4y)^{-\ell}\frac{\Gamma(s{+}\ell)}{\ell! \Gamma(s{-}\ell)} \sum_{n=1}^{\infty}\frac{a(n)}{n^{s+\ell}} (q^n \pm \bar{q}^n) \label{eq:IterDeltaE0}\\
& = -\frac12  \sum_{\ell=0}^{s-1} (4y)^{-\ell} \frac{\Gamma(s{+}\ell)}{\ell! \Gamma(s{-}\ell)} \Big[{\cal E}_0({\Delta_{2s}},0^{s-1+\ell};\tau) \pm \cc\Big]\,, \notag
\end{align} 
which is the direct analogue of the second line in \eqref{eq:EII} for $\rm{E}_s$ and the definition
of $\rm{E}_s^{(-)}$ in section 5.5 of Part I. Given that the transcendental weight of  
iterated integrals ${\cal E}_0({\Delta_{2s}},0^{p};\tau)$ is $p{+}1$ in our conventions, their combinations 
with $y = \pi  \Im \tau$ of weight $1$ in (\ref{eq:IterDeltaE0}) assign transcendentality $s$ to $\Hpm_{\Delta_{2s}}$.

The sum over Fourier modes in the first line of (\ref{eq:IterDeltaE0}) can also be 
recast in terms of Bessel functions using~\eqref{eq:BesselLap} as
\begin{align}
\label{eq:HpmBessel}
\Hpm_{\Delta_{2s}}(\tau) =\sum_{n=1}^\infty a(n) n^{-s} \sqrt{\frac{ny}{\pi}}K_{s-1/2}(2 n y) \left(e^{2\pi i n \Re\tau} \pm e^{-2\pi i n \Re\tau} \right)\,.
\end{align}
Even though the functions $\Hpm_{\Delta_{2s}}$ and the modular invariant functions defined in~\cite{Brown:III} are both real-analytic and are both obtained from iterated integrals of holomorphic cusp forms $\Delta_{2s}$, they differ crucially in their modular properties. In particular, as we shall show next, the $\Hpm_{\Delta_{2s}}$ are not invariant under S-modular transformations.

\subsubsection{Modular properties}

For studying modular transformations of functions $F(\tau)$ on the upper half-plane, it is convenient to introduce the following cocycles under the generating T- and S-transformations of $\SLtwoZ$:
\begin{align}
\delta_{\rm T} F(\tau) = F(\tau) - F(\tau{+}1) \, , \ \ \ \ \ \ 
\delta_{\rm S} F(\tau) = F(\tau) - F(- \tfrac{1}{\tau})\,.
\label{modtrfs}
\end{align}
When both of them vanish, $F$ is invariant under modular transformations and in general there is a connection to the group cohomology of $\SLtwoZ$~\cite{Zagier:2000}.

From the explicit $q$-series in~\eqref{eq:IterDeltaE0}, it is clear that the functions $\Hpm_{\Delta_{2s}}$ have a vanishing T-cocycle
\begin{equation}
\delta_{\rm T} \Hpm_{\Delta_{2s}}(\tau) = \Hpm_{\Delta_{2s}}(\tau)-\Hpm_{\Delta_{2s}}(\tau{+}1)=0\,.
\end{equation}
However, the functions $\Hpm_{\Delta_{2s}}$ have a non-trivial cocycle under the S-transformation
\begin{align}
\label{eq:Scocy}
\delta_{\rm S} \Hpm_{\Delta_{2s}}(\tau) &= \Hpm_{\Delta_{2s}}(\tau) - \Hpm_{\Delta_{2s}}(-\tfrac1{\tau}) 
= \frac{(-i\pi)^{2s-1} }{\Gamma(s)} y^{1-s} \int_0^{i\infty} \!  \dd\tau_1 \,(\tau{-}\tau_1)^{s-1} (\bar\tau{-}\tau_1)^{s-1} {\Delta_{2s}}(\tau_1) \pm \cc\nn\\
&=  (-1)^{s+1} \frac{\pi^{2s-1} }{\Gamma(s)}   y^{1-s} \!\!\sum_{a,b=0}^{s-1}(- i)^{a+b} \,\binom{s{-}1}{a}\binom{s{-}1}{b} \,\tau^{s-1-a} \bar\tau^{s-1-b} \Lambda({\Delta_{2s}},a{+}b{+}1) \pm \text{c.c.}\nn\\
&=\left\{\begin{array}{cl} \! 2(-1)^{s+1} \frac{\pi^{2s-1}}{\Gamma(s)}y^{1-s} \!\!\sum\limits_{\substack{a,b=0\\a+b \in 2\mathbb{Z}}}^{s-1} (-1)^{(a+b)/2} \binom{s{-}1}{a}\binom{s{-}1}{b} \tau^{s-1-a} \bar\tau^{s-1-b} \Lambda({\Delta_{2s}},a{+}b{+}1) \\[7mm] 
\! 2(-1)^{s}\frac{i\pi^{2s-1}}{\Gamma(s)}y^{1-s} \!\!\!\!\sum\limits_{\substack{a,b=0\\a+b \in 2\mathbb{Z}+1}}^{s-1} (-1)^{(a+b-1)/2} \binom{s{-}1}{a}\binom{s{-}1}{b} \tau^{s-1-a} \bar\tau^{s-1-b} \Lambda({\Delta_{2s}},a{+}b{+}1) \, ,
\end{array}\right.
\end{align}
where in the last equality the upper line is for $\Hp_{\Delta_{2s}}$ and the lower line for $\Hm_{\Delta_{2s}}$. 
The function $\Lambda({\Delta_{2s}},t)$ appearing in the above expressions is real-valued for real $t$ and corresponds to the completed L-function of the cusp form ${\Delta_{2s}}(\tau)=\sum_{n=1}^\infty a(n) q^n$ of weight $2s$ defined by
\begin{align}
\Lambda({\Delta_{2s}},t) = (2\pi)^{-t} \Gamma(t) \sum_{n=1}^\infty \frac{a(n)}{n^t} = (-i)^{t} \int_0^{i\infty}  \dd\tau_1 \, \tau_1^{t-1} {\Delta_{2s}}(\tau_1)\,,
\end{align}
where the  sum converges absolutely for $\Re(t){>}s{+}\tfrac12$, using~\cite{Hecke:1937} and the improved growth of the Fourier coefficients of cusp forms following from~\cite{Deligne:1974}.\footnote{We thank Nils Matthes for correspondence on this point.} From its integral Mellin-form representation, it is well-known that the completed L-function enjoys an analytic continuation to 
the complex plane and satisfies the functional relation
\begin{align}
\label{eq:Lrefl}
\Lambda({\Delta_{2s}},t) = (-1)^s \Lambda({\Delta_{2s}}, 2s{-}t)\,.
\end{align}
The interval $t\in(0,2s)$ is called the critical strip, and (\ref{eq:Scocy}) implies that the failure of modularity of $\Hpm_{\Delta_{2s}}$ involves the completed L-function evaluated at integers inside the critical strip, with only odd integer arguments $t$ contributing to $\Hp_{\Delta_{2s}}$ and only even integers for $\Hm_{\Delta_{2s}}$.

\subsubsection{Integrals of Hecke normalised holomorphic cusp forms}

If ${\Delta_{2s}}$ is a normalised eigenform of all Hecke operators $T_n$ and of weight $2s$, i.e., 
\begin{align}
\big[T_n {\Delta_{2s}}\big] (\tau) =  n^{2s-1}\sum_{d|n} d^{-2s} \sum_{b=0}^{d-1} {\Delta_{2s}}\!\left(\frac{n\tau+bd}{d^2}\right) = a(n) {\Delta_{2s}}(\tau)\label{eq:HeckeEigen}
\end{align}
for all $n>0$, implying Hecke normalisation $a(1)=1$, then it is moreover known that the values $\Lambda({\Delta_{2s}},t)$ for all even $t$ inside the critical strip are related, as are all the values for odd 
$t$~\cite{Eichler:1957,Shimura:1959,Manin:1973}. The ratios between the even (or the odd) values must belong to the number-field extension of $\mathbb{Q}$ defined by the Fourier coefficients $\{ a(n), n \in \mathbb N\}$ of ${\Delta_{2s}}$. The first time a non-trivial extension arises is for cusp forms of weight $2s=24$ where there are two linearly independent cusp forms and the number field is $\mathbb{Q}(\sqrt{144169})$.
The non-trivial Galois automorphism of the number field exchanges the two independent Hecke eigenforms. 

Therefore, for (normalised) Hecke eigenforms ${\Delta_{2s}}$ of weight $2s$ we find that we can rearrange (\ref{eq:Scocy}) to
\begin{subequations}
\label{eq:Hnoninv}
\begin{align}
\delta_{\rm S} \Hp_{\Delta_{2s}}(\tau)  &=  \frac{2(-1)^{s+1}\pi^{2s-1}}{\Gamma(s) } y^{1-s}  \Lambda({\Delta_{2s}},2s{-}1)\!\! \sum\limits_{\substack{a,b=0\\a+b \in 2\mathbb{Z}}}^{s-1} \! (-1)^{\frac{a+b}{2}} \binom{s{-}1}{a}\binom{s{-}1}{b} \tau^{s-1-a} \bar\tau^{s-1-b} c^+_{a+b}\,,\\
\delta_{\rm S} \Hm_{\Delta_{2s}}(\tau)  &=\frac{  2(-1)^{s}i\pi^{2s-1}}{\Gamma(s)}  y^{1-s} \Lambda({\Delta_{2s}},2s{-}2) \!\! \! \sum\limits_{\substack{a,b=0\\a+b \in 2\mathbb{Z}+1}}^{s-1} \! \!\! (-1)^{\frac{a+b-1}{2}} \binom{s{-}1}{a}\binom{s{-}1}{b} \tau^{s-1-a} \bar\tau^{s-1-b} c^-_{a+b}\, ,
\end{align}
\end{subequations}
with coefficients $c^{\pm}_{\ell}$ from the number field associated with ${\Delta_{2s}}$ defined by
\beq
 \Lambda({\Delta_{2s}},\ell{+}1)  =  \left\{ \begin{array}{cl} 
c^+_{\ell}  \Lambda({\Delta_{2s}},2s{-}1)  &: \ \ell \ \te{even} \, ,\\
c^-_{\ell} \Lambda({\Delta_{2s}},2s{-}2)  &: \ \ell \ \te{odd}\, ,
 \end{array} \right.
 \label{defcpm}
\eeq
for $\ell= 0,1,\ldots,2s{-}2$. The polynomials arising in (\ref{eq:Hnoninv}) have also appeared in~\cite{Brown:mmv,Brown2019} and are related to period polynomials as will be also discussed 
in section~\ref{sec:4.1.1}.

We had argued above that $\Hpm_{\Delta_{2s}}$ should be assigned transcendentality $s$ and this 
is consistent with the fact that there is no transcendentality carried by L-values inside the critical 
strip $(0,2s)$ like $\Lambda({\Delta_{2s}},2s{-}1)$ and $\Lambda({\Delta_{2s}},2s{-}2)$. 
As we will see later on, the value $\Lambda({\Delta_{2s}},2s{+}m)$ has transcendentality $(m{+}1)$, so that just at the end of the critical strip we have $\Lambda({\Delta_{2s}},2s)$ of transcendentality one. 
This is analogous to the Riemann zeta function whose transcendental weight grows
in the same way from the upper end of its critical strip $(0,1)$. 
The transcendentality of the cocycles (\ref{eq:Hnoninv}) is therefore determined by the prefactor, including $y=\pi\Im\tau$, and is also given by $s$, consistent with that of $\Hpm_{\Delta_{2s}}$ itself.

Note that with our definition (\ref{eq:defH}) for the iterated integral of a cusp form, the S-cocycles in (\ref{eq:Scocy}) or (\ref{eq:Hnoninv}) are two-variable generalisations of the classic period polynomial associated to the same cusp form\cite{ZagierF}, with $\delta_{\rm S} \Hpm_{\Delta_{2s}}(\tau)$ playing the roles of the even/odd part of said period polynomial. Furthermore, along the same lines as Manin's original work \cite{Manin:1973}, we have that $\delta_{\rm S} \Hpm_{\Delta_{2s}}(\tau)$ satisfies the two cocycle conditions \cite{Zagier:2000}
\begin{subequations}
\begin{align}
\left[\delta_{\rm S} \Hpm_{\Delta_{2s}}(\tau) \right]\Big\vert_{1+{\rm S}} &\label{eq:CocyCond1}= 0\,,\\
\left[\delta_{\rm S} \Hpm_{\Delta_{2s}}(\tau) \right]\Big\vert_{1+{\rm U}+{\rm U}^2} &\label{eq:CocyCond2}= 0\,,
\end{align}
\end{subequations}
where $\vert$ denotes the $\SLtwoZ$-action on $\tau$ and ${\rm U} ={\rm T}{\rm S}$ is an order $3$ generator of $\SLtwoZ$.

When applied to (\ref{eq:Scocy}), the first cocycle condition (\ref{eq:CocyCond1}) is equivalent to the reflection formula~\eqref{eq:Lrefl}
for the even/odd values inside the critical strip.
Similarly the second condition (\ref{eq:CocyCond2}), together with the Hecke condition (\ref{eq:HeckeEigen}), is equivalent to the statement that all the ratios between the even (or odd) critical values must be in the number field generated by the Fourier coefficients.
In (\ref{eq:Hnoninv}) we chose to factorise out $\Lambda(\Delta_{2s},2s{-}1)$ and $ \Lambda(\Delta_{2s},2s{-}2)$, respectively, thus making a particular choice for what are usually called the (holomorphic) periods of the cusp form $\Delta_{2s}$, sometimes denoted by $\omega^\pm_{\Delta_{2s}}$ \cite{Brown2019}.

\subsubsection{Example with $s=6$}

As a concrete example we can study the cusp form of lowest weight $2s=12$, i.e.\ the Ramanujan cusp form ${\Delta_{2s}} = \Delta_{12}=\sum_{n=1}^{\infty}\tau(n)q^n$ with $\tau(1)=1$.
Since the vector space of cusp forms at weight $2s=12$ is one-dimensional we trivially have that $\Delta_{12}$ is a normalised Hecke eigenform and obviously the associated number field is simply $\mathbb{Q}$, i.e.\ $\tau(n)\in\mathbb{Q}$ for all $n>0$.
Following \cite{Manin:1973} we have the following number-field relations amongst the completed L-values, even and odd, inside the critical strip:
\begin{align}
\Lambda(\Delta_{12},6):\Lambda(\Delta_{12},8): \Lambda(\Delta_{12},10) & = \frac{5}{12}: \frac{25}{48}:1\,,\label{ratios12}\\
\Lambda(\Delta_{12},7) :  \Lambda(\Delta_{12},9 ) : \Lambda(\Delta_{12},11)  &= \frac{691}{2520}: \frac{691}{1620}:1
\, ,
\notag
\end{align} 
where the remaining values can be obtained via the reflection formula $\Lambda(\Delta_{12},12-t) = \Lambda(\Delta_{12},t)$.

If we compute the cocycles \eqref{eq:Hnoninv} defined above we obtain
\begin{subequations}
\begin{align}
\Hp_{\Delta_{12}}(\tau) - \Hp_{\Delta_{12}}(-\tfrac1{\tau})  &= \frac{ \pi^{11} \Lambda(\Delta_{12},11) (1{-}\tau \bar \tau) }{60 y^5} \bigg\{
1 +  \tau^4 \bar \tau^4 - 
 \frac{691 }{162} (  \tau^2 {+} \bar \tau^2) (1 {+}  \tau^2 \bar \tau^2)  \label{eq:D12Pcocycle} \\
 & \ \ \ \ + 
\frac{ 691 }{504} (\tau^4 {+} \bar \tau^4) - \frac{3131 \tau  \bar \tau}{ 324} (1 {+} \tau^2 \bar \tau^2 ) 
+ \frac{21421  \tau \bar \tau}{  2268} (  \tau^2 {+} \bar \tau^2 ) + \frac{40273  \tau^2 \bar \tau^2}{2268}
  \bigg\}
\notag \\
\Hm_{\Delta_{12}}(\tau) - \Hm_{\Delta_{12}}(-\tfrac1{\tau})  &= \frac{ i \pi^{11}  \Lambda(\Delta_{12},10) (\tau{+}\bar \tau)}{12 y^5} \bigg\{
1 +  \tau^4 \bar \tau^4 - \frac{25}{24} (  \tau^2{+} \bar \tau^2) (1 {+} \tau^2 \bar \tau^2 )  \label{eq:D12Mcocycle} \\
 & \ \ \ \ +  \frac{1}{12} (\tau^4 {+} \bar \tau^4) 
 - \frac{25 }{6}  \tau \bar \tau (1 {+}  \tau^2 \bar \tau^2) + 
 2  \tau \bar \tau ( \tau^2 {+} \bar \tau^2) + \frac{19  \tau^2 \bar \tau^2}{3} \bigg\} \, . \notag 
\end{align}
\end{subequations}
Similar expressions for the cocycles of $\Hpm_{\Delta_{16}}$ and $\Hpm_{\Delta_{18}}$ can be
found in appendix \ref{app:morecocyc}, while in the ancillary file the expressions are given up to modular weight $2s=26$. 

In summary, we can construct even and odd solutions $\Hpm_{\Delta_{2s}}$ of the homogeneous Laplace equation~\eqref{eq:Laps} whenever we have a holomorphic cusp form ${\Delta_{2s}}$ of weight $2s$. These homogeneous solutions are expressible through iterated integrals of ${\Delta_{2s}}$ of depth one, see (\ref{eq:defH}). They are not modular invariant but their failure of modularity is characterised by a single number that is a value of the completed L-function inside the critical strip. This number is expected to be independent over $\mathbb{Q}$ from the set of multiple zeta values.

%%%%%%%
\subsection{\texorpdfstring{Properties of multiple modular values and the $\bsv$}{Properties of multiple modular values and the betasv}}
\label{sec:mmvS}
%%%%%%%

We now study the multiple modular values defined in~\eqref{eq:defmmvs} in more detail and also present expressions for the S-modular transformation of the $\bsv$ introduced in~\eqref{eq:beta}.

From their definition~\eqref{eq:defmmvs} the multiple modular values inherit the shuffle relations of the
(regularised) iterated integrals in (\ref{eq:defEE})
\begin{align}
\MM{j_1}{k_1} \MM{j_2}{k_2} = \MM{j_1&j_2}{k_1&k_2}  + \MM{j_2&j_1}{k_2&k_1} \,.
\label{mmvshuffle}
\end{align}
By applying another S-transformation to~\eqref{eq:EESmod} one can show the reflection properties
\begin{subequations}
\label{reflectMMV}
\begin{align}
\MM{j}{k} &= - (-1)^j \MM{k-2-j}{k}\,,\\
\MM{j_1&j_2}{k_1&k_2} &= -(-1)^{j_1+j_2} \MM{k_1-2-j_1&k_2-2-j_2}{k_1&k_2} + \MM{j_1}{k_1}\MM{j_2}{k_2} \notag \\
&= (-1)^{j_1+j_2} \MM{k_2-2-j_2 &k_1-2-j_1}{k_2&k_1}
\,.
\end{align}
\end{subequations}
Under complex conjugation they satisfy
\begin{align}
\label{eq:MMVcc}
\overline{\MM{j}{k}} = (-1)^{j+1} \MM{j}{k}\,,\quad
\overline{\MM{j_1&j_2}{k_1&k_2}} = (-1)^{j_1+j_2+2} \MM{j_1&j_2}{k_1&k_2}\,.
\end{align}

The transcendentality of the multiple modular values as defined in~\eqref{eq:defmmvs} is given by $\sum_i k_i$. For the depth-one case~\eqref{eq:MMV1} this is evident from the fact that $(2\pi i)^{k-j-1}$ has transcendentality $k{-}j{-}1$, $\zeta_{j+1}$ has transcendentality $j{+}1$  and $\zeta_{j+2-k}$ has transcendentality zero.\footnote{The last statement is true since $j\leq k{-}2$ and zeta values at non-positive integers are either zero or given by rational numbers (expressible through Bernoulli numbers).}
In the general case, this follows from the definition~\eqref{eq:defEE} by realising that $\GG_k$ in our convention has transcendentality $k$, see~\eqref{eq:GG}. The iterated integrals~\eqref{eq:E0depth1} and \eqref{eq:E0depth2} therefore have transcendentality $p{+}1$ and $p_1{+}p_2{+}2$, respectively.

\subsubsection{Reduced multiple modular values}

While the multiple modular values~\eqref{eq:defmmvs} appear in the S-transformation of the holomorphic iterated integrals~\eqref{eq:defEE}, the S-transformation of the $\bsv$ only contains the following specific combinations of them:
\begin{subequations}
\label{svmmod} 
\begin{align}
\MMsv{j}{k} &= \MM{j}{k} - \overline{\MM{j}{k} } 
= \left(1+(-1)^j\right) \MM{j}{k}\,,\\
\MMsv{j_1&j_2}{k_1&k_2} &= 
\MM{j_2&j_1}{k_2&k_1} - \MM{j_2}{k_2} \overline{ \MM{j_1}{k_1} }
+ \overline{ \MM{j_1&j_2}{k_1&k_2} }\nn\\
&= (-1)^{j_1}\left(1+(-1)^{j_2}\right) \MM{j_1&j_2}{k_1&k_2} 
+ \left(1+(-1)^{j_1}\right) \MM{j_2&j_1}{k_2&k_1}\nn\\
&= \left((-1)^{j_1+j_2} -1\right) \MM{j_1&j_2}{k_1&k_2}  + \left(1+(-1)^{j_1}\right) \MM{j_1}{k_1}\MM{j_2}{k_2}\,.
\end{align}
\end{subequations}
We refer to the combinations $\textMMsv{j}{k}$ and $\textMMsv{j_1&j_2}{k_1&k_2}$ as {\it reduced multiple modular values} of depth one and two, respectively. The simplifications at depth two are based on \eqref{mmvshuffle} and \eqref{eq:MMVcc}, and they show that reduced multiple modular values vanish if both  of $j_1$ and $j_2$ are odd, whereas cases with both $j_1$ and $j_2$
even yield the product $2 \textMM{j_1}{k_1}\textMM{j_2}{k_2}$. Moreover, reduced multiple modular values inherit the shuffle property and so satisfy
\beq
\MMsv{j_1}{k_1} \MMsv{j_2}{k_2} = \MMsv{j_1&j_2}{k_1&k_2} + \MMsv{j_2&j_1}{k_2&k_1}\,.
\eeq

\subsubsection{Depth one reduced multiple modular values and $\bsv$ modular transformations}

At depth one, \eqref{eq:MMV1} leads to the following explicit expressions
\beq
\label{eq:svMMV1}
\MMsv{j}{k} = \left\{\begin{array}{cl}
\displaystyle - \frac{4\pi i \zeta_{k-1} }{k{-}1} & \, j = 0 \,, \\[2mm]
\Bigg. \displaystyle + \frac{4\pi i \zeta_{k-1} }{k{-}1} \Bigg.& \, j = k{-}2\,, \\[2mm]
0& \, {\rm otherwise}\,, 
\end{array}\right.
\eeq
where the vanishing of all cases with $j=1,2,\ldots,k{-}3$ is in agreement with their 
occurrence in certain coboundary polynomials~\cite{Brown:mmv, Brown:II, Brown2019}.
They appear in the transformation of the depth-one $\bsv$ according to
\begin{align}
\bsvBR{j }{k}{-\tfrac{1}{\tau}} &= \bar \tau^{k-2-2j}  \bsvBR{j }{k}{\tau} 
- \frac{ (\tau \bar \tau)^{k-2-j}  }{2\pi i (4y)^{k-2-j}} \nn
\\
&\ \ \times \sum_{A=0}^{k-2-j} \sum_{B=0}^{j} 
\binom{k{-}2{-}j}{A}  \binom{j}{B} \bigg( {-}\frac{1}{\tau}\bigg)^{A}
 \bigg( {-}\frac{1}{\bar \tau}\bigg)^{B}\MMsv{A+B}{k} \label{eq:Sbsv1}
 \\
&=\bar \tau^{k-2-2j}  \bsvBR{j }{k}{\tau} 
+ \frac{2 \zeta_{k-1}}{(k{-}1) (4y)^{k-2-j}}  \Big\{   (\tau \bar \tau)^{k-2-j} - \bar\tau^{k-2-2j} \Big\}\,. \nn
\end{align}

\subsubsection{Modular transformation of $\beta^{\rm sv}$ at depth two}

In the same way the depth-one reduced multiple modular values arise in the modular transformations of the depth-one $\bsv$, the depth-two $\textMMsv{j_1&j_2}{k_1&k_2}$ arise in the modular transformation of the $\bsv$ at depth two.
Performing the calculation
based on the integral representation~\eqref{eq:bsv2} one can show that
\begin{align}
\label{eq:Sbsv2}
\bsvBR{j_1 &j_2}{k_1 &k_2}{-\tfrac{1}{\tau}} &=  \bar \tau^{k_1+k_2-4-2j_1-2j_2} \bsvBR{j_1 &j_2}{k_1 &k_2}{\tau}  \notag \\
&\quad + \bar \tau^{k_2-2- 2 j_2}  \bsvBR{j_2 }{k_2}{\tau}  \frac{ 2 \zeta_{k_1-1} \Big\{
(\tau \bar \tau)^{k_1-2-j_1} -\bar\tau^{k_1-2-2j_1} 
\Big\}}{(k_1{-}1) (4y)^{k_1-2-j_1}} 
\notag \\
&\quad+ \frac{ (\tau \bar \tau)^{k_1+k_2-4-j_1-j_2} }{(2\pi i)^2 (4y)^{k_1+k_2-4-j_1-j_2}}  \sum_{A_1=0}^{k_1-2-j_1} \sum_{A_2=0}^{k_2-2-j_2} \sum_{B_1=0}^{j_1} \sum_{B_2=0}^{j_2} 
\binom{k_1{-}2{-}j_1}{A_1} \binom{k_2{-}2{-}j_2}{A_2} \notag \\
&\quad \ \ \ \ \times  \binom{j_1}{ B_1}
\binom{j_2}{B_2} \bigg( {-}\frac{1}{\tau}\bigg)^{A_1+A_2}
 \bigg( {-}\frac{1}{\bar \tau}\bigg)^{B_1+B_2}\MMsv{A_1+B_1 &A_2+B_2}{k_1 &k_2}
\notag \\
&\quad + \frac{  \bar \tau^{k_1+k_2-4-2j_1-2j_2} }{(2\pi i)^2 (4y)^{k_1+k_2-4-j_1-j_2} }
C\left[ \smallmatrix j_1 &j_2 \\ k_1 &k_2 \endsmallmatrix;\tau \right]\,.
\end{align}
Here, the $C[\cdots]$ are pure depth-zero terms, i.e.\ rational functions of $\tau$ and $\bar \tau$ multiplied by rational combinations of odd zeta values and powers of $\pi$, that can be traced back to the modular transformation of the $\overline{\alpha[\cdots]}$. Their definition is most conveniently given in terms of the shorthand
\beq
\betasvtaualpha{j_1 &j_2 \\ k_1 &k_2} = 
\sum_{p_1=0}^{k_1{-}2{-}j_1} \sum_{p_2=0}^{k_2{-}2{-}j_2} \frac{\binom{k_1{-}2{-}j_1}{p_1}\binom{k_2{-}2{-}j_2}{p_2}}{(4y)^{p_1+p_2}} \overline{\alphaBR{j_1 +p_1&j_2+p_2}{k_1 &k_2}{\tau}}
\eeq
for the contributions of the antiholomorphic $\overline{\alpha[\cdots]}$ to (\ref{eq:bsv2}), 
namely
\begin{align}
C\left[ \smallmatrix j_1 &j_2 \\ k_1 &k_2 \endsmallmatrix;\tau \right] &= 
\frac{ (2\pi i)^2 (4y)^{k_1+k_2-4-j_1-j_2} }{   \bar \tau^{k_1+k_2-4-2j_1-2j_2} }
\betasvStaualpha{j_1 &j_2 \\ k_1 &k_2} -
 (2\pi i)^2 (4y)^{k_1+k_2-4-j_1-j_2}  \betasvtaualpha{j_1 &j_2 \\ k_1 &k_2} \notag\\
& \ \ \ \ +2\pi i \, \bigg\{
\frac{2 \zeta_{k_2-1} }{k_2{-}1}(1- \tau^{k_2-j_2-2} \bar \tau^{j_2})  
\int\limits_{\bar\tau}^{-i\infty}  \dd\bar\tau_1 (\tau{-}\bar\tau_1)^{k_1-2-j_1} (\bar\tau{-}\bar\tau_1)^{j_1} \overline{\GG_{k_1}(\tau_1)} \label{defccs}  \\
&\ \ \ \ \ \ \ \ \ \ \ \, -
\frac{2 \zeta_{k_1-1} }{k_1{-}1}(1- \tau^{k_1-j_1-2} \bar \tau^{j_1})  
\int\limits_{\bar\tau}^{-i\infty}  \dd\bar\tau_1 (\tau{-}\bar\tau_1)^{k_2-2-j_2} (\bar\tau{-}\bar\tau_1)^{j_2} \overline{\GG_{k_2}(\tau_1)} 
\bigg\}\, . \notag 
\end{align}
Then, the known expressions for $\overline{\alpha[\cdots]}$ \cite{Gerken:2020yii, PartI} and the
modular transformations of the ${\cal E}_0(k,0^p)$ (see for instance section \ref{sec:trfe0})
lead to representative examples such as\footnote{Here, we have chosen to replace $\tau$ by $\bar\tau+ \frac{2i}{\pi} y$ for slightly  more compact expressions.}
{\allowdisplaybreaks
\begin{align}
C\left[ \smallmatrix 0& 1\\4& 4 \endsmallmatrix \right]  &=
 \frac{32 \bar \tau^2 \pi^4 y \zeta_3 }{405}   -\frac{4 \pi^4 y \zeta_3}{135}  + 
 \frac{128 i \bar \tau \pi^3 y^2 \zeta_3}{405}  - \frac{16 \pi^2 y^3 \zeta_3}{81}  - 
 \frac{16 \pi^2 y^3 \zeta_3}{135 \bar \tau^2} + 
 \frac{16 i \bar \tau \pi y \zeta_3^2}{9}  - \frac{32y^2 \zeta_3^2}{9} 
\,,\notag\\
C\left[ \smallmatrix 0&2\\4& 4 \endsmallmatrix \right]  &= -\frac{8 \pi^4 y \zeta_3}{135}  +\frac{ 64 \bar\tau^2 \pi^4 y \zeta_3}{405}  + 
 \frac{128 i \bar \tau \pi^3 y^2 \zeta_3}{405}  + 
 \frac{32 i \bar \tau \pi y \zeta_3^2}{9}  - \frac{32 y^2 \zeta_3^2}{9} \,,
\\
C\left[ \smallmatrix 1 &2 \\ 4& 4 \endsmallmatrix \right] & = -\frac{4 \pi^4 y \zeta_3}{135}  + \frac{32 \bar \tau^2 \pi^4 y \zeta_3}{405}  + 
 \frac{16  i \bar \tau \pi y \zeta_3^2}{9}
\,,\notag \\
C\left[ \smallmatrix 4 & 2 \\6& 4 \endsmallmatrix \right] &= 
 \frac{ 4 i \bar \tau^3 \pi^7 \zeta_3}{2025}
 - \frac{8 i \bar \tau \pi^7 \zeta_3}{14175}  
 + \frac{ 4 i \bar \tau^5 \pi^7 \zeta_3}{2835} + 
 \frac{8  i \bar \tau^3 \pi^5 \zeta_5}{675} -  \frac{4  i  \bar \tau^5 \pi^5 \zeta_5}{135} - 
 \frac{8 \bar \tau^2 \pi^2 \zeta_3 \zeta_5}{15}  +  \frac{8 \bar \tau^4 \pi^2 \zeta_3 \zeta_5}{15} \,.
 \notag
\end{align} }%
While the antisymmetry of $\beta^{\rm sv}_{(\alpha)}$ in $(j_1,k_1) \leftrightarrow (j_2,k_2)$ clearly propagates to
\beq
C\left[ \smallmatrix j_1 &j_2 \\ k_1 &k_2 \endsmallmatrix \right] = - C\left[ \smallmatrix j_2 &j_1 \\ k_2 &k_1 \endsmallmatrix \right]\,,
\eeq
it is not immediately obvious from the definition (\ref{defccs}) that the depth-one terms cancel. In fact, one may view
the dropout of $\overline{{\cal E}_0(k,0^p)}$ from (\ref{defccs}) as a defining property of $\overline{\alpha[\cdots]}$.

\subsubsection{Examples at depth two expressible via zeta values}

For reduced multiple modular values at depth two, no analogue of the closed formula~\eqref{eq:svMMV1} is known. We begin with a few illustrative examples. In the $(\GG_4,\GG_4)$ sector we have~\cite{Brown2019}
\begin{align}
\MMsv{0& 0}{4& 4} &= -\frac{8\pi^2 \zeta_3^2}{9}  \, ,
&\MMsv{0& 1}{4& 4} &= - \frac{ 16i  \pi^5 \zeta_3}{405} + \frac{ 10i \pi^3 \zeta_5}{27 } \, ,
&\MMsv{1& 0}{4& 4} &= \frac{ 16 i \pi^5 \zeta_3}{405}  -  \frac{10 i  \pi^3 \zeta_5}{27} \, ,\notag \\
\MMsv{1& 1}{4& 4} &= 0 \, ,
&\MMsv{0& 2}{4& 4} &= \frac{8\pi^2 \zeta_3^2}{9}  \, ,
&\MMsv{2& 0}{4& 4} &= \frac{8\pi^2 \zeta_3^2}{9} \, ,
\\
\MMsv{2& 2}{4& 4} &= -\frac{8\pi^2 \zeta_3^2}{9}  \, ,
&\MMsv{1& 2}{4& 4} &= \frac{2i \pi^5 \zeta_3}{135}   - \frac{10 i \pi^3 \zeta_5}{27}  \, ,
&\MMsv{2& 1}{4& 4} &= -\frac{ 2 i \pi^5 \zeta_3}{135}  +  \frac{10i \pi^3 \zeta_5}{27 }  
\,. \notag
\end{align}
We note that all rational multiples of $\pi^8$ present in the 
individual terms such as~\cite{Brown2019}
\begin{align}
\MM{0& 2}{4& 4} = \frac{209 \pi^8}{364500}\,,\quad
\MM{2& 0}{4& 4} = -\frac{209 \pi^8}{364500} +\frac{4  \pi^2 \zeta_3^2}{9}\,,
\end{align}
disappear in the combination~\eqref{svmmod}.

More remarkable instances of such simplifications occur for higher weight, for instance in the
reduced multiple modular values of the $({\rm G}_4,{\rm G}_6)$ sector~\cite{Brown2019}
\begin{align}
\MMsv{0& 0}{4& 6} &= -\frac{8 \pi^2 \zeta_3 \zeta_5}{15}  \,,
&\MMsv{0& 1}{4& 6} &= -\frac{i  \pi^7 \zeta_3}{1575} +  \frac{7i  \pi^3 \zeta_7}{90} \,, \notag \\
\MMsv{0& 2}{4& 6} &=0 \,, 
&\MMsv{0& 3}{4& 6} &=  \frac{ i \pi^7 \zeta_3}{4725}  -  \frac{4 i \pi^5 \zeta_5}{675}   \,,\notag \\
\MMsv{0& 4}{4& 6} &= \frac{ 8 \pi^2 \zeta_3 \zeta_5}{15} \,, 
&\MMsv{1& 0}{4& 6} &=  \frac{8 i  \pi^7 \zeta_3}{14175} + \frac{ 2i \pi^5 \zeta_5}{135 }  - 
\frac{ 7i \pi^3 \zeta_7}{45 } \,, \notag \\
\MMsv{1& 1}{4& 6} &=0 \,,
&\MMsv{1& 2}{4& 6} &= - \frac{ 4 i \pi^7 \zeta_3}{8505}    + \frac{ 2i \pi^5 \zeta_5}{675 } \,, \\
\MMsv{1& 3}{4& 6} &= 0 \,,
&\MMsv{1& 4}{4& 6} &=  \frac{ 8 i \pi^7 \zeta_3}{14175}   - \frac{ 7i\pi^3 \zeta_7 }{45 } \,, \notag \\
\MMsv{2& 0}{4& 6} &=  \frac{8 \pi^2 \zeta_3 \zeta_5}{15} \,,
&\MMsv{2& 1}{4& 6} &=  \frac{ 2i  \pi^7 \zeta_3}{2835}  -  \frac{4i \pi^5 \zeta_5}{675}  \,,\notag \\
\MMsv{2& 2}{4& 6} &=0 \,,
&\MMsv{2& 3}{4& 6} &= -\frac{ 2 i \pi^7 \zeta_3}{14175}  + \frac{ 7i  \pi^3 \zeta_7}{90 }\,, \notag \\
\MMsv{2& 4}{4& 6} &= - \frac{8\pi^2 \zeta_3 \zeta_5}{15} \,. \notag
\end{align}
Rational multiples of $\pi^{10}$ and $ \pi^2 \zeta_{3, 5}$ drop out from all the reduced counterparts $\textMMsv{j_1& j_2}{4& 6} $ even though they appear in individual multiple modular values such as~\cite{Brown2019}
\begin{align}
\MM{0 &0}{4 &6} &= -\frac{503 \pi^{10}}{25515000} +   \frac{4\pi^2 }{75} \zeta_{3, 5} \,,\nn\\
\MM{0 &0}{6 &4} &= \frac{503 \pi^{10}}{25515000} -   \frac{4\pi^2 }{75} \zeta_{3, 5} - \frac{ 4 \pi^2}{15}  \zeta_3 \zeta_5\, .
\end{align}
We expect that more generally, the double zeta values $\zeta_{n_1,n_2}$ present in individual 
$\textMM{j_1 &j_2}{k_1 &k_2}$ \cite{Saad:2020mzv} will drop out in the combination to their reduced 
counterparts at arbitrary weight. 

\subsubsection{Examples at depth two involving L-values}

Individual $\textMM{j_1 &j_2}{k_1 &k_2}$ at weight $k_1{+}k_2\geq 14$
involve certain ``new numbers'' \cite{Brown2019} such as $c(\Delta;12)$ and L-values of holomorphic cusp forms outside the critical strip. However, the reduced combinations~\eqref{svmmod} are conjectured to feature only single zeta values, L-values of cusp forms and powers of $\pi$. This can be checked from the 
$\textMMsv{j_1& j_2}{k_1& k_2}$ provided in the ancillary file up to $k_1{+}k_2\leq 28$
and the examples presented in this work.

The simplest examples of reduced multiple modular values involving non-critical L-values
occur in the $({\rm G}_4, {\rm G}_{10})$ and $({\rm G}_6, {\rm G}_{8})$ sectors 
\cite{Brown2019},\footnote{We are indebted to Francis Brown for correspondence on his work~\cite{FBpriv} and making many explicit expressions available to us, such as~\eqref{410sec},~\eqref{68sec} and the first three lines of~\eqref{alsoodd}.}
\begin{align}
\MMsv{0& 1}{4& 10}&= - \frac{ 43 i \pi^{11} \zeta_3}{25259850} + 
   \frac{11 i \pi^3 \zeta_{11}}{540}  +  \frac{  256 i \pi^{13} \Lambda(\Delta_{12},12)}{1913625}\,,
\notag \\
\MMsv{0& 3}{4& 10} &=   \frac{17 i \pi^{11} \zeta_3}{265228425} 
-  \frac{  i \pi^5 \zeta_9}{17010} -  \frac{  16 i \pi^{13} \Lambda(\Delta_{12},12)}{229635}\,,
\notag \\
\MMsv{1& 0}{4& 10} &=  \frac{8 i \pi^{11} \zeta_3}{2525985} + 
   \frac{2 i \pi^5 \zeta_9}{243}  -  \frac{11 i \pi^3 \zeta_{11}}{135}  + 
   \frac{256 i \pi^{13} \Lambda(\Delta_{12},12)}{1913625}\,,
   \label{410sec}\\
\MMsv{1& 2}{4& 10} &= -  \frac{4 i \pi^{11} \zeta_3}{10609137} 
+  \frac{  i \pi^5 \zeta_9}{5670} -  \frac{  16 i \pi^{13} \Lambda(\Delta_{12},12)}{229635}\,,
\notag\\
\MMsv{1& 4}{4& 10} &= - \frac{4 i \pi^{11} \zeta_3}{37889775} + 
 \frac{  64 i \pi^{13} \Lambda(\Delta_{12},12)}{1148175}\,,
 \notag
\end{align}
as well as
\begin{align}
\MMsv{0& 1}{6& 8} &= - \frac{4 i \pi^9 \zeta_5}{297675} + 
 \frac{ 22 i \pi^3 \zeta_{11}}{1575} - 
 \frac{ 256 i \pi^{13} \Lambda(\Delta_{12},12)}{826875}\,,
 \notag \\
\MMsv{0& 3}{6& 8} &= - \frac{2 i \pi^9 \zeta_5}{496125} + 
 \frac{ i \pi^5 \zeta_9}{2250} + \frac{
  16 i \pi^{13} \Lambda(\Delta_{12},12)}{99225}\,,
  \notag \\
\MMsv{1& 0}{6& 8} &= \frac{i \pi^7 \zeta_7}{4725} - 
  \frac{11 i \pi^3 \zeta_{11}}{525}  - 
  \frac{256 i \pi^{13} \Lambda(\Delta_{12},12)}{826875}\,,
 \label{68sec} \\
\MMsv{1& 2}{6& 8} &=  \frac{ 4 i \pi^9 \zeta_5}{2480625} - 
 \frac{ i \pi^5 \zeta_9}{2250} + 
 \frac{ 16 i \pi^{13} \Lambda(\Delta_{12},12)}{99225}\,,
\notag \\
\MMsv{2& 1}{6& 8} &= - \frac{2 i \pi^9 \zeta_5}{496125} 
+ \frac{  i \pi^5 \zeta_9}{1350} + 
  \frac{16 i \pi^{13} \Lambda(\Delta_{12},12)}{99225}\,,
\notag \\
\MMsv{2& 3}{6& 8} &= \frac{i \pi^9 \zeta_5}{165375} - 
 \frac{ 4 i \pi^7 \zeta_7}{99225} - 
 \frac{ 64 i \pi^{13} \Lambda(\Delta_{12},12)}{496125}
\notag  \, .
\end{align}
All other cases with $(k_1,k_2) = (4,10)$ or $(6,8)$ are determined from
(\ref{svmmod}), the depth-one results (\ref{eq:MMV1}) and the reflection properties
(\ref{reflectMMV}). Starting from $k_1{+}k_2=16$, we also find L-values
$\Lambda(\Delta_{2s},t)$ at odd $t$ such as
\begin{align}
\MMsv{1 &6}{4 &12} &= -\frac{2764 i \pi^{13} \zeta_3}{   1005657778125}
+ \frac{64 i \pi^{14} \Lambda(\Delta_{12}, 13)}{  245581875}\,,
\notag \\
\MMsv{3&4}{6&10} &= - \frac{2i\pi^{11} \zeta_5}{21049875} +  \frac{i\pi^7\zeta_9}{297675} +  \frac{8i\pi^{14}\Lambda(\Delta_{12},13)}{2679075} \,,\notag \\
\MMsv{2 &5}{8 &8} &=  - \frac{2 i \pi^9 \zeta_7}{496125} 
+ \frac{ 4 i \pi^7 \zeta_9}{83349} + \frac{  32 i \pi^{14} \Lambda(\Delta_{12},13)}{3472875}\,,
 \label{alsoodd}  \\
\MMsv{1&6}{10&10} &=
  \frac{i\pi^{11} \zeta_9}{5893965} -  \frac{i\pi^9 \zeta_{11}}{612360} -  \frac{4i\pi^{16} \Lambda(\Delta_{12},15)}{104483925} - \frac{8i \pi^{18}\Lambda(\Delta_{16},17)}{1316497455}\,, \notag\\
\MMsv{0& 1}{4& 16} &= - \frac{31034 i \pi^{17} \zeta_3}{51288546684375} 
+ \frac{680 i \pi^3 \zeta_{17}}{97659} 
- \frac{  256 i \pi^{18} \Lambda(\Delta_{16}, 17)}{138175277625} 
- \frac{ 512 i \pi^{19} \Lambda(\Delta_{18}, 18)}{28733079375}\, .
  \notag
\end{align}
In the ancillary file accompanying the arXiv submission and journal publication of this work and Part~I, we present the complete list of reduced multiple modular values at depth two up to $k_1{+}k_2=28$. 
The values there were fixed by knowing on which numbers to expand the reduced multiple modular value~\cite{Brown2019} and fitting the rational coefficients via numerical evaluation.\footnote{Numerical approximations of the L-values can be obtained efficiently using PARI/GP~\cite{PARI2,computeL}. In the ancillary file, we have collected the numerical values of relevance to this work, meaning up to modular weight $26$ and the first integer values outside critical strip.}
Since we assign transcendental weight $m{+}1$ to $\Lambda(\Delta_{2s},2s{+}m)$, the explicit expressions are compatible with the transcendentality $k_1{+}k_2$ of $\textMMsv{j_1&j_2}{k_1&k_2}$.

%%%%%%
\section{Modular properties of solutions to the Laplace equations}
\label{sec:3}
%%%%%%

The problem that motivates this work is to find modular invariant solutions $\FFpm{s}{m}{k}$
to the Laplace equations~\eqref{eq:Laps}. In Part~I, we showed how a leading-depth solution could be constructed in terms of the $\bsv$ and how to complete it by including lower-depth $\bsv$ terms. We already pointed out in Part~I that the modular invariance of the resulting function $\cFFpm{s}{m}{k}$ is not guaranteed by the construction that was only tailored to produce an exact solution of~\eqref{eq:Laps} in terms of various~$\bsv$. Since the $\bsv$ have the more involved modular properties presented above, this does not necessarily entail modular invariance of $\cFFpm{s}{m}{k}$.

As we argued, failure of modular invariance can and will arise whenever the space of $\FFpm{s}{m}{k}$ is larger than the space of modular graph forms constructed  from $\bsv$ at depth $\leq 2$. The explicit counting done in section 3.6 of Part~I showed that this can happen at Laplace eigenvalue $s(s{-}1)$ whenever there are holomorphic cusp forms at modular weight $2s$. This is not surprising since the generating series of MGFs~\cite{Gerken:2020yii} only contains combinations of $\bsv$ that are compatible with the relations in Tsunogai's derivation algebra, see section~\ref{sec:select} for a more detailed discussion of the corresponding `dropouts' from the $\bsv$. 
As the relations in the derivation algebra are triggered by holomorphic cusp forms~\cite{Pollack} we have a consistent picture that iterated integrals of such cusp forms should arise. They also feature naturally in the space of real-analytic modular functions studied in~\cite{Brown:I,Brown:II,Brown:III,Diamantis:2020,Drewitt:2021}.

For every holomorphic cusp form ${\Delta_{2s}}$ of modular weight $2s$, we have constructed even and odd homogeneous solutions $\Hpm_{\Delta_{2s}}$ to the Laplace equation in~\eqref{eq:defH} and we have also shown that they are not modular invariant, see~\eqref{eq:Hnoninv}. Therefore, if the combination $\cFFpm{s}{m}{k}$ of $\bsv$ is not modular invariant but solves the correct inhomogeneous Laplace equation, we can consider 
\begin{align}
\label{eq:GenLinear}
\FFpm{s}{m}{k}(\tau) =\cFFpm{s}{m}{k}(\tau) + \sum_{\Delta_{2s}\in \mathcal{S}_{2s}} a_{\Delta_{2s},m,k}^{\pm} \Hpm_{{\Delta_{2s}}}(\tau)\,,
\end{align}
where the sum runs over the space $\mathcal{S}_{2s}$ of holomorphic cusp forms of weight $2s$. Then we can ask whether a suitable choice of constants $a_{\Delta_{2s},m,k}^{\pm}$  renders this new solution $\FFpm{s}{m}{k}$ to the same Laplace equation modular invariant. 

To answer this question we have to determine the modular transformation of $\cFFpm{s}{m}{k}$. As this is a combination of $\bsv$, potentially multiplied by powers of $y$, we have to use the S-modular transformation of the $\bsv$ discussed in section~\ref{sec:mmvS}. 
As is evident from~\eqref{eq:Sbsv1} and~\eqref{eq:Sbsv2}, the modular transformation generates special combinations of multiple modular values and additional depth-zero terms $C[\cdots]$, which can be derived from the relevant $\overline{\alpha[\cdots]}$ appearing in~\eqref{eq:bsv2}. Ultimately, we obtain the explicit S-modular transformation of $\cFFpm{s}{m}{k}$ in terms of these multiple modular values. In order for a failure of modularity of $\cFFpm{s}{m}{k}$ to be cancelled by that of $\Hpm_{\Delta_{2s}}$, one must obtain very specific combinations of multiple modular values that ultimately are proportional to the same polynomial in $\tau$ and $\bar\tau$ given in~\eqref{eq:Hnoninv}. This is true in all examples and should follow from the general analysis in~\cite{Brown:mmv, Brown:I}. We exemplify the mechanism for a variety of weights and Laplace eigenvalues.
In all cases, their construction from L-values assigns transcendental weight $m{+}k{-}s$ to the constants $a_{\Delta_{2s},m,k}^{\pm}$ in (\ref{eq:GenLinear}) such that their combination with $\Hpm_{{\Delta_{2s}}}$ of weight $s$ matches the transcendentality 
$m{+}k$ of $\FFpm{s}{m}{k}$.

The occurrence of these extra terms $\Hpm_{\Delta_{2s}}$ was also argued for on general grounds from the Cauchy--Riemann equation in Part~I. We shall initially focus on the case when $\mathcal{S}_{2s}$ is one-dimensional and defer more general cases to sections~\ref{sec:Higher} and~\ref{sec:GeneralHigher}.
We also note that the presence of $\Hpm_{\Delta_{2s}}$ in the modular-invariant solution $\FFpm{s}{m}{k}$ of the Laplace equation~\eqref{eq:Laps} implies that, inspecting Fourier by Fourier mode, there are non-zero coefficients for the homogeneous solutions (provided by Bessel functions) to the Laplace equations, see~\eqref{eq:HpmBessel}. This is in contrast to what was observed in examples in an $\SLtwoZ$ U-duality context in~\cite{Green:2014yxa,Klinger:2018, Minprogress}.

\subsection{Examples involving the Ramanujan cusp form}

The cusp form of lowest weight is the Ramanujan cusp form $\Delta_{12}=\sum_{n=1}^{\infty} \tau(n) q^n$. Since it has holomorphic modular weight $12$, its iterated integrals can arise as the modular completion of $\cFFpm{s}{m}{k}$ at eigenvalue $s=6$. According to the spectrum~\eqref{eq:Spectrum} this happens first for $m{+}k=7$ in the odd sector and for $m{+}k=8$ in the even sector.

\subsubsection{Odd functions for $(m,k)=(2,5)$ and $(m,k)=(3,4)$}
\label{sec:3.1.1}

As presented in Part~I, we can carry out the procedure for obtaining a solution to the Laplace equations
(\ref{eq:Laps}), 
and the simplest odd functions where modular invariance at the level of the $\bsv$ breaks down are given by
\begin{align}
\cFFm{6}{2}{5} &= -1890 \betasv{1& 4\\4& 10} - 1512 \betasv{2& 3\\4& 10} + 
 1890 \betasv{4& 1\\10& 4} + 1512 \betasv{5& 0\\10& 4} + 
 1008\zeta_3 \betasv{3\\10} \nn\\
 &\quad + \frac{315\zeta_3}{y}\betasv{4\\10}  -  \frac{21\zeta_9}{4y^3} \betasv{0\\4} - \frac{105\zeta_9}{64 y^4}\betasv{1\\4} + \frac{7\zeta_9}{1920y^2} \,, \nn\\
 \cFFm{6}{3}{4} &= -2100 \betasv{2& 3\\6& 8} - 2100 \betasv{3& 2\\6 & 8} + 
 2100 \betasv{3& 2\\8 & 6} - 210 \betasv{4& 1\\6& 8} + 
 2100 \betasv{4& 1\\8& 6}  \label{bsvex25}\\
 &\quad + 210 \betasv{5& 0\\8& 6} + 
 84 \zeta_5 \betasv{1\\8} + \frac{210 \zeta_5}{y}\betasv{2\\8}+ \frac{105 \zeta_5}{2y^2} \betasv{3\\8}  -\frac{15\zeta_7}y \betasv{0\\6}\nn\\
 &\quad - \frac{75\zeta_7}{2y^2}\betasv{1\\6}- \frac{74 \zeta_7}{8y^3}\betasv{2\\6} -\frac{\zeta_7}{15120}\,.
 \notag
\end{align}
The modular transformations of these expressions can be obtained by the methods of section~\ref{sec:mmvS} 
where we also use the knowledge of the $\overline{\alpha[\cdots]}$ whose values were determined in Part~I, building on~\cite{Gerken:2020yii}. The modular transformation  (\ref{modtrfs}) of the functions above yields
\begin{align}
\delta_{\rm S} \cFFm{6}{2}{5}(\tau) &= \left(
\frac{1701 i}{8\pi^7} \MMsv{0&1}{4&10}+ \frac{1701 i}{32 \pi^7} \MMsv{1&0}{4&10} - \frac{23\pi^4 \zeta_3}{118800} + \frac{7 \zeta_9}{16\pi^2}\right) \frac{\tau{+}\bar\tau} {(\tau {-}\bar\tau)^5} + \ldots \,, \nn\\
\delta_{\rm S} \cFFm{6}{3}{4}(\tau)  &= \left(
\frac{6615  i}{32\pi^7} \MMsv{0&1}{6&8}+\frac{2205 i}{16 \pi^7} \MMsv{1&0}{6&8} -\frac{\pi^2 \zeta_5}{360}+ \frac{7\zeta_7}{240} \right) \frac{\tau{+}\bar\tau} {(\tau{-}\bar\tau)^5} + \ldots \,,
\label{Scocw7}
\end{align}
where the dots denote similar terms that consist of multiple modular values and polynomials in $\tau$ and $\bar\tau$ over the same denominator $(\tau{-}\bar\tau)^5$. Even without further knowledge of the exact multiple modular values one can check numerically, using \eqref{eq:MMVnum}, that neither of the above S-transformations vanishes. 
Hence, we are indeed in one of the cases where the combination $\cFFm{6}{2}{5}$ of $\bsv$ that solves the Laplace equation fails to be modular invariant.

Fortunately, further knowledge of the multiple modular values is available through the beautiful work of Brown~\cite{Brown:mmv}, and the examples relevant to (\ref{Scocw7}) can be found in (\ref{410sec})
and (\ref{68sec}). We emphasise that the multiple modular values are not built out of multiple zeta values alone but also contain L-values of cusp forms. In general the individual multiple modular values also contain other `new numbers' such as the object $c(\Delta;12)$ in~\cite{Brown2019}, but these do not enter in the examples above and cancel in all reduced multiple modular values (\ref{svmmod}) that we have encountered.

Using these and similar results one can show that
\begin{align}
\delta_{\rm S}\cFFm{6}{2}{5}(\tau)  &=
 \frac{ i \pi^{11}  \Lambda(\Delta_{12},12)  }{900 y^5} p_{12}^{-}(\tau,\bar \tau)  \label{f625mod} \\
\delta_{\rm S} \cFFm{6}{3}{4}(\tau) &=
- \frac{ i \pi^{11}  \Lambda(\Delta_{12},12) }{300 y^5}  p_{12}^{-}(\tau,\bar \tau)
\notag
 \end{align}
with the same polynomial in $\tau,\bar \tau$ in both cases
\begin{align}
p_{12}^{-}(\tau,\bar \tau) &=  (\tau{+}\bar \tau) \bigg\{
1 +  \tau^4 \bar \tau^4 - \frac{25}{24} (  \tau^2{+} \bar \tau^2) (1 {+} \tau^2 \bar \tau^2 ) 
+  \frac{1}{12} (\tau^4 {+} \bar \tau^4) \label{p12min} \\
 & \ \  \ \  \ \  \ \   \ \   \ \   \ \ 
 - \frac{25 }{6}  \tau \bar \tau (1 {+}  \tau^2 \bar \tau^2) + 
 2  \tau \bar \tau ( \tau^2 {+} \bar \tau^2) + \frac{19  }{3} \tau^2 \bar \tau^2 \bigg\} \, .
 \notag
\end{align}
These cocycles turn out to be proportional to those of $\Hm_{\Delta_{12}}$ in (\ref{eq:D12Mcocycle}), 
\begin{align}
\cFFm{6}{2}{5}(\tau) - \cFFm{6}{2}{5}(-\tfrac1{\tau}) & = \frac1{75}  \frac{\Lambda(\Delta_{12},12)}{\Lambda(\Delta_{12},10)} \left( \Hm_{\Delta_{12}}(\tau) - \Hm_{\Delta_{12}}(-\tfrac{1}\tau)\right)\,, \label{eq:Im6coc}\\
\cFFm{6}{3}{4}(\tau) - \cFFm{6}{3}{4}(-\tfrac1{\tau}) & = - \frac1{25}  \frac{\Lambda(\Delta_{12},12)}{\Lambda(\Delta_{12},10)} \left( \Hm_{\Delta_{12}}(\tau) - \Hm_{\Delta_{12}}(-\tfrac{1}\tau)\right) \, . \notag
\end{align}
We can therefore form the linear combinations
\begin{align}
\FFm{6}{2}{5} &= \cFFm{6}{2}{5} -  \frac1{75}  \frac{\Lambda(\Delta_{12},12)}{\Lambda(\Delta_{12},10)} \Hm_{\Delta_{12}}\,,
\label{eq:FcFm6}\\
\FFm{6}{3}{4} &= \cFFm{6}{3}{4} +  \frac1{25}  \frac{\Lambda(\Delta_{12},12)}{\Lambda(\Delta_{12},10)} \Hm_{\Delta_{12}}\, ,
\notag
\end{align}
that are modular invariant thanks to \eqref{eq:Im6coc}
and solve the Laplace equation~\eqref{eq:Laps:odd}.
The ratio of L-values appearing here might appear a bit surprising since in all  previous expressions of this work and Part~I, (multiple) zeta values only appear in the numerator. The reason for the ratio showing up here lies in our normalisation of the homogeneous solution in~\eqref{eq:defH} as a simple iterated integral. This definition leads to the explicit appearance of an L-value in the S-cocycle, see~(\ref{eq:D12Mcocycle}), which needs to be cancelled by the denominator in (\ref{eq:FcFm6}). Choosing a different normalisation for $\Hpm_{\Delta_{2s}}$ one could turn the quotient in~\eqref{eq:FcFm6} into a multiplication by an L-value only. 
Note that the transcendental weight six of $\Hm_{\Delta_{12}}$, weight zero of $\Lambda(\Delta_{12},10)$ and weight one of $\Lambda(\Delta_{12},12)$ ensure that both terms on the right-hand sides of (\ref{eq:FcFm6}) have the expected weight seven.

We note that the combination
\begin{equation}
\label{eq:3.8}
3\FFm{6}{2}{5} + \FFm{6}{3}{4}  = 3 \cFFm{6}{2}{5} +  \cFFm{6}{3}{4}\,,
\end{equation}
is an eigenfunction of the Laplacian with eigenvalue $30$
which is perfectly modular invariant on its own without the need of adding any iterated integral of $\Delta_{12}$.
This is one of the examples of modular objects analysed in Part~I that are expressible in terms of $\bsv,y$ and
odd zeta values.

\subsubsection{Even functions for $(m,k)=(2,6)$, $(m,k)=(3,5)$ and $(m,k)=(4,4)$}

In the even sector, the first occurrence of the eigenvalue $s=6$ is for $\FFp{6}{2}{6}$, $\FFp{6}{3}{5}$ and $\FFp{6}{4}{4}$.
The $\bsv$ representations of the associated $\cFFp{6}{m}{8-m}$ take a form similar to their odd counterparts in (\ref{bsvex25})
and can be found in section 4.5 of Part~I. Performing the S-transformations on the combinations constructed only out of the $\bsv$ we find
{ \allowdisplaybreaks
\begin{align}
\delta_{\rm S} \cFFp{6}{2}{6}(\tau)  &= \left( \frac{693}{64\pi^8} ( \MMsv{0&1}{4&12} -  \MMsv{1&0}{4&12}) + \frac{823 i \pi^5 \zeta_3}{202702500} + \frac{7i \zeta_{11}}{96\pi^3} -\frac{9555 i \zeta_{13}}{11056 \pi^5} \right)\frac{1}{(\tau{-}\bar\tau)^5} + \ldots\nn\\*
&= - \frac{\pi^{11} \Lambda(\Delta_{12},13)}{518250 y^5}\, p^{+}_{12}(\tau,\bar \tau) \, , \notag
\\
\delta_{\rm S}\cFFp{6}{3}{5}(\tau)  &=\left( \frac{945}{64\pi^8}  (\MMsv{0&1}{6&10} -
 \MMsv{1&0}{6&10})+\frac{i\pi^3 \zeta_5}{86400} + \frac{7 i  \zeta_9}{2880\pi } - \frac{63063 i \zeta_{13}}{176896\pi^5}  \right)\frac{1}{(\tau{-}\bar\tau)^5} + \ldots\nn\\*
&=  \frac{ \pi^{11} \Lambda(\Delta_{12},13)}{132672 y^5} \, p^{+}_{12}(\tau,\bar \tau) \, ,
\label{f8plustrf} \\
\delta_{\rm S} \cFFp{6}{4}{4}(\tau) &= \left( \frac{1029}{32\pi^8}  \MMsv{0&1}{8&8} +\frac{i\pi\zeta_7}{3240}  - \frac{5005 i  \zeta_{13}}{16584\pi^5 } \right)\frac{1}{(\tau{-}\bar\tau)^5} + \ldots \notag \\*
&= 
- \frac{ 7 \pi^{11}  \Lambda(\Delta_{12},13) }{621900y^5} \, p^{+}_{12}(\tau,\bar \tau)\, . \nn
\end{align}}%
In the second steps, we have inserted the relevant expressions for the reduced multiple modular values 
and thereby arrived at the polynomial
\begin{align}
p^{+}_{12}(\tau,\bar \tau) &= (1{-}\tau \bar \tau)  \bigg\{
1 +  \tau^4 \bar \tau^4 -  \frac{691 }{162} (  \tau^2 {+} \bar \tau^2) (1 {+}  \tau^2 \bar \tau^2)  
+ \frac{ 691 }{504} (\tau^4 {+} \bar \tau^4) \label{poly12}\\
 & \ \ \ \ \ \ \ \ - \frac{3131}{ 324}   \tau  \bar \tau (1 {+} \tau^2 \bar \tau^2 ) 
+ \frac{21421  }{  2268}  \tau \bar \tau(  \tau^2 {+} \bar \tau^2 ) + \frac{40273  }{2268} \tau^2 \bar \tau^2
  \bigg\}
\notag
\end{align}
which determines the ellipses in the first steps of (\ref{f8plustrf}).
Since the corresponding S-cocycle of the homogeneous solution $\Hp_{\Delta_{12}}$
in \eqref{eq:D12Pcocycle} is proportional to the same polynomial in (\ref{poly12}), 
the following combinations are modular invariant:
\begin{align}
\FFp{6}{2}{6} &= \cFFp{6}{2}{6} + \frac{2}{17275} \frac{\Lambda(\Delta_{12},13)}{\Lambda(\Delta_{12},11)} \Hp_{\Delta_{12}}\,, \notag\\
\FFp{6}{3}{5} &= \cFFp{6}{3}{5} - \frac{5}{11056} \frac{\Lambda(\Delta_{12},13)}{\Lambda(\Delta_{12},11)} \Hp_{\Delta_{12}}\,,
\label{fplusw8}\\
\FFp{6}{4}{4} &= \cFFp{6}{4}{4} + \frac{7}{10365} \frac{\Lambda(\Delta_{12},13)}{\Lambda(\Delta_{12},11)} \Hp_{\Delta_{12}}\,.
\notag
\end{align}
From this we also see that the following combinations are modular invariant without the inclusions of an iterated integral of a
holomorphic cusp form,
\begin{align}
\FFp{6}{2}{6} -\frac{6}{35} \FFp{6}{4}{4}  &= \cFFp{6}{2}{6} -\frac{6}{35} \cFFp{6}{4}{4}  \,,\label{fpluswb}\\
\FFp{6}{3}{5} +\frac{75}{112} \FFp{6}{4}{4}  &= \cFFp{6}{3}{5} +\frac{75}{112} \cFFp{6}{4}{4}  \,, \notag
\end{align}
as already discussed in section 4.5.2 of Part~I.
There is a two-dimensional subspace of the three modular invariants
$\{ \FFp{6}{2}{6}, \FFp{6}{3}{5}, \FFp{6}{4}{4}\}$ that does not require a cusp form -- in 
agreement with our counting of MGFs in section 3.6 of Part~I.

Similar to the odd case, the right-hand sides of (\ref{fplusw8}) have uniform 
transcendental weight eight by combining weight six of $\Hp_{\Delta_{12}}$ with weight zero of $\Lambda(\Delta_{12},11)$
and weight two of $\Lambda(\Delta_{12},13)$.

\subsection{Examples with cusp forms of higher weight}

We have performed the same analysis as in the previous section for all $\FFpm{s}{m}{k}$ with ${m{+}k\leq 14}$. 
The combinations $\cFFpm{s}{m}{k}$ of $\bsv$ require additional cusp forms for their modular invariant completion whenever $s$ 
is half the weight of a holomorphic cusp form, i.e.\ $s\in \{6,8,9,\ldots\}$.
We only list those cases here, a full list of the $\cFFpm{s}{m}{k}$ with $m{+}k \leq 14$ in terms of $\bsv$ can be found in the ancillary file along with their modular completions.

Besides the examples of the previous section, the Ramanujan cusp form occurs in the following modular invariant functions
at transcendental weight $m{+}k=9,10$,
\begin{align}
\FFm{6}{2}{7}  &= \cFFm{6}{2}{7} + \frac1{51825} \frac{\Lambda(\Delta_{12},14)}{\Lambda(\Delta_{12},10)} \Hm_{\Delta_{12}}\,,\nn\\
\FFm{6}{3}{6}  &= \cFFm{6}{3}{6} - \frac1{4500} \frac{\Lambda(\Delta_{12},14)}{\Lambda(\Delta_{12},10)} \Hm_{\Delta_{12}}\,,\nn\\
\FFm{6}{4}{5}  &= \cFFm{6}{4}{5} + \frac1{1800} \frac{\Lambda(\Delta_{12},14)}{\Lambda(\Delta_{12},10)} \Hm_{\Delta_{12}}\,,
\label{fplushigher}\\
\FFp{6}{3}{7}  &= \cFFp{6}{3}{7} + \frac1{725550} \frac{\Lambda(\Delta_{12},15)}{\Lambda(\Delta_{12},11)} \Hp_{\Delta_{12}}\,,\nn\\
\FFp{6}{4}{6}  &= \cFFp{6}{4}{6} - \frac1{232176} \frac{\Lambda(\Delta_{12},15)}{\Lambda(\Delta_{12},11)} \Hp_{\Delta_{12}}\,,\nn\\
\FFp{6}{5}{5}  &= \cFFp{6}{5}{5} + \frac1{165840} \frac{\Lambda(\Delta_{12},15)}{\Lambda(\Delta_{12},11)} \Hp_{\Delta_{12}}\,,\nn
\end{align}
where $\Lambda(\Delta_{12},14)$ and $\Lambda(\Delta_{12},15)$ carry three and four
units of transcendental weight, respectively.
The Hecke normalised cusp form of weight $16$ arises in (also see appendix \ref{subappendixA})
\begin{align}
\FFm{8}{2}{7} &= \cFFm{8}{2}{7} - \frac{13}{7350} \frac{\Lambda(\Delta_{16},16)}{\Lambda(\Delta_{16},14)} \Hm_{\Delta_{16}}\,,\nn\\
\FFm{8}{3}{6} &= \cFFm{8}{3}{6} + \frac{13}{2100} \frac{\Lambda(\Delta_{16},16)}{\Lambda(\Delta_{16},14)} \Hm_{\Delta_{16}}\,,\nn\\
\FFm{8}{4}{5} &= \cFFm{8}{4}{5} - \frac{143}{14700} \frac{\Lambda(\Delta_{16},16)}{\Lambda(\Delta_{16},14)} \Hm_{\Delta_{16}}\,,\nn\\
\FFp{8}{2}{8} &= \cFFp{8}{2}{8} + \frac{39}{3544660} \frac{\Lambda(\Delta_{16},17)}{\Lambda(\Delta_{16},15)} \Hp_{\Delta_{16}}\,,
\label{fminhigher}\\
\FFp{8}{3}{7} &= \cFFp{8}{3}{7} - \frac{91}{2083392} \frac{\Lambda(\Delta_{16},17)}{\Lambda(\Delta_{16},15)} \Hp_{\Delta_{16}}\,,\nn\\
\FFp{8}{4}{6} &= \cFFp{8}{4}{6} + \frac{1001}{13021200} \frac{\Lambda(\Delta_{16},17)}{\Lambda(\Delta_{16},15)} \Hp_{\Delta_{16}}\,,\nn\\
\FFp{8}{5}{5} &= \cFFp{8}{5}{5} - \frac{143}{1620416} \frac{\Lambda(\Delta_{16},17)}{\Lambda(\Delta_{16},15)} \Hp_{\Delta_{16}}\nn\,,
\end{align}
with $\Lambda(\Delta_{16},16)$ and $\Lambda(\Delta_{16},17)$ of transcendental 
weight one and two, respectively. The Hecke normalised cusp form of weight $18$ arises in 
(also see appendix \ref{subappendixB})
\begin{align}
\FFm{9}{2}{8} &= \cFFm{9}{2}{8} + \frac{1}{1260} \frac{\Lambda(\Delta_{18},18)}{\Lambda(\Delta_{18},16)} \Hm_{\Delta_{18}}\,,
\notag\\
\FFm{9}{3}{7} &= \cFFm{9}{3}{7} - \frac{5}{2016} \frac{\Lambda(\Delta_{18},18)}{\Lambda(\Delta_{18},16)} \Hm_{\Delta_{18}}\,,
\label{fminw10}\\
\FFm{9}{4}{6} &= \cFFm{9}{4}{6} + \frac{13}{5040} \frac{\Lambda(\Delta_{18},18)}{\Lambda(\Delta_{18},16)} \Hm_{\Delta_{18}}\,,
\notag
\end{align}
with $\Lambda(\Delta_{18},18)$ of transcendental weight one.
Just as discussed in the previous examples, it is possible to find suitable rational linear combinations of these objects to produce modular invariant  functions for which the homogeneous solutions $\Hpm_{{\Delta_{2s}}}$ cancel out, thus properly living in the realm of MGFs.

%%%%%%%%%%%%%%%%%%%%%%%%%%%%%%%%%%%%%%%%%%%%%%%%%%%%%%%%%%%
\subsection{\texorpdfstring{An example involving the two weight $24$ cusp forms}{An example involving the two weight 24 cusp forms} }\label{sec:Higher}
%%%%%%%%%%%%%%%%%%%%%%%%%%%%%%%%%%%%%%%%%%%%%%%%%%%%%%%%%%%

As explained in Part~I~\cite{PartI}, as we increase the total transcendental weight $w=m{+}k$ we encounter higher and higher eigenvalues $s \leq k{+}m{-}1$ in the spectrum, see~\eqref{eq:Spectrum}. This in turn means that the obstructions to finding modular solutions
to the Laplace systems \eqref{eq:Laps} are related to iterated integrals \eqref{eq:defH} of cusp forms ${\Delta_{2s}}$ of higher and higher modular weight $2s$. Denoting by $\mathcal{S}_{2s}$ the vector space of holomorphic cusp forms for $\SLtwoZ$ with even integer modular weight ${2s}$, we have the classic result~\cite{ApostolTomM1976MfaD}
\begin{equation}
\dim \mathcal{S}_{2s} = \left\lbrace \begin{array}{lc}
\left\lfloor \frac{2s}{12}\right\rfloor -1 & 2s\equiv 2 \!\!\!\mod 12\,,\\[2mm]
 \left\lfloor \frac{2s}{12}\right\rfloor \phantom{-1}& \mbox{otherwise}\,.
\end{array}\right.\,
\label{dimcusps}
\end{equation}

Hence, we see that starting with $2s = 24$ we should in general expect the space of obstructions to have dimension greater than one since $\dim\mathcal{S}_{24} =2$. From~\eqref{eq:Spectrum} we know that the first instance for which the eigenvalue $s=12$ appears is for the odd sector and with transcendental weight $w = m{+}k = 13$.
We are then led to expect that a modular invariant solution to \eqref{eq:Laps} in this sector must take the form
\begin{equation}\label{eq:dim2Ex}
\FFm{12}{m}{k} = \cFFm{12}{m}{k} + a^-_{\Delta_{24,i},m,k}  \Hm_{\Delta_{24,i}}+a^-_{\Delta_{24,ii},m,k} \Hm_{\Delta_{24,ii}}\,,
\end{equation}
with $m{+}k=13$ and $m,k\geq 2$, where we use a basis of Hecke eigenforms subject to \eqref{eq:HeckeEigen}
and denoted by $\Delta_{24,i}, \Delta_{24,ii}$.

Such a basis can be constructed by considering the linear combination $\alpha \Delta_{12}^2 +\beta \Delta_{12} {\rm G}_{12}$, i.e.\ the most general holomorphic cusp form of weight $2s = 24$. Then, the real coefficients $\alpha,\beta$ for Hecke eigenforms are obtained by imposing the resulting Fourier coefficients to be multiplicative, i.e.\ $a(m) a(n) = a(m{\cdot }n )$ for $m,n$ coprime: ${\rm gcd}(m,n)=1$.
This procedure constructs the two Hecke eigenforms\footnote{Alternative expressions in terms of the ring
generators ${\rm G}_{4},{\rm G}_{6}$ read
\begin{align*}
\Delta_{24,i}(\tau) &= \bigg( \frac{131}{248832} - \frac{ \sqrt{144169}}{248832} \bigg) \bigg( \frac{ {\rm G}_{4}(\tau) }{2 \zeta_{4} } \bigg)^6 
- \bigg( \frac{13}{248832} + \frac{ \sqrt{ 144169} }{248832}
\bigg) \bigg( \frac{ {\rm G}_{6}(\tau) }{2 \zeta_{6} } \bigg)^4  \\
& \ \ \ \ + 
\bigg(  {-}\frac{59}{124416} +  \frac{ \sqrt{  144169}}{124416} \bigg)
\bigg( \frac{ {\rm G}_{4}(\tau) }{2 \zeta_{4} } \bigg)^3  \bigg( \frac{ {\rm G}_{6}(\tau) }{2 \zeta_{6} } \bigg)^2 
\end{align*}
and the analogous combination with $\sqrt{144169} \rightarrow - \sqrt{144169}$ for $\Delta_{24,ii}(\tau) $.}
\begin{subequations}
\begin{align}
\Delta_{24,i}(\tau) &= \bigg( \frac{324204}{691} - 12 \sqrt{144169} \bigg)  \Delta_{12}(\tau)^2 + 
\frac{ {\rm G}_{12}(\tau) }{2 \zeta_{12} } \Delta_{12}(\tau)\, ,  \\
\Delta_{24,ii}(\tau) &= \bigg( \frac{324204}{691} + 12 \sqrt{144169} \bigg)  \Delta_{12}(\tau)^2 + 
\frac{ {\rm G}_{12}(\tau) }{2 \zeta_{12} } \Delta_{12}(\tau)\, ,
\end{align}
\end{subequations}
whose Fourier coefficients lie in the number field $\mathbb{Q}(\sqrt{144169})$:
\begin{subequations}
\begin{align}
\Delta_{24,i}(\tau) & = q^1+( 540 - 12 \sqrt{144169})q^2+( 169740 + 576 \sqrt{144169})q^3+O(q^4) \,,\\
\Delta_{24,ii}(\tau) & = q^1+( 540 + 12 \sqrt{144169})q^2+( 169740 - 576 \sqrt{144169})q^3+O(q^4)\,.
\end{align}
\end{subequations}
The number field generated by the above Fourier coefficients has a non-trivial Galois automorphism $\sigma \in \mbox{Aut}_{\mathbb{Q}}\Big(\mathbb{Q}(\sqrt{144169})\Big)$, which acts as $\sigma:\sqrt{144169}\to - \sqrt{144169}$ and under which the two basis elements are exchanged, i.e.\ $\sigma(\Delta_{24,i}) = \Delta_{24,ii}$.

In the basis of Hecke eigenforms we have that all the even/odd completed L-values inside the critical strip are $\mathbb{Q}(\sqrt{144169})$ multiples of one another \cite{Manin:1973} and the S-cocycles for $\Delta_{24,i}$ and $\Delta_{24,ii}$ can be put in the form \eqref{eq:Hnoninv}. The Galois automorphism exchanges the two cocycles, as well as the completed L-values. 

Following the same types of arguments that led to \eqref{eq:Im6coc}, we see that in general the S-cocycle for $ \cFFm{12}{m}{k}$ does not vanish and involves the completed L-values $ \Lambda(\Delta_{24,i},t), \Lambda(\Delta_{24,ii},t)$ of the two Hecke eigenforms through the reduced multiple modular values.
For example, we find 
\begin{subequations}
\begin{align}
\MMsv{0&1}{4&22} &= -\frac{24438334 i \pi^{23} \zeta_3}{77661373832214046875} +\frac{4016053 i \pi^3 \zeta_{23}}{1153592550} +\frac{4096 i\pi^{24} \Lambda(\Delta_{22},23)}{7791418647627615} \nn\\
&\quad + \left(\frac{39121664 i \pi^{25}}{7428479236124821875} + \frac{1279322368 i \pi^{25} \sqrt{144169}}{1070956422992879444896875}\right) \Lambda(\Delta_{24,i},24)\\
&\quad 
 + \left(\frac{39121664 i \pi^{25}}{7428479236124821875} - \frac{1279322368 i \pi^{25} \sqrt{144169}}{1070956422992879444896875}\right) \Lambda(\Delta_{24,ii},24)\,,\nn\\
\MMsv{1&0}{4&22} &= \frac{1242928 i \pi^{23} \zeta_3}{847214987260516875} +\frac{2i \pi^5 \zeta_{21}}{567} +\frac{4016053 i \pi^3 \zeta_{23}}{115359255} -\frac{8192 i\pi^{24} \Lambda(\Delta_{22},23)}{1558283729525523} \nn\\
&\quad + \left(\frac{39121664 i \pi^{25}}{7428479236124821875} + \frac{1279322368 i \pi^{25} \sqrt{144169}}{1070956422992879444896875}\right) \Lambda(\Delta_{24,i},24)\\
&\quad 
 + \left(\frac{39121664 i \pi^{25}}{7428479236124821875} - \frac{1279322368 i \pi^{25} \sqrt{144169}}{1070956422992879444896875}\right) \Lambda(\Delta_{24,ii},24)\,.\nn
\end{align}
\end{subequations}

Using these and similar expressions for the other $\textMMsv{j_1&j_2}{k_1&k_2}$ with $k_1{+}k_2=26$, there is a unique choice of the constants $a^-_{\Delta_{24,i},m,k} ,a^-_{\Delta_{24,ii},m,k}$ such that the combinations \eqref{eq:dim2Ex} are modular invariant. Explicitly,
\begin{align}
a^-_{\Delta_{24,i},2,11}
 &=  \frac{1}{7200  } \left(-\frac{152819}{76230} -\frac{4997353 \sqrt{144169}}{10990002870}  \right) \frac{\Lambda(\Delta_{24,i},24)}{ \Lambda(\Delta_{24,i},22)}\,, \nn \\
a^-_{\Delta_{24,i},3,10}
 &=-\frac{19}{4320} \left( -\frac{18289}{87780} - \frac{554243 \sqrt{144169}}{12655154820}  \right)\frac{\Lambda(\Delta_{24,i},24)}{ \Lambda(\Delta_{24,i},22)}\,, \nn \\
a^-_{\Delta_{24,i},4,9}
&=  \frac{323 }{7200}\left( -\frac{2117}{78540} -\frac{44479 \sqrt{144169}}{11323033260} \right) \,\frac{\Lambda(\Delta_{24,i},24)}{ \Lambda(\Delta_{24,i},22)}\,, \label{aafor24}\\
a^-_{\Delta_{24,i},5,8}
&=    -\frac{323 }{1680} \left( -\frac{1}{220} + \frac{13 \sqrt{144169}}{31717180}  \right)\frac{\Lambda(\Delta_{24,i},24)}{ \Lambda(\Delta_{24,i},22)}\,, \nn \\
a^-_{\Delta_{24,i},6,7}
&=  \frac{4199 }{10800}\left( -\frac{823}{660660}+\frac{82699 \sqrt{144169}}{95246691540}\right)\frac{\Lambda(\Delta_{24,i},24)}{ \Lambda(\Delta_{24,i},22)}\,,
\nn
\end{align}
where we have split the result into three factors: The ratio of L-values carries transcendental weight one, the middle factor is valued in the number field $\mathbb{Q}(\sqrt{144169})$ and corresponds to the inverse of the Petersson--Haberland pairing~\cite{Haberland} between two  properly normalised polynomials associated with the cusp from ${\Delta_{2s}}$. The first factor is a rational number multiplying the vector in the number field and  we refer to \cite{Brown:mmv} for why this splitting occurs. 
The constants $a^-_{\Delta_{24,ii},m,k}$ for the second cusp form can be directly obtained by the application of the Galois automorphism to $a^-_{\Delta_{24,i},m,k}$, i.e.\ $a^-_{\Delta_{24,ii},m,k}= \sigma(a^-_{\Delta_{24,i},m,k})$ that acts on the number-field-valued middle factor by the Galois action and on the L-values by $\sigma(\Lambda(\Delta_{24,i},t) ) = \Lambda(\Delta_{24,ii},t)$.

With the above values we obtain that the combinations \eqref{eq:dim2Ex} are then the unique modular invariant solutions to the Laplace system \eqref{eq:Laps}. Furthermore, we can easily check that, for some particular rational linear combinations of $\FFm{12}{m}{k}$, the iterated integrals of the cusp forms $\Hm_{\Delta_{24,i}}(\tau)$ and $\Hm_{\Delta_{24,ii}}(\tau)$ drop out, thus producing the modular objects discussed in Part~I, for example:
\begin{align}
\frac{1862}{103} \FFm{12}{2}{11} - \frac{470}{103} \FFm{12}{4}{9} + \FFm{12}{6}{7} &=\frac{1862}{103} \cFFm{12}{2}{11} - \frac{470}{103} \cFFm{12}{4}{9} + \cFFm{12}{6}{7}\,, \notag \\
\frac{171}{49}\FFm{12}{3}{10} + \frac{165}{49}\FFm{12}{4}{9} +\FFm{12}{5}{8} &=\frac{171}{49}\cFFm{12}{3}{10} + \frac{165}{49}\cFFm{12}{4}{9} +\cFFm{12}{5}{8}\,, \label{fforw24}\\  
- \frac{76}{11} \FFm{12}{3}{10}- \frac{62}{11} \FFm{12}{4}{9} +\FFm{12}{6}{7}&=- \frac{76}{11} \cFFm{12}{3}{10}- \frac{62}{11} \cFFm{12}{4}{9} +\cFFm{12}{6}{7}\,. \notag
\end{align}
In section~\ref{sec:select}, we discuss the relation between linear combinations of this type and Tsunogai's derivation algebra. 

The complete list of modular completions $\cFFpm{s}{m}{k}\to \FFpm{s}{m}{k}$ with $m{+}k\leq 14$ can be found in the ancillary file.

%%%%%%%%%%%%%%%%%%%%%%%%%%%%%%%%%%%%%%%%%%%%%%%%%%%%%%%%%%%
\subsection{Structure for general weight}
\label{sec:GeneralHigher}
%%%%%%%%%%%%%%%%%%%%%%%%%%%%%%%%%%%%%%%%%%%%%%%%%%%%%%%%%%%

For general transcendental weight $w=m{+}k$ and eigenvalue $s$ in the spectrum (\ref{eq:Spectrum}), 
the modular invariant solutions $\FFpm{s}{m}{k}$ to the Laplace system \eqref{eq:Laps} must take the 
form given in~\eqref{eq:GenLinear}.

From their general form given in \eqref{eq:Hnoninv} we know that the cocycles $\delta_{\rm S} \Hpm_{{\Delta_{2s}}}$ can be normalised so that we have factorised out the completed L-values $\Lambda({\Delta_{2s}}, 2s{-}1)$ in the even case and $\Lambda({\Delta_{2s}}, 2s{-}2)$ in the odd case, times a rational function in $\tau,\bar{\tau}$ with coefficients in $K_{\Delta_{2s}}$, the number field generated by the Fourier coefficients of ${\Delta_{2s}}$.

To understand the generic structure of the coefficients $a^\pm_{{\Delta_{2s}},m,k}$ we can analyse in more depth their transcendentality properties. Since the iterated-integral representation \eqref{eq:IterDeltaE0} assigns
transcendentality $s$ to $\Hpm_{{\Delta_{2s}}}$, the coefficients $a^\pm_{{\Delta_{2s}},m,k}$
must have weight $m{+}k{-}s$ in order to arrive at combinations $\FFpm{s}{m}{k}$ 
of weight $w=m{+}k$ in \eqref{eq:GenLinear}.
On these grounds, by the transcendentality $\ell$ of $\Lambda({\Delta_{2s}},2s{+}\ell{-}1)$ outside 
the critical strip $\ell\geq 1$, we are led to conclude that
\begin{equation}
a^\pm_{{\Delta_{2s}},m,k} =\left\lbrace\begin{matrix} \displaystyle \Bigg. q^+_{{\Delta_{2s}}, m,k} \, \kappa^+_{{\Delta_{2s}},m,k} \frac{\Lambda({\Delta_{2s}},m{+}k{+}s{-}1)}{\Lambda({\Delta_{2s}},2s{-}1)} \, , \\ \displaystyle \Bigg.
q^-_{{\Delta_{2s}}, m,k} \, \kappa^-_{{\Delta_{2s}},m,k} \frac{\Lambda({\Delta_{2s}},m{+}k{+}s{-}1)}{\Lambda({\Delta_{2s}},2s{-}2)} \, , \end{matrix}\right.
\label{Lnumerator}
\end{equation}
where we have $ q^\pm_{{\Delta_{2s}}, m,k}\in\mathbb{Q}$, while $\kappa^\pm_{{\Delta_{2s}},m,k} \in K_{\Delta_{2s}}$ is given by the inverse of the Petersson-Haberland pairing between the two cocycles $\delta_{\rm S} \Hp_{{\Delta_{2s}}}(\tau)$ and $\delta_{\rm S} \Hm_{{\Delta_{2s}}}(\tau)$ properly normalised, see \cite{Brown:mmv}.  Given that $m{+}k{+}s$ is even (odd) for even functions $\FFp{s}{m}{k}$ (odd functions $\FFm{s}{m}{k}$), the L-function $\Lambda({\Delta_{2s}},m{+}k{+}s{-}1)$ in the numerator of (\ref{Lnumerator}) is evaluated at odd integers in the modular invariant completion of $\FFp{s}{m}{k}$ (at even integers in the case of $\FFm{s}{m}{k}$).

Note that, as discussed previously, to determine the number field $K_{\Delta_{2s}}$, which contains the Fourier coefficient of ${\Delta_{2s}}\in \mathcal{S}_{2s}$, one has to diagonalise the Hecke operators \eqref{eq:HeckeEigen} in $\mathcal{S}_{2s}$. The nature of this number field $K_{\Delta_{2s}}$ is clarified\footnote{We would like to thank Herbert Gangl for related discussions.} by the \textit{Maeda conjecture} \cite{Maeda}, which states that the characteristic polynomials of the Hecke operator $T_n$ are irreducible over $\mathbb{Q}$. In a certain sense the number field $K_{\Delta_{2s}}$ is ``maximal'' in that the Maeda conjecture suggests that the associated Galois group is the full symmetric group $\mathfrak{S}_d$ with $d = \mbox{dim}\,\mathcal{S}_{2s}$.

For example at weight $2s=28$ with $d = \mbox{dim}\,\mathcal{S}_{28} = 2$ we have $K_{\Delta_{28}} = \mathbb{Q}(\sqrt{18209})$,  for $2s=30$ we have once more $d=2$ and now $K_{\Delta_{30}}=\mathbb{Q}(\sqrt{51349})$, while moving to weight $2s=36$, the lowest weight for which $d=3$, we have
$K_{\Delta_{36}} = \mathbb{Q}[x] \slash (x^{3} - 12422194 x - 2645665785 )$, which means that the algebraic number-field extension of $\mathbb{Q}$ contains all three roots of the polynomial.
 The Maeda conjecture, although still unproven, has been extensively tested for all modular weights up to $2s=12000$, see \cite{Ghitza} where strong evidence is presented to support its validity. More examples of such number fields can be found on the very comprehensive LMFDB database \cite{LMFDB} of L-functions and modular forms.

As a final comment we stress that the action of a non-trivial element, $\sigma$, of the Galois automorphism group $\mbox{Aut}_\mathbb{Q}(K_{\Delta_{2s}})$ allows us to relate the constants $a^\pm_{{\Delta_{2s}},m,k}$ for different Hecke eigenforms $\sigma(a^\pm_{{\Delta_{2s}},m,k}) = a^\pm_{\sigma({\Delta_{2s}}),m,k}$, hence if Maeda's conjecture were to be true it would be enough to find one such number $a^\pm_{{\Delta_{2s}},m,k}$ for a single Hecke cusp ${\Delta_{2s}}$ to deduce all the others.

%%%%%%%%%%%%%%%%%%%%%%%%%%%%%%%%%%%%%%%%%%
\section{\texorpdfstring{Selection rules on $\beta^{\rm sv}$ from Tsunogai's derivation algebra}{Selection rules on betasv from Tsunogai's derivation algebra}}
\label{sec:select}

In this section, we study the interplay between the generating series of MGFs introduced in~\cite{Gerken:2019cxz}, the modular invariant functions $\FFpm{s}{m}{k}$ and an abstract algebra on generating derivations $\epsilon_k$ introduced by Tsunogai~\cite{Tsunogai} that is related to holomorphic cusp forms~\cite{Pollack}. This connection will clarify why some instances of $\cFFpm{s}{m}{k}$ are not modular invariant. The commutation relations
among Tsunogai's derivations $\epsilon_k $ with $k = 0,4,6,8,\ldots$ govern which linear combinations of the modular completions $\FFpm{s}{m}{k}$ appear as MGFs.

The generating series of MGFs introduced in~\cite{Gerken:2019cxz} captures the structure of the $\alpha'$-expansion of certain genus-one integrals in closed-string amplitudes that eventually comprise all MGFs when integrating over sufficiently many torus punctures. More specifically, the first-order differential equations in $\tau$ of this generating series is solved by 
\begin{align}
\series^{\tau} &= 1 + \sum_{k=4}^\infty \sum_{j=0}^{k-2} \frac{(-1)^j (k{-}1) }{(k{-}2{-}j)!}
\betasv{j\\k} \epsilon_{k}^{(k-j-2)} 
\label{select.1} \\
&\ \ +  \sum_{k_1=4}^\infty  \sum_{k_2=4}^\infty \sum_{j_1=0}^{k_1-2}   \sum_{j_2=0}^{k_2-2} \frac{(-1)^{j_1+j_2} (k_1{-}1) (k_2{-}1) }{(k_1{-}2{-}j_1)!(k_2{-}2{-}j_2)!}
\betasv{j_1 &j_2 \\k_1 &k_2} \epsilon_{k_2}^{(k_2-j_2-2)} \epsilon_{k_1}^{(k_1-j_1-2)}  + O(\epsilon_k^3)
\notag
\end{align}
with conjectural matrix representations of $\epsilon_k$ acting on suitable initial values
as $\tau \rightarrow i\infty$ that are series in zeta values \cite{Gerken:2020yii}.\footnote{The generating series
of genus-one integrals are denoted by $Y^\tau$ in \cite{Gerken:2020yii} and given by
$\series^{\tau} \exp(- \frac{ \ep_0}{4y}) \widehat Y^{i\infty}$, where the initial values $\widehat Y^{i\infty}$
comprise all the (conjecturally single-valued \cite{Zerbini:2015rss, DHoker:2015wxz}) MZVs in the expansion of the resulting MGFs around the cusp.}
In~\cite{Gerken:2019cxz,Gerken:2020yii} the $\epsilon_k$ were explicit finite-dimensional matrix operators that were checked at low orders to obey the relations of Tsuongai's derivation algebra, and this is conjectured to hold to all orders. Here, we think of the $\epsilon_k$ as the abstract generators of Tsuongai's derivation algebra. In~\eqref{select.1} and the following we make use of the convenient shorthand notation
\beq
\epsilon^{(j)}_k =   ({\rm ad}_{\epsilon_0})^j (\epsilon_k)\, ,\label{select.2} 
\eeq
for the repeated adjoint actions ${\rm ad}_{\epsilon_0}(\ast) = [\epsilon_0,\ast]$. The suppressed terms $O(\epsilon_k^3)$ in (\ref{select.1}) are all multiplied by $\bsv$ of depth $\geq 3$.

%%%%%%%%%%%%%%%%%%%%%%%%%%%%%%%%%%%%%%%%%%
%%%%%%%%%%%%%%%%%%%%%%%%%%%%%%%%%%%%%%%%%%
%%%%%%%%%%%%%%%%%%%%%%%%%%%%%%%%%%%%%%%%%%

\subsection{\texorpdfstring{Overview of $\epsilon_k$ relations at depth two}{Overview of epsilon(k) relations at depth two}}
\label{sec:select.1}

Tsunogai's derivations satisfy a wealth of commutator relations. First of all,
$[\ep_2,\ep_k] = 0 $ with $k = 0,2,4,\ldots$ identifies $\epsilon_2$ to be a central
element which does not occur in the series (\ref{select.1}). On the remaining
derivations, ${\rm ad}_{\ep_0}$ enjoys the nilpotency properties
\beq
{\rm ad}_{\ep_0}^{k-1}(\ep_k) = \ep_k^{(k-1)} = 0 \, , \ \ \ k=4,6,8,\ldots 
\label{nilpotrel}
\eeq
such that the $j_i = 0$ terms in (\ref{select.1}) exhaust the maximal non-vanishing nested
commutators $\ep_k^{(k-2)}$. Apart from the relations (\ref{nilpotrel}) that take a simple and universal form for all $k$,
commutators of $\ep_{k_1},\ep_{k_2},\ldots$ at weight $k_1{+}k_2{+}\ldots \geq 14$ obey
more involved identities starting with \cite{LNT, Pollack, Broedel:2015hia}
\begin{align}
0 &=[\ep_{10},\ep_4]-3[\ep_{8},\ep_6] \,, \notag \\
0&= [\ep_{14},\ep_4] - \frac{7}{2}[\ep_{12},\ep_6] + \frac{11}{2} [\ep_{10},\ep_8]  \,, \label{select.3}  \\
0 &=  [\ep_{16},\ep_{4}]- \frac{25}{8} [\ep_{14},\ep_{6}]+ \frac{13}{4}[\ep_{12},\ep_{8}]\, .
\notag
\end{align}
Nested commutators of three and more derivations obey corollaries of
these relations obtained from action of ${\rm ad}_{\ep_k}$. At the same
time, there are infinite families of {\it indecomposable} relations among three or more derivations, i.e.\ relations
that cannot be obtained by repeatedly acting with ${\rm ad}_{\ep_k}$ on simpler ones. 
The simplest indecomposable relation trilinear in derivations reads
\begin{align}
0&=80[\ep_{12},[\ep_4,\ep_{0}]] + 16 [\ep_4,[\ep_{12},\ep_0]] - 250 [\ep_{10},[\ep_6,\ep_0]]
- 125 [\ep_6,[\ep_{10},\ep_0]] + 280 [\ep_8,[\ep_8,\ep_0]] \notag \\
& \ \ \ \  - 462 [\ep_4,[\ep_4,\ep_8]] - 1725 [\ep_6,[\ep_6,\ep_4]] 
\label{select.4}
\end{align}
and already illustrates a generic feature: When referring to the number of $\ep_{k \neq 0}$
in a nested commutator as its {\it depth},\footnote{This terminology differs from the work of
Pollack \cite{Pollack} where also $\epsilon_0$ is assigned depth one.}  indecomposable relations involving
more than two $\ep_k$ usually mix terms of different depth. The depth-two terms
in the first line of (\ref{select.4}) affect the appearance of $\beta^{\rm sv}$ of depth two in (\ref{select.1})
while the depth-three terms in the second line are related to $\beta^{\rm sv}$ of depth three that are 
part of the suppressed terms $O(\epsilon_k^3)$ in (\ref{select.1}) and beyond the scope of this work. 

Relations in the derivation algebra are also assigned a notion of depth according to
the maximal depth of the nested commutator therein, e.g.\ (\ref{select.4}) is said to have depth three.
As will be reviewed in the remainder of this subsection, Pollack determined the depth-two terms 
in indecomposable relations of arbitrary depth in closed form. More precisely,
their rational coefficients are determined from the period polynomials of holomorphic cusp forms in \cite{Pollack}.

\subsubsection{Cusp forms and depth-two relations}
\label{sec:4.1.1}

In order to concisely relate the coefficients in relations like (\ref{select.3}) or (\ref{select.4}) to holomorphic
cusp forms $\Delta_{2s}$ of modular weight $2s$,  we follow the conventions of \cite{Pollack} for period polynomials 
\begin{align}
r_{\Delta_{2s}}(X,Y) &= \int_0^{i\infty} \dd \tau \, \Delta_{2s}(\tau) (X-\tau Y)^{2s-2}
\label{seceq4.1} \\
&= i \sum_{k=0}^{2s-2} \binom{2s{-}2}{k} X^{2s-2-k} (-iY)^k \Lambda(\Delta_{2s}, k{+}1) \, , \notag
\end{align}
where the arguments $t$ of the L-function $\Lambda(\Delta_{2s}, t)$ are all within the critical strip $t \in (0,2s)$.
Moreover, we introduce the even and odd parts of the period polynomials (\ref{seceq4.1}) via
\begin{align}
r^{\pm}_{\Delta_{2s}}(X,Y) &= \frac{1}{2} \big[
r_{\Delta_{2s}}(X,Y)  \pm r_{\Delta_{2s}}(X,-Y) 
\big]
\notag \\
&= \left\{ \begin{array}{ll}
 i\, \sum\limits_{\substack{k=0 \\ k \ {\rm even}}}^{2s-2} \, {\binom{2s{-}2}{ k}} X^{2s-2-k} (-iY)^k \Lambda(\Delta_{2s}, k{+}1)  &: \ r_{\Delta_{2s}}^+
 \\[7mm]
  i \, \sum\limits_{\substack{k=0\\  k \ {\rm odd}}}^{2s-2} \,{\binom{2s{-}2}{ k}}X^{2s-2-k} (-iY)^k \Lambda(\Delta_{2s}, k{+}1)  &: \ r_{\Delta_{2s}}^- \, .
\end{array} \right.   \label{seceq4.2}
\end{align}
{\allowdisplaybreaks The ratios of L-values at $2s=12,16$ and $18$ noted in \eqref{ratios12}, \eqref{ratios16} and \eqref{ratios18}
are equivalent to
\begin{align}
r_{\Delta_{12}}^+(X,Y) &=  - \frac{ 691 i \Lambda(\Delta_{12}, 11)}{36}
 \bigg\{ 
  X^8 Y^2 - X^2 Y^8 - 3 (X^6 Y^4 - X^4 Y^6) - \frac{36 }{691} ( X^{10} - Y^{10} )\bigg\}\,, \notag \\
r_{\Delta_{16}}^+(X,Y) &= -\frac{ 3617 i \Lambda(\Delta_{16}, 15)}{  180 } 
\bigg\{  X^{12} Y^2 - X^2 Y^{12} - \frac{ 7}{2} (X^{10} Y^4 - X^4 Y^{10})  \notag \\*
&\hspace{3.7cm} + 
  \frac{ 11}{2} (X^8 Y^6 - X^6 Y^8) - \frac{180 }{3617} (X^{14} - Y^{14})\bigg\} \,,
\label{seceq4.3} \\
r_{\Delta_{18}}^+(X,Y) &= \frac{ 43867 i \Lambda(\Delta_{18}, 17)}{  2250 }
 \bigg\{ X^{14} Y^2 - X^2 Y^{14} - \frac{ 25}{8} (X^{12} Y^4 - X^4 Y^{12}) \notag \\*
 &\hspace{3.7cm} + 
   \frac{13}{4} (X^{10} Y^6 - X^6 Y^{10}) - \frac{ 2250}{43867} (X^{16} - Y^{16}) \bigg\}\, . \notag 
\end{align}}%
The rational coefficients $r_{\Delta_{2s}}^+(X,Y)  \, \big|_{X^a Y^b}$ of the non-zero powers $X^aY^b$ with $a,b>0$ are easily seen
to match those in the depth-two relations (\ref{select.3}) among $[\ep_{k_1},\ep_{k_2}]$
with $k_1{+}k_2=2s{+}2$. More generally, the depth-two relations in the derivation
algebra are given by
\beq
0 = \sum_{a+b=2s-2} r_{\Delta_{2s}}^+(X,Y)  \, \big|_{X^a Y^b} [\ep_{a+2},\ep_{b+2}]
\label{pollackd2}
\eeq
in terms of the even parts of period polynomials (\ref{seceq4.1}) of cusp forms \cite{Pollack}. The extremal
terms $\sim (X^{2s-2} - Y^{2s-2})$ in (\ref{seceq4.3}) are mapped to coefficients of the vanishing
commutators $[\ep_2,\ast]$ in (\ref{pollackd2}), that is why their more involved coefficients
$ \frac{36 }{691},\frac{180 }{3617}$ and $\frac{ 2250}{43867}$ do not enter the depth-two relations~(\ref{select.3}).

\subsubsection{Cusp forms and higher-depth relations}

Also for higher-depth relations such as (\ref{select.4}), the coefficients of the
depth-two terms $[\ep_{k_1}^{(j_1)},\ep_{k_2}^{(j_2)}] = [{\rm ad}_{\ep_0}^{j_1}(\ep_{k_1}),{\rm ad}_{\ep_0}^{j_2}(\ep_{k_2})]$ are determined by the period polynomials (\ref{seceq4.1}). More specifically, 
relations of even (odd) depth are governed by
the even part $r_{\Delta_{2s}}^+$ (the odd part $r_{\Delta_{2s}}^-$) in (\ref{seceq4.2}).
At weight $2s=12,16,18$, the coefficients in the odd counterpart of (\ref{seceq4.3})
\begin{align}
r_{\Delta_{12}}^-(X,Y) &= 10 \Lambda(\Delta_{12}, 10) \bigg\{  X^9 Y + X Y^9 - \frac{25}{4} (X^7 Y^3 + X^3 Y^7) + \frac{21}{2} X^5 Y^5 \bigg\} \,,\notag \\
r_{\Delta_{16}}^-(X,Y) &= 14 \Lambda(\Delta_{16}, 14)  \bigg\{  X^{13} Y + X Y^{13} - \frac{ 245}{36} (X^{11} Y^3 + X^3 Y^{11})  \notag \\
&\hspace{3cm} + \frac{ 539}{36} (X^9 Y^5 + X^5 Y^9) - \frac{55}{3} X^7 Y^7\bigg\}\,,
\label{seceq4.4} \\
r_{\Delta_{18}}^-(X,Y) &= -16 \Lambda(\Delta_{18}, 16)  \bigg\{  X^{15} Y + X Y^{15}
 - \frac{ 77}{12} (X^{13} Y^3 + X^3 Y^{13}) \notag \\
 &\hspace{3cm}  + \frac{ 91}{8} (X^{11} Y^5 + X^5 Y^{11}) 
 - \frac{143}{24} (X^9 Y^7 + X^7 Y^9) \bigg\} \,, \notag 
\end{align}
enter relations of depth three, five, $\ldots$ such as (\ref{select.4}). By rewriting the first line
of (\ref{select.4}) as
\beq
-160 \bigg\{ 
\frac{ [\ep_{12} , \ep_4^{(1)}] }{2} + \frac{ [\ep_{4} , \ep_{12}^{(1)}] }{10} 
- \frac{25}{4} \bigg( \frac{ [\ep_{10} , \ep_6^{(1)}] }{4} + \frac{ [\ep_{6} , \ep_{10}^{(1)}] }{8}  \bigg)
+ \frac{21}{2} \frac{ [\ep_{8} , \ep_8^{(1)}] }{6}  
\bigg\} = 0 \ \te{mod depth 3} \, ,
\label{seceq4.21}
\eeq
one can identify the relative factors of $-\frac{25}{4}$ and $\frac{21}{2}$ inside the curly
bracket with those in the expression (\ref{seceq4.4}) for the odd part $r_{\Delta_{12}}^-(X,Y)$. 
The additional denominators of the combinations $\frac{ [\ep_{k_1} , \ep_{k_2}^{(1)}] }{k_2-2}$ line 
up with the $d=3$ instance of the general depth-two combination\footnote{We depart from Pollack's conventions
for the commutators (\ref{seceq4.22}) to ensure that the subscripts of $\ourh^d_{p,q}$
line up with the $\ep_p, \ep_q$ in their definition. The commutators $h^d_{p,q}$ in
\cite{Pollack} are reproduced by $\ourh^d_{p,q}= h^d_{p-2,q-2}$.}
\beq
\ourh^d_{p,q} = (d{-}2)! \sum_{i=0}^{d-2} (-1)^i \frac{ (p{-}2{-}i)! (q{-}d{+}i)! }{i!(p{-}2)! (d{-}2{-}i)! (q{-}2)!} [\epsilon_p^{(i)} , \epsilon_q^{(d-2-i)} ] \,,
\label{seceq4.22}
\eeq
with $d\geq 2$ subject to alternating symmetry properties $\ourh^d_{p,q} = (-1)^{d-1} \ourh^d_{q,p}$.
The ratios of factorials in (\ref{seceq4.22}) are engineered such that $\ourh^d_{p,q}$ is firstly annhilated by $p{+}q{-}2d{+}1$
powers of ${\rm ad}_{\ep_0}$, i.e.\
\beq
{\rm ad}_{\ep_0}^{p{+}q{-}2d{+}1}(\ourh^d_{p,q}) = 0 \, , \ \ \ \ \ \ 
{\rm ad}_{\ep_0}^{p{+}q{-}2d}(\ourh^d_{p,q}) \neq 0 \, .
\label{seceq4.23}
\eeq
In the simplest instances of (\ref{seceq4.22}),
\begin{align}
\ourh^2_{p,q} &= [\ep_{p},\ep_{q}]\,, \notag \\
\ourh^3_{p,q} &= \frac{ [\ep_{p}, \ep_{q}^{(1)}] }{q{-}2} + \frac{ [\ep_{q}, \ep_{p}^{(1)}] }{p{-}2} 
\label{seceq4.24alt} \,,\\
\ourh^4_{p,q} &= \frac{ [\ep_{p},\ep_{q}^{(2)}] }{(q{-}2)(q{-}3)} 
- \frac{2  [\ep_{p}^{(1)},  \ep_{q}^{(1)}]}{(p{-}2)(q{-}2)}
+ \frac{ [\ep_{p}^{(2)},\ep_{q}] }{(p{-}2)(p{-}3)}\, ,
 \notag 
\end{align}
the relative factors ensure that $\ourh^3_{p,q}$ and $\ourh^4_{p,q}$ are annihilated
by $p{+}q{-}5$ and $p{+}q{-}7$ powers of ${\rm ad}_{\ep_0}$, respectively, e.g.\ that terms of
the form  $[\ep_p^{(p-2)}, \ep_q^{(q-2)}]$ cancel
from ${\rm ad}^{p+q-5}_{\ep_0}(\ourh^3_{p,q})$. One can therefore view the $\ourh^d_{p,q}$
at different values of $d$ (together with their non-vanishing ${\rm ad}^j_{\ep_0}$ actions at
$j=0,1,\ldots, p{+}q{-}2d$) as spanning different ${\rm SL}(2)$ representations contained
in the tensor product of 
$\{ \ep^{(j)}_{p}, \ j=0,1,\ldots,p{-}2\}$ 
and
$\{ \ep_{q}^{(j)}, \ j=0,1,\ldots,q{-}2\}$.

With the combinations of commutators in (\ref{seceq4.22}) and the even or odd parts
of period polynomials in (\ref{seceq4.2}), Pollack's result for the depth-two coefficients in indecomposable
relations takes the compact form \cite{Pollack}
\beq
\begin{array}{ll}
\displaystyle
\bigg. 0 = \sum_{a+b=2s-2} r_{\Delta_{2s}}^+(X,Y)  \, \big|_{X^a Y^b} \ourh^d_{a+d,b+d} \ \te{mod depth $\geq 3$} 
&:\ d \ \te{even} \, ,\\
\displaystyle
\bigg. 0 = \sum_{a+b=2s-2} r_{\Delta_{2s}}^-(X,Y)  \, \big|_{X^a Y^b} \ourh^d_{a+d,b+d} \ \te{mod depth $\geq 3$} 
&:\ d \ \te{odd} \, .
\end{array}
\label{pollackdn}
\eeq
In other words, the ratios of odd L-values $\Lambda(\Delta_{2s},2n{+}1)$ (with $n{\in}\NN$) in the even parts $ r_{\Delta_{2s}}^+$ 
determine the coefficients in indecomposable relations among $\ourh^d_{p,q}$ at even $d$. Conversely,
$\ourh^d_{p,q}$ at odd $d$ are related by even L-values $\Lambda(\Delta_{2s},2n)$ (with $n{\in}\NN$) in the odd 
parts $ r_{\Delta_{2s}}^-$. In both cases, the L-values are within the critical strip, and the coefficients of higher-depth terms such as the second line of (\ref{select.4}) remain undetermined by (\ref{pollackdn}). The $\epsilon$-weight of the $[\ep^{(j_1)}_{k_1},\ep^{(j_2)}_{k_2}]$
in a relation (\ref{pollackdn}) due to cusp forms of modular weight $2s$ is given by $k_1{+}k_2=2s{+}2d{-}2$.

\subsubsection{Examples}

The depth-two relations (\ref{pollackd2}) are the special case $d=2$ of (\ref{pollackdn}),
i.e.\ the examples in (\ref{select.3}) are reproduced by the first three lines of
\begin{align}
r^+_{\Delta_{12}} \ &\Rightarrow \ 0 =
-2 \ourh_{4, 10}^{2} + 6 \ourh_{6, 8}^{2}\,,
\notag \\
r^+_{\Delta_{16}} \ &\Rightarrow \ 0 =
-2 \ourh_{4, 14}^{2} + 7 \ourh_{6, 12}^{2} - 11 \ourh_{8, 10}^{2}\,,
\notag \\
r^+_{\Delta_{18}} \ &\Rightarrow \ 0 =
-2 \ourh_{4, 16}^{2} + \frac{25}{4} \ourh_{6, 14}^{2} - \frac{13}{2} \ourh_{8, 12}^{2}\,,
\label{seceq4.24} \\
r^+_{\Delta_{20}} \ &\Rightarrow \ 0 =
-2 \ourh_{4, 18}^{2} + \frac{20}{3} \ourh_{6, 16}^{2} - \frac{28}{3} \ourh_{8, 14}^{2} + 
 \frac{26}{3} \ourh_{10, 12}^{2}\,,
\notag \\
r^+_{\Delta_{22}} \ &\Rightarrow \ 0 =
-2 \ourh_{4, 20}^{2} + \frac{105}{16} \ourh_{6, 18}^{2} - \frac{17}{2} \ourh_{8, 16}^{2} + 
 \frac{85}{16} \ourh_{10, 14}^{2}\, ,
\notag
\end{align}
where $\ourh_{2, q}^{d}=0$ since $\epsilon_2$ commutes with all the $\ep_k$. 
Indecomposable relations at depth three due to (\ref{pollackdn}) at $d=3$ include
\begin{align}
r^-_{\Delta_{12}} \ &\Rightarrow \ 0 =
2 \ourh_{4, 12}^{3} - \frac{25}{2} \ourh_{6, 10}^{3} + \frac{21}{2} \ourh_{8, 8}^{3} \ \te{mod depth $\geq 3$} \,,
\notag \\
r^-_{\Delta_{16}} \ &\Rightarrow \ 0 =
2 \ourh_{4, 16}^{3} - \frac{245}{18} \ourh_{6, 14}^{3} + \frac{539}{18} \ourh_{8, 12}^{3} - 
\frac{ 55}{3} \ourh_{10, 10}^{3} \ \te{mod depth $\geq 3$}  \,,
\notag \\
r^-_{\Delta_{18}} \ &\Rightarrow \ 0 =
2 \ourh_{4, 18}^{3} - \frac{77}{6} \ourh_{6, 16}^{3} + \frac{91}{4} \ourh_{8, 14}^{3} - 
 \frac{143}{12} \ourh_{10, 12}^{3}\ \te{mod depth $\geq 3$}  \,,
\label{seceq4.27} \\
r^-_{\Delta_{20}} \ &\Rightarrow \ 0 =
2 \ourh_{4, 20}^{3} - \frac{715}{54} \ourh_{6, 18}^{3} + \frac{242}{9} \ourh_{8, 16}^{3} - 
 \frac{260}{9} \ourh_{10, 14}^{3} + \frac{715}{54} \ourh_{12, 12}^{3}\ \te{mod depth $\geq 3$}  \,,
\notag \\
r^-_{\Delta_{22}} \ &\Rightarrow \ 0 =
2 \ourh_{4, 22}^{3} - \frac{1183}{90} \ourh_{6, 20}^{3} + \frac{1547}{60} \ourh_{8, 18}^{3} - 
 \frac{697}{30} \ourh_{10, 16}^{3} + \frac{1547}{180} \ourh_{12, 14}^{3}\ \te{mod depth $\geq 3$}  \,,
\notag
\end{align}
whose first line due to $r^-_{\Delta_{12}}$ is equivalent to (\ref{seceq4.21}).
The relations at higher $d \geq 4$ start with
\begin{align}
r^+_{\Delta_{12}} \ &\Rightarrow \ 0 = 
\frac{72}{691} \ourh_{4, 14}^{4} - 2 \ourh_{6, 12}^{4} + 6 \ourh_{8, 10}^{4}
\ \te{mod depth $\geq 3$}  \,,
\label{seceq4.28} \\
r^+_{\Delta_{16}} \ &\Rightarrow \ 0 =
\frac{360}{3617} \ourh_{4, 18}^{4}  - 2 \ourh_{6, 16}^{4} + 7 \ourh_{8, 14}^{4} - 
 11 \ourh_{10, 12}^{4}
 \ \te{mod depth $\geq 3$}  \,,
\notag 
\end{align}
as well as
\begin{align}
r^-_{\Delta_{12}} \ &\Rightarrow \ 0 =
2 \ourh_{6, 14}^{5} - \frac{25}{2} \ourh_{8, 12}^{5} + \frac{21}{2} \ourh_{10, 10}^{5} \ \te{mod depth $\geq 3$}  \,,
\label{seceq4.29} \\
r^-_{\Delta_{16}} \ &\Rightarrow \ 0 =
2 \ourh_{6, 18}^{5} - \frac{245}{18} \ourh_{8, 16}^{5} + \frac{539}{18} \ourh_{10, 14}^{5} - 
 \frac{55}{3} \ourh_{12, 12}^{5}\ \te{mod depth $\geq 3$}  \,,
\notag 
\end{align}
and feed into relations among $[[\ldots [[ \ep_{k_1},\ep_{k_2}],\ep_{k_3}],\ldots  ], \ep_{k_\ell} ]$
of increasing $\epsilon$-weight $k_1{+}k_2{+}\ldots{+}k_\ell$.
Table~\ref{select.5} summarises the counting of indecomposable higher-depth relations 
along with their $\epsilon_k$-weight (which is $p{+}q$ in case of $t^d_{p,q}$ for arbitrary $d\geq 2$) due to 
given holomorphic cusp forms. 
\begin{table}
\centering
\begin{tabular}{c||c|c|c|c}
 depth &2 &3 &4 &5
\\\hline\hline
$\Delta_{12}$ &14 &16 &18 &20
\\\hline
$\Delta_{16}$ &18 &20 &22 &24
\\\hline
$\Delta_{18}$ &20 &22 &24 &26
\\\hline
$\Delta_{20}$ &22 &24 &26 &28
\\\hline
$\Delta_{22}$ &24 &26 &28 &30
\\\hline
$\Delta_{24,i},\, \Delta_{24,ii}$ &2$\times$ 26 &2$\times$ 28 &2$\times$ 30 &2$\times$ 32
\end{tabular}
\caption{\label{select.5}\textit{The number given corresponds to the $\epsilon$-weight of the indecomposable relation of the given depth triggered by the cusp form $\Delta_{2s}$, so that for instance $\Delta_{12}$ induces one indecomposable relation of depth $2$ at weight $14$. There are two indecomposable relations of depth $2,3,4,\ldots$ at weight $26,28,30,\ldots$, respectively,} due to the two cusp forms of modular weight $24$.}
\end{table}

\subsubsection{Comparison with the eMZV datamine}

The explicit form of various indecomposable relations up to depth 5 can be downloaded from the datamine \cite{WWWe}
of relations among elliptic multiple zeta values (eMZVs) \cite{Enriquez:Emzv}. The datamine uses the notation {\tt rRel[w, r]} for relations of $\epsilon$-weight $w$ and depth $r$.
The depth-two parts of these relations {\tt rRel[w, r]}
are usually combinations of several relations (\ref{pollackdn}) at different~$d\leq r$. 
For instance, {\tt rRel[20, 3]} at depth three mixes the specialisations of (\ref{pollackdn}) to $(r^-_{\Delta_{16}},d{=}3)$
with ${\rm ad}_{\ep_0}$ acting on the relation from $(r^+_{\Delta_{18}},d{=}2)$.
Similarly, the depth-two terms of {\tt rRel[18, 4]} can be reconstructed from 
$(r^+_{\Delta_{12}},d{=}4)$ and ${\rm ad}^2_{\ep_0}$ action on $(r^+_{\Delta_{16}},d{=}2)$ whereas {\tt rRel[20, 5]}
mixes input from all of $(r^-_{\Delta_{12}},d{=}5), {\rm ad}^2_{\ep_0}(r^-_{\Delta_{16}},d{=}3)$ and ${\rm ad}_{\ep_0}^3(r^+_{\Delta_{18}},d{=}2)$.

%%%%%%%%%%%%%%%%%%%%%%%%%%%%%%%%%%%%%%%%%%
%%%%%%%%%%%%%%%%%%%%%%%%%%%%%%%%%%%%%%%%%%
%%%%%%%%%%%%%%%%%%%%%%%%%%%%%%%%%%%%%%%%%%

\subsection{\texorpdfstring{Modular graph forms and $\ep_k$ relations at depth two}{Modular graph forms and epsilon(k) relations at depth two}}
\label{sec:select.d2}

Based on the relations among depth-two commutators
$[\ep_{k_1}^{(j_1)} ,\ep_{k_2}^{(j_2)}  ] $ reviewed above, we shall now describe the dropouts
of iterated Eisenstein integrals of depth two from the generating series (\ref{select.1}) of MGFs.
We will be interested in the modular invariant cases where the entries of 
$\beta^{\rm sv}[\begin{smallmatrix} j_1 &j_2\\ k_1 &k_2 \end{smallmatrix}]$ obey
$2j_1{+}2j_2{+}4 = k_1{+}k_2$, see (\ref{eq:Sbsv2}). These cases of $\beta^{\rm sv}$
are related to the $\FFpm{s}{m}{k}$ with $2m{+}2k = k_1{+}k_2$, modulo shuffles
$\beta^{\rm sv}[\begin{smallmatrix} j_1 \\ k_1   \end{smallmatrix}] \beta^{\rm sv}[\begin{smallmatrix} j_2\\ k_2 \end{smallmatrix}]$ and lower-depth terms. The derivations 
$[\ep_{k_1}^{(j_1)} ,\ep_{k_2}^{(j_2)}  ] $ relevant to modular invariant terms in (\ref{select.1})
involve a total of $j_1{+}j_2= \frac{1}{2}(k_1{+}k_2) {-}2$ powers of ${\rm ad}_{\ep_0}$.

In the following, we shall rewrite the shuffle-irreducible modular invariants with $\beta^{\rm sv}$ 
at depth two in terms of $\FFpm{s}{m}{k}$. In this way, the $[\ep_{k_1}^{(j_1)} ,\ep_{k_2}^{(j_2)}  ] $
conspire to the commutators $\ourh^d_{p,q}$ defined in (\ref{seceq4.22})
and their images under ${\rm ad}_{\ep_0}^N$ with $N\leq p{+}q{-}2d$,
\beq
{\rm ad}_{\ep_0}^N [\ep_{k_1}^{(j_1)} ,\ep_{k_2}^{(j_2)}  ] = \sum_{j=0}^N \binom{N}{j} 
[\ep_{k_1}^{(j_1+j)} ,\ep_{k_2}^{(j_2+N-j)}  ] \, .
\label{select.10}
\eeq 
We will decompose the generating series (\ref{select.1}) into depth-two sectors $\series^{\tau}(k_1,k_2)$ 
associated with double integrals of given $({\rm G}_{k_1},{\rm G}_{k_2})$.
The modular invariant contributions are isolated by means of the delta function in  
\begin{align}
\series^{\tau}(k_1,k_2) &= (k_1{-}1) (k_2{-}1)
\sum_{j_1=0}^{k_1-2}   \sum_{j_2=0}^{k_2-2} \frac{(-1)^{j_1+j_2} \delta(2j_1{+}2j_2{+}4{-}k_1{-}k_2)  }{(k_1{-}2{-}j_1)!(k_2{-}2{-}j_2)!} \notag \\
&\ \ \ \ \ \ \ \ \ \ \ \ \ \ \ \ \ \ \ \ \ \ \ \ \times
\betasv{j_1 &j_2 \\k_1 &k_2} \epsilon_{k_2}^{(k_2-j_2-2)} \epsilon_{k_1}^{(k_1-j_1-2)}  \, ,
\label{fixsector}
\end{align}
and the depth-two modular-invariant part of the generating series~\eqref{select.1} is obtained by summing over $k_1,k_2 \in 2\mathbb N {+} 2$.
The shuffle-irreducible modular invariants of the $({\rm G}_{4}, {\rm G}_{6})$ sector are 
for instance encoded in
\begin{align}
\series^{\tau}(4,6)  + \series^{\tau}(6,4)  &=
-\frac{5}{2} (\betasv{1& 2\\6& 4} \ep_4 \ep_{6}^{(3)}
+    3 \betasv{2& 1\\6& 4} \ep_{4}^{(1)} \ep_{6}^{(2)} 
+    3 \betasv{0& 3\\4& 6} \ep_{6}^{(1)} \ep_{4}^{(2)}    \label{g4g6poe.1}\\
& \ \ \ \ \
+    3 \betasv{3& 0\\6& 4} \ep_{4}^{(2)} \ep_{6}^{(1)} 
+    3 \betasv{1& 2\\4& 6} \ep_{6}^{(2)}  \ep_{4}^{(1)}
+    \betasv{2& 1\\4& 6}  \ep_{6}^{(3)} \ep_4 ) \, .
\notag
\end{align}
Using~\eqref{eq:bsvshuffle} we can rewrite $
\beta^{\rm sv}[\begin{smallmatrix} j_1 &j_2\\ k_1 &k_2 \end{smallmatrix}]
= - \beta^{\rm sv}[\begin{smallmatrix} j_2 &j_1\\ k_2 &k_1 \end{smallmatrix}] \ \te{mod} \ \shuffle$, where $\shuffle$ refers to shuffle products
(\ref{eq:bsvshuffle}) of depth-one $\bsv$. Picking the basis\footnote{The relation between
(\ref{g4g6bsv}) and MGFs in their lattice-sum representations is discussed in section 5.4 of
Part~I and section 5 of \cite{Gerken:2020yii}.}
\begin{align}
\FFp{3}{2}{3} &= 30 \betasv{2& 1\\4& 6} + 30 \betasv{3& 0\\6& 4} 
\ \te{mod lower depth} \,,
 \notag \\
\FFm{2}{2}{3} &= -90 \betasv{1& 2\\4& 6} + 90 \betasv{2& 1\\4& 6} + 
 90 \betasv{2& 1\\6& 4} - 90 \betasv{3& 0\\6& 4}
\ \te{mod lower depth} \,,
\label{g4g6bsv} \\
\FFm{4}{2}{3} &=
-90 \betasv{1& 2\\4& 6} - 60 \betasv{2& 1\\4& 6} + 
 90 \betasv{2& 1\\6& 4} + 60 \betasv{3& 0\\6& 4}
\ \te{mod lower depth}  \,,
\notag
\end{align}
of shuffle irreducibles, the derivations in (\ref{g4g6poe.1}) conspire to 
the following ${\rm ad}_{\ep_0}$-images of the 
commutators $\ourh^d_{p,q}$ in (\ref{seceq4.22}):
\begin{align}
\series^{\tau}(4,6)  + \series^{\tau}(6,4)  &=
\frac{1}{10} \FFm{2}{2}{3}  {\rm ad}_{\ep_0}( \ourh^4_{4,6})
 - \frac{1}{120} \FFm{4}{2}{3}  {\rm ad}_{\ep_0}^3 ( \ourh^2_{4,6} )   \label{g4g6poe.2} \\
&\ \ + \frac{1}{6} \FFp{3}{2}{3}  {\rm ad}_{\ep_0}^2 ( \ourh^3_{4,6} )
\ \te{mod lower depth}\ \& \ \shuffle\,.
\notag
\end{align}
The analogous expressions at weights $m{+}k \geq7$ will in the first place involve
the combinations $\cFFpm{s}{m}{k}$ of $\beta^{\rm sv}$ rather than the full
modular invariants $\FFpm{s}{m}{k}$: The iterated integrals (\ref{eq:defH})
of cusp forms are consistently absent from the generating series
(\ref{select.1}). In the matrix representations of (\ref{select.1}) 
relevant to closed-string genus-one integrals \cite{Gerken:2020yii}, the combinations
of $\betasv{j \\ k}$ contributing to MGFs at depth two can be recovered
by the initial conditions $\tau \rightarrow i \infty$ that the derivations in
$\series^{\tau}$ act on. 

However, the initial conditions do not allow us to retrieve
the iterated integrals of cusp forms in (\ref{eq:defH}): They do not have
any known realisation in closed-string integrals over torus punctures since Cauchy--Riemann
derivatives of MGFs \cite{DHoker:2016mwo} or their generating series \cite{Gerken:2019cxz} do not introduce
any holomorphic cusp forms.
Many of the subsequent equations will hold modulo lower depth and shuffles
as in (\ref{g4g6poe.2}), and we will indicate by using $\cong$ in the place of
$=$ that shuffles, $\beta^{\rm sv}$ of depth one and depth-zero terms have been dropped
while depth-one integrals of holomorphic cusp forms are still tracked.

We will exemplify in the following sections that the depth-two terms
(\ref{pollackdn}) of $\epsilon_k$-relations are sufficient to effectively
replace all the $\cFFpm{s}{m}{k}$ in $\series^{\tau}$ by 
$\FFpm{s}{m}{k}$. In all cases up to and including $m{+}k=14$,
the coefficient of $\cFFpm{s}{m}{k}$ is checked to be a $\mathbb Q$-multiple
of ${\rm ad}_{\epsilon_0}^{s-1}(\ourh_{2m, 2k}^{m+k-s+1} )$.
Given that $m{+}k{-}s$ is even for $\cFFp{s}{m}{k}$ and odd for $\cFFm{s}{m}{k}$,
the transition to $\FFp{s}{m}{k}$ (to $\FFm{s}{m}{k}$) by adding cusp-form
contributions is governed by $\ourh_{p, q}^{d}$ at odd values of $d$ (even values of $d$).

\subsubsection{Weight 7}

At weight $m{+}k=7$, the analogue of (\ref{g4g6poe.2}) for the shuffle-irreducible modular invariants reads
\begin{align}
\series^{\tau}(4,10) +\series^{\tau}(10,4)&\cong
 \frac{  \cFFm{4}{2}{ 5} }{540 }  {\rm ad}_{\epsilon_0}^{3}(\ourh_{4, 10}^{4} )
- \frac{ \cFFm{6}{2}{ 5}}{30240   }  {\rm ad}_{\epsilon_0}^{5}(\ourh_{4, 10}^{2} ) 
+ \frac{ \cFFp{5}{2}{ 5} }{420 }  {\rm ad}_{\epsilon_0}^{4}(\ourh_{4, 10}^{3} )  \,,
\notag \\
\series^{\tau}(6,8) +\series^{\tau}(8,6)&\cong
- \frac{3 \cFFm{2}{3}{ 4}}{140 }  {\rm ad}_{\epsilon_0}(\ourh_{6, 8}^{6} )  + 
\frac{ \cFFm{4}{3}{ 4} }{360 }  {\rm ad}_{\epsilon_0}^{3}(\ourh_{6, 8}^{4} ) 
  - \frac{\cFFm{6}{3}{ 4}}{30240 }  {\rm ad}_{\epsilon_0}^{5}(\ourh_{6, 8}^{2} ) \label{seceq4.40}\\
  &\quad - 
 \frac{\cFFp{3}{3}{ 4}}{14 }  {\rm ad}_{\epsilon_0}^{2}(\ourh_{6, 8}^{5} ) + 
\frac{ \cFFp{5}{3}{ 4}}{280 }  {\rm ad}_{\epsilon_0}^{4}(\ourh_{6, 8}^{3} ) \, .
\notag
\end{align}
There are no cusp-form contributions to the even functions at this weight, $\cFFp{s}{m}{7-m}=\FFp{s}{m}{7-m}$,
but the individual differences $\cFFm{6}{m}{7-m}-\FFm{6}{m}{7-m}$ in (\ref{eq:FcFm6}) involve the
odd combination $\Hm_{\Delta_{12}}$. This leads to the extra terms $\sim \Hm_{\Delta_{12}}$ in the last line of
\begin{align}
\series^{\tau}(4,10) &+\series^{\tau}(10,4)+\series^{\tau}(6,8) +\series^{\tau}(8,6) \cong
 \frac{  \FFm{4}{2}{ 5} }{540 }  {\rm ad}_{\epsilon_0}^{3}(\ourh_{4, 10}^{4} )
- \frac{ \FFm{6}{2}{ 5}}{30240   }  {\rm ad}_{\epsilon_0}^{5}(\ourh_{4, 10}^{2} ) 
\notag \\
& \ \
+ \frac{ \FFp{5}{2}{ 5} }{420 }  {\rm ad}_{\epsilon_0}^{4}(\ourh_{4, 10}^{3} ) 
- \frac{3 \FFm{2}{3}{ 4}}{140 }  {\rm ad}_{\epsilon_0}(\ourh_{6, 8}^{6} )  + 
\frac{ \FFm{4}{3}{ 4} }{360 }  {\rm ad}_{\epsilon_0}^{3}(\ourh_{6, 8}^{4} ) 
  - \frac{\FFm{6}{3}{ 4}}{30240 }  {\rm ad}_{\epsilon_0}^{5}(\ourh_{6, 8}^{2} ) \label{seceq4.41}\\
  &\ \ - 
 \frac{\FFp{3}{3}{ 4}}{14 }  {\rm ad}_{\epsilon_0}^{2}(\ourh_{6, 8}^{5} ) + 
\frac{ \FFp{5}{3}{ 4}}{280 }  {\rm ad}_{\epsilon_0}^{4}(\ourh_{6, 8}^{3} )  - \frac{ \Lambda(\Delta_{12},12) \Hm_{\Delta_{12}}}{  2268000 \Lambda(\Delta_{12},10) } \,  {\rm ad}_{\epsilon_0}^{5}( \ourh_{4, 10}^{2} - 3 \ourh_{6, 8}^{2} ) \,, \notag
\end{align}
which vanish due to ${\rm ad}_{\epsilon_0}^{5}( \ourh_{4, 10}^{2} - 3 \ourh_{6, 8}^{2} )=0$,
i.e.\ due to the $\epsilon_k$-relation in the first line of (\ref{seceq4.24}). Using the same relation also for the 
coefficients of $\FFm{6}{m}{k}$ shows that only the linear combination 
$3\FFm{6}{2}{5}+\FFm{6}{3}{4}$ seen in~\eqref{eq:3.8} appears as an MGF, but not the individual
modular invariant functions $\FFm{6}{2}{5}$ or $\FFm{6}{3}{4}$.

\subsubsection{Weight 8}

At weight $m{+}k=8$, we can similarly start from
\begin{align}
\series^{\tau}(4,12) +\series^{\tau}(12,4)&\cong
- \frac{\cFFm{5}{2}{ 6}}{7392 }  {\rm ad}_{\epsilon_0}^{4}(\ourh_{4, 12}^{4} ) 
+\frac{ \cFFm{7}{2}{ 6}}{665280  }  {\rm ad}_{\epsilon_0}^{6}(\ourh_{4, 12}^{2} ) 
- \frac{\cFFp{6}{2}{ 6}}{6048 }  {\rm ad}_{\epsilon_0}^{5}(\ourh_{4, 12}^{3} )  \,,
\notag \\
\series^{\tau}(6,10) +\series^{\tau}(10,6)&\cong
\frac{\cFFm{3}{3}{ 5}  }{252 }  {\rm ad}_{\epsilon_0}^{2}(\ourh_{6, 10}^{6} ) 
-\frac{ \cFFm{5}{3}{ 5} }{4620  }  {\rm ad}_{\epsilon_0}^{4}(\ourh_{6, 10}^{4} )  
+\frac{ \cFFm{7}{3}{ 5}}{665280 }  {\rm ad}_{\epsilon_0}^{6}(\ourh_{6, 10}^{2} ) \notag \\
& \quad+ 
 \frac{\cFFp{4}{3}{ 5} }{90 }  {\rm ad}_{\epsilon_0}^{3}(\ourh_{6, 10}^{5} )  
  - \frac{\cFFp{6}{3}{ 5}}{3780  }  {\rm ad}_{\epsilon_0}^{5}(\ourh_{6, 10}^{3})  \,,
\label{seceq4.42} \\
 \series^{\tau}(8,8)&\cong
- \frac{ 9 \cFFp{2}{4}{ 4}}{280 }  {\rm ad}_{\epsilon_0}(\ourh_{8, 8}^{7} ) 
+ \frac{ \cFFp{4}{4}{ 4}}{120 }  {\rm ad}_{\epsilon_0}^{3}(\ourh_{8, 8}^{5} ) 
- \frac{\cFFp{6}{4}{ 4}}{6720 }  {\rm ad}_{\epsilon_0}^{5}(\ourh_{8, 8}^{3} )  \,,
\notag
\end{align}
and immediately identify $\cFFm{s}{m}{8-m}=\FFm{s}{m}{8-m}$ since all
of these odd depth-two combinations are modular invariant without any need for cusp forms.
However, relating $\cFFp{6}{m}{8-m}$ to $\FFp{6}{m}{8-m}$ introduces the
even combination $\Hp_{\Delta_{12}}$  via (\ref{fplusw8}), leading to 
extra terms in the last line of
\begin{align}
&\series^{\tau}(4,12) +\series^{\tau}(12,4) + \series^{\tau}(6,10) +\series^{\tau}(10,6) +  \series^{\tau}(8,8) \notag \\
&\ \cong
- \frac{\FFm{5}{2}{ 6}}{7392 }  {\rm ad}_{\epsilon_0}^{4}(\ourh_{4, 12}^{4} )
+\frac{ \FFm{7}{2}{ 6}}{665280  }  {\rm ad}_{\epsilon_0}^{6}(\ourh_{4, 12}^{2} ) 
- \frac{\FFp{6}{2}{ 6}}{6048 }  {\rm ad}_{\epsilon_0}^{5}(\ourh_{4, 12}^{3} ) 
+\frac{\FFm{3}{3}{ 5}  }{252 }  {\rm ad}_{\epsilon_0}^{2}(\ourh_{6, 10}^{6} ) 
\notag \\
&\ \quad
-\frac{ \FFm{5}{3}{ 5} }{4620  }  {\rm ad}_{\epsilon_0}^{4}(\ourh_{6, 10}^{4} )  
+\frac{ \FFm{7}{3}{ 5}}{665280 }  {\rm ad}_{\epsilon_0}^{6}(\ourh_{6, 10}^{2} )  + 
 \frac{\FFp{4}{3}{ 5} }{90 }  {\rm ad}_{\epsilon_0}^{3}(\ourh_{6, 10}^{5} )  
  - \frac{\FFp{6}{3}{ 5}}{3780  }  {\rm ad}_{\epsilon_0}^{5}(\ourh_{6, 10}^{3}) 
\label{seceq4.43} \\
&\ \quad
- \frac{ 9 \FFp{2}{4}{ 4}}{280 }  {\rm ad}_{\epsilon_0}(\ourh_{8, 8}^{7} ) 
+ \frac{ \FFp{4}{4}{ 4}}{120 }  {\rm ad}_{\epsilon_0}^{3}(\ourh_{8, 8}^{5} ) 
- \frac{\FFp{6}{4}{ 4}}{6720 }  {\rm ad}_{\epsilon_0}^{5}(\ourh_{8, 8}^{3} )  \notag \\
&\ \quad
+ \frac{   \Lambda(\Delta_{12},13) \Hp_{\Delta_{12}} }{208958400 \Lambda(\Delta_{12},11) }
{\rm ad}^5_{\ep_0}( 4 \ourh_{4, 12}^{3} - 25 \ourh_{6, 10}^{3} + 
 21 \ourh_{8, 8}^{3})\, .
\notag
\end{align}
In this case, it is the depth-two terms in first line of (\ref{seceq4.27}) which
imply the vanishing of ${\rm ad}^5_{\ep_0}( 4 \ourh_{4, 12}^{3} - 25 \ourh_{6, 10}^{3} + 
 21 \ourh_{8, 8}^{3})$  and therefore the dropout of $ \Hp_{\Delta_{12}}$ modulo higher-depth terms that are given in the second line of~\eqref{select.4}.
The higher-depth commutators of $\ep_k$ in the second line of (\ref{select.4}) will be associated with higher-depth $\bsv$, and their modular invariant completions must also contain the iterated integrals of $\Hp_{\Delta_{12}}$ of depth one. Using the first line of (\ref{seceq4.27}) also for the coefficients of the 
functions $\FFp{6}{m}{k}$ in (\ref{seceq4.43}) reproduces the linear combinations appearing in~\eqref{fpluswb} (modulo commutators of $\epsilon_k$ of higher depth).

\subsubsection{Weight 9}

Similar to (\ref{seceq4.41}) at weight seven, the even functions $\cFFp{s}{m}{ 9-m}$
at weight $m{+}k=9$ are all identical to the modular invariants $\FFp{s}{m}{ 9-m}$.
The odd functions in turn introduce iterated integrals of the cusp forms $\Delta_{12}$
and $\Delta_{16}$ from the Laplace eigenspaces with $s=6$ and $s=8$, respectively:
\begin{align}
&\series^{\tau}(4,14) +\series^{\tau}(14,4)
+\series^{\tau}(6,12) +\series^{\tau}(12,6)
+ \series^{\tau}(8,10) +  \series^{\tau}(10,8) \notag \\
& \  \cong \frac{ \FFm{6}{2}{ 7}}{131040 }  {\rm ad}_{\epsilon_0}^{5}(\ourh_{4, 14}^{4} ) 
- \frac{\FFm{8}{2}{ 7}}{17297280  }  {\rm ad}_{\epsilon_0}^{7}(\ourh_{4, 14}^{2} ) 
+\frac{ \FFp{7}{2}{ 7}}{110880  }  {\rm ad}_{\epsilon_0}^{6}(\ourh_{4, 14}^{3} ) 
- \frac{\FFm{4}{3}{ 6}}{2376 }  {\rm ad}_{\epsilon_0}^{3}(\ourh_{6, 12}^{6} )  \notag \\
  & \ \quad   +\frac{ \FFm{6}{3}{ 6}}{78624 }  {\rm ad}_{\epsilon_0}^{5}(\ourh_{6, 12}^{4} ) 
  -\frac{ \FFm{8}{3}{ 6}}{17297280 }  {\rm ad}_{\epsilon_0}^{7}(\ourh_{6, 12}^{2} )  
   - \frac{ \FFp{5}{3}{ 6}}{924 }  {\rm ad}_{\epsilon_0}^{4}(\ourh_{6, 12}^{5} ) 
  +\frac{ \FFp{7}{3}{ 6}}{66528 }  {\rm ad}_{\epsilon_0}^{6}(\ourh_{6, 12}^{3} )  \notag \\
& \ \quad + \frac{\FFm{2}{4}{ 5} }{210 }  {\rm ad}_{\epsilon_0}(\ourh_{8, 10}^{8} ) 
 -\frac{ \FFm{4}{4}{ 5} }{1320  }  {\rm ad}_{\epsilon_0}^{3}(\ourh_{8, 10}^{6} ) 
 + \frac{\FFm{6}{4}{ 5}}{65520 }  {\rm ad}_{\epsilon_0}^{5}(\ourh_{8, 10}^{4} )  
  - \frac{\FFm{8}{4}{ 5}}{17297280 }  {\rm ad}_{\epsilon_0}^{7}(\ourh_{8, 10}^{2} ) \label{seceq4.44} \\
  & \ \quad  +  \frac{\FFp{3}{4}{ 5} }{42 }  {\rm ad}_{\epsilon_0}^{2}(\ourh_{8, 10}^{7} ) 
  -\frac{ 3 \FFp{5}{4}{ 5} }{1540 }  {\rm ad}_{\epsilon_0}^{4}(\ourh_{8, 10}^{5} ) 
  +\frac{ \FFp{7}{4}{ 5}}{55440 }  {\rm ad}_{\epsilon_0}^{6}(\ourh_{8, 10}^{3} ) \notag \\
  & \ \quad- \frac{691 \Lambda(\Delta_{12},14) \Hm_{\Delta_{12}} }{244481328000 \Lambda(\Delta_{12},10)}  \,{\rm ad}^5_{\ep_0} 
  \bigg( \frac{ 36 }{691} \ourh_{4, 14}^4 - \ourh_{6, 12}^4 +  3 \ourh_{8, 10}^4 \bigg) \notag \\
  & \ \quad - \frac{ \Lambda(\Delta_{16},16) \Hm_{\Delta_{16}} }{19559232000 \Lambda(\Delta_{16},14)}
  \,{\rm ad}^7_{\ep_0}   (2  \ourh_{4, 14}^2 - 7 \ourh_{6, 12}^{2} + 11 \ourh_{8, 10}^2)\,. \notag
\end{align}
Both combinations $\frac{ 36 }{691} \ourh_{4, 14}^4 - \ourh_{6, 12}^4 +  3 \ourh_{8, 10}^4$ and 
$2  \ourh_{4, 14}^2 - 7 \ourh_{6, 12}^{2} + 11 \ourh_{8, 10}^2$ of $\epsilon_k$ commutators vanish 
by (\ref{seceq4.28}) and (\ref{seceq4.24}), respectively (modulo higher-depth commutators
in the case of the $\ourh_{p, q}^4$).

\subsubsection{Weight 10}

Starting from weight 10, both even and odd functions $\cFFpm{s}{m}{ k}$
involve Laplace eigenvalues $s=6,8,9,\ldots$ where cusp forms are required in
the modular completion to $\FFpm{s}{m}{ k}$. All of $\Delta_{12},\Delta_{16}$
and $\Delta_{18}$ feature in the corrections in the lowest lines of the weight-10
terms at depth two,
\begin{align}
&\series^{\tau}(4,16) +\series^{\tau}(16,4)
+\series^{\tau}(6,14) +\series^{\tau}(14,6)
+\series^{\tau}(8,12) +\series^{\tau}(12,8) 
+\series^{\tau}(10,10)
\notag \\
&\ \ 
\cong
\frac{ \FFp{2}{5}{ 5}}{105 }  {\rm ad}_{\epsilon_0}(\ourh_{10, 10}^{9} ) 
 - \frac{ \FFp{4}{5}{ 5}}{330 }  {\rm ad}_{\epsilon_0}^{3}(\ourh_{10, 10}^{7} )  
+\frac{ \FFp{6}{5}{ 5}}{10920 }  {\rm ad}_{\epsilon_0}^{5}(\ourh_{10, 10}^{5} ) 
 -\frac{ \FFp{8}{5}{ 5}}{2162160} {\rm ad}_{\epsilon_0}^{7}(\ourh_{10, 10}^{3} ) 
\notag \\
& \ \ \ \
-\frac{ \FFm{7}{2}{ 8}}{2851200 }  {\rm ad}_{\epsilon_0}^{6}(\ourh_{4, 16}^{4} ) 
+\frac{ \FFm{9}{2}{ 8}}{518918400 }  {\rm ad}_{\epsilon_0}^{8}(\ourh_{4, 16}^{2} ) 
-\frac{ \FFp{8}{2}{ 8}}{2471040 }  {\rm ad}_{\epsilon_0}^{7}(\ourh_{4, 16}^{3} ) 
+\frac{ \FFm{5}{3}{ 7}}{32032 }  {\rm ad}_{\epsilon_0}^{4}(\ourh_{6, 14}^{6} ) 
\notag \\
& \ \ \ \
-\frac{ \FFm{7}{3}{ 7}}{1663200 }  {\rm ad}_{\epsilon_0}^{6}(\ourh_{6, 14}^{4} ) 
+\frac{ \FFm{9}{3}{ 7}}{518918400  }  {\rm ad}_{\epsilon_0}^{8}(\ourh_{6, 14}^{2} )  
+\frac{ \FFp{6}{3}{ 7}}{13104  }  {\rm ad}_{\epsilon_0}^{5}(\ourh_{6, 14}^{5} ) 
- \frac{\FFp{8}{3}{ 7}}{1441440 }  {\rm ad}_{\epsilon_0}^{7}(\ourh_{6, 14}^{3} ) 
 \notag \\
& \ \ \ \
-\frac{ 5 \FFm{3}{4}{ 6} }{5544  }  {\rm ad}_{\epsilon_0}^{2}(\ourh_{8, 12}^{8} )  
+\frac{ \FFm{5}{4}{ 6}}{16016 }  {\rm ad}_{\epsilon_0}^{4}(\ourh_{8, 12}^{6} )   
-\frac{ \FFm{7}{4}{ 6}}{1330560  }  {\rm ad}_{\epsilon_0}^{6}(\ourh_{8, 12}^{4} )  
+\frac{ \FFm{9}{4}{ 6}}{518918400 }  {\rm ad}_{\epsilon_0}^{8}(\ourh_{8, 12}^{2} )  \label{seceq4.45} \\
&\ \ \ \
- \frac{ \FFp{4}{4}{ 6}  }{264 }  {\rm ad}_{\epsilon_0}^{3}(\ourh_{8, 12}^{7} )  
+\frac{ \FFp{6}{4}{ 6}  }{6552 }  {\rm ad}_{\epsilon_0}^{5}(\ourh_{8, 12}^{5} ) 
  -\frac{ \FFp{8}{4}{ 6}}{1153152 }  {\rm ad}_{\epsilon_0}^{7}(\ourh_{8, 12}^{3} ) 
\notag \\
& \ \ \ \ - \frac{ \Lambda(\Delta_{12},15) \Hp_{\Delta_{12}} }{38030428800 \Lambda(\Delta_{12},11)} \,
{\rm ad}_{\ep_0}^5
(4 \ourh_{6, 14}^{5} - 25 \ourh_{8, 12}^{5} +  21 \ourh_{10, 10}^5)
    \notag \\
& \ \ \ \  + \frac{ \Lambda(\Delta_{16},17)  \Hp_{\Delta_{16}} }{ 8085227673600 \Lambda(\Delta_{16},15)} \,
{\rm ad}_{\ep_0}^7(36 \ourh_{4, 16}^3 - 
   245 \ourh_{6, 14}^3 + 539 \ourh_{8, 12}^3 - 
   330 \ourh_{10, 10}^3)
\notag \\
&\ \ \ \ - \frac{ \Lambda(\Delta_{18},18) \Hm_{\Delta_{18}}  }{5230697472000 \Lambda(\Delta_{18},16)} 
{\rm ad}_{\ep_0}^8 (8 \ourh_{4, 16}^2 - 25 \ourh_{6, 14}^2 +  26 \ourh_{8, 12}^2) \, .
\notag
\end{align}
Again, all the $ \Hpm_{\Delta_{2s}}$ are accompanied by combinations $4 \ourh_{6, 14}^{5} - 25 \ourh_{8, 12}^{5} +  21 \ourh_{10, 10}^5$ as well as $36 \ourh_{4, 16}^3 - 
   245 \ourh_{6, 14}^3 + 539 \ourh_{8, 12}^3 - 
   330 \ourh_{10, 10}^3$ and $8 \ourh_{4, 16}^2 - 25 \ourh_{6, 14}^2 +  26 \ourh_{8, 12}^2$
which vanish by
(\ref{seceq4.29}), (\ref{seceq4.27}) and (\ref{seceq4.24}), respectively (modulo higher-depth commutators in case of the $\ourh_{p, q}^5$ and $\ourh_{p, q}^3$).

\subsubsection{Summary}

In this section, we have demonstrated in detail how (specific linear combinations of) the modular functions $\FFpm{s}{m}{k}$ appear
in the generating series of MGFs. On the one hand, the representations of MGFs 
as lattice sums or integrals over torus punctures manifest that they are modular forms; on the other hand, 
their differential equations \cite{DHoker:2016mwo, Gerken:2019cxz}
rule out iterated integrals of holomorphic cusp forms. These requirements have been explicitly confirmed
for the $\beta^{\rm sv}$-contributions (\ref{select.1}) to modular invariant MGFs at depth two and a wide
range of weights.

By reorganising the shuffle irreducible $\beta^{\rm sv}$ of depth two in terms of $\cFFpm{s}{m}{k}$,
the accompanying derivations in (\ref{select.1}) conspire to specific combinations $\ourh^d_{p,q}$ of
commutators defined in (\ref{seceq4.22}) that are singled out by representation theory of $\SLtwoZ$.
More importantly, these combinations $\ourh^d_{p,q}$ were identified by Pollack \cite{Pollack} to streamline
relations in the derivation algebra. By rewriting parts of the generating series (\ref{select.1}) in terms
of $\cFFpm{s}{m}{k}$ and $\ourh^d_{p,q}$, we have exhibited the interplay between Pollack's relations and the dropout of the modular completions $\Hpm_{\Delta_{2s}}$ from the generating series of MGFs at depth two. These findings are consistent with the main result of this work that not all linear combinations of $\FFpm{s}{m}{k}$ can be represented by $\bsv$ of depth one and two, powers of $y$ and odd zeta values.

%%%%%%%%%%%%%%%%%%%%%%%%%%%%%%%%%%%%%%%%%%%
%%%%%%%%%%%%%%%%%%%%%%%%%%%%%%%%%%%%%%%%%%%
%%%%%%%%%%%%%%%%%%%%%%%%%%%%%%%%%%%%%%%%%%%

\section{\texorpdfstring{Kloosterman sums and the Fourier expansion of the $\FFpm{s}{m}{k}$}{Kloosterman sums and the Fourier expansion of the  Fpm(s)(m)(k)}}
\label{sec:5}

In this last section we want to highlight some of the ``side-effects'' due to the presence of iterated integrals of holomorphic cusp forms in the generic expression \eqref{eq:GenLinear} for $\FFpm{s}{m}{k}$. One of the consequences will be to provide a connection between L-values and Kloosterman sums that come out of the Poincar\'e-series representation of the $\FFpm{s}{m}{k}$ as anticipated in Part~I.

The addition of the iterated integrals over cusp forms has consequences for the Fourier expansion of the $\FFpm{s}{m}{k}$: In the following, we will compare different approaches to determining the coefficients of $q^a \bar q^b$ in the expansion around the cusp
\beq
\FFpm{s}{m}{k}(\tau)= \sum_{a,b=0}^\infty d_{a,b}(y) q^a \bar q^b \, .
\label{qbarqexp}
\eeq
The coefficients $d_{a,b}(y) $ are Laurent polynomials in $y=\pi \Im \tau$ 
(with powers ranging from $y^{m+k}$ to $y^{-m-k+2}$) which will be referred
to as $\FFpm{s}{m}{k} \big|_{q^a \bar q^b}$ in the rest of this section. 
They are straightforwardly determined from the representation of 
$\FFpm{s}{m}{k}$ in terms of $\beta^{\rm sv}$ and $\Hpm_{\Delta_{2s}}$ and carry
highly non-trivial information on (infinite combinations of) Kloosterman sums by comparing
with the Poincar\'e-series representations (\ref{eq:PSsol}) of $\FFpm{s}{m}{k}$.
As detailed in Part~I, the $\bsv$-part $\cFFpm{s}{m}{k}$ only introduces 
single-valued zeta values and rational numbers into $d_{a,b}(y)$. 
However, the novelty is that now the addition of $\Hpm_{\Delta_{2s}}$ 
also introduces L-values into the Fourier coefficients in (\ref{qbarqexp}).

\subsection{Odd example}

As a first example we assemble the order $q^1 \bar{q}^0$ term in $\FFm{6}{2}{5}$ from~\eqref{bsvex25} and~\eqref{eq:FcFm6},
\begin{align}
\FFm{6}{2}{5}\big|_{q^1\bar{q}^0} &=  \frac{2   y^5}{93555}-\frac{y^2}{1080} -\frac{y}{360} +\frac{19}{720} +\frac{2}{9 y}  +\frac{301}{384 y^2}+\frac{301}{192 y^3}+\frac{903}{512 y^4}+\frac{903}{1024 y^5} \nn\\
   &\quad  
    - \zeta_3 \bigg(\frac{1}{24 y}+\frac{1}{4 y^2}+\frac{21}{32 y^3} +\frac{7  }{8 y^4} +\frac{63  }{128 y^5}\bigg)
    + \zeta_9 \bigg(\frac{35}{64 y^4}+\frac{63}{128  y^5} \bigg) \label{eq:1InstFFm}\\
&\quad    - \frac{\Lambda(\Delta_{12},12)}{\Lambda(\Delta_{12},10)} \bigg(\frac{1}{150} +\frac{1}{
 20 y}+\frac{7}{40y^2}+\frac{7}{20y^3} +\frac{63}{160y^4}+\frac{63}{320 y^5} \bigg) \,.\notag
\end{align}
Here, we can see clearly the separate contributions from $\cFFm{6}{2}{5}$ in the first two lines, containing only rational coefficients and single odd zetas, and $\Hm_{\Delta_{12}}$ in the last line, 
which is instead multiplied by the ratio of L-values. 
Note that the complete Fourier mode $e^{2\pi i \Re\tau}$ receives an infinite series of additional
contributions beyond the $q^1 \bar{q}^0$ term in (\ref{eq:1InstFFm}): All the exponentially 
suppressed corrections $q (q\bar{q})^n$ for $n>0$ share the factor of $e^{2\pi i \Re\tau}$
and are multiplied by Laurent polynomials in $y$ with rational coefficients,
see section 7.1 of Part~I for their precise form.

At the same time, we have found a Poincar\'e-series representation of $\FFm{6}{2}{5}$ in Part~I, see section 5.2 there. Specialising the general seed formula, recapped in \eqref{eq:seedm}, to the case at hand yields
\begin{align}
\FFm{6}{2}{5}(\tau)  &=  \PS \left[  {-}\frac{ 4iy^5 }{31185} \Im {\cal E}_0(4,0;\tau) \right]_\gamma \label{eq:PoincFFm}
\\
&=  \PS \left[ \frac{2 y^5}{93555}\sum_{n=1}^{\infty} n^{-2} \sigma_3(n) (q^n-\bar{q}^n) \right]_\gamma\, ,
\notag
\end{align}
where the $\SLtwoZ$ element $\gamma$ acts on the $\tau$-dependence via both $q,\bar q$ and $y$.

The general formula for obtaining the Fourier series of a Poincar\'e sum from the Fourier series of its seed function~\cite{Iwaniec:2002,Fleig:2015vky}, see also appendix A of Part~I, then leads to the following identity
\begin{align}
\FFm{6}{2}{5}(\tau) \big|_{q^1 \bar q^0} + \sum_{n=1}^\infty \FFm{6}{2}{5}(\tau) \big|_{q^{n+1} \bar{q}^n} &= \frac{2y^5 }{93555} +  e^{2 y}\sum_{d=1}^\infty \sum_{\ell \in \ZZ} S(\ell,1;d) \label{eq:qExpFFm}
\\
& \quad \times \int_{\RR} e^{-2\pi i \omega - 2\pi i \ell \frac{\omega}{d^2( (\Im\tau)^2+\omega^2)}} c_\ell \left(\frac{\Im\tau}{d^2((\Im\tau)^2+\omega^2)}\right)\dd \omega\, ,
\notag
\end{align}
where $c_\ell(\Im\tau)$ is the $\ell^{\rm{th}}$ Fourier mode of the Poincar\'e seed in \eqref{eq:PoincFFm}, 
i.e.\ $c_0(\Im\tau)=0$ and 
\beq
c_\ell(\Im\tau) =\frac{2\,\mbox{sign}(\ell)}{93555}    |\ell|^{-2} \sigma_3(|\ell|) (\pi \Im\tau)^5 e^{-2\pi |\ell| \Im\tau} \quad\text{for $\ell\neq 0$}\,,
\eeq 
where the $\mbox{sign}(\ell) $ arises since we are considering an odd modular invariant.
Moreover, $S(\ell,1;d)$ denotes a particular instance of the Kloosterman sum
\begin{align}
S(\ell,n;d) = \sum_{r\in (\ZZ/ d\ZZ)^\times} 
\exp\bigg(\frac{ 2\pi i }{d} \bigg[\ell r+\frac{n}{r}\bigg] \bigg)\,,
\label{Kloosum}
\end{align}
where $0\leq r \leq d$ is coprime to $d$, such that $r$ has the multiplicative inverse $r^{-1}$ in $(\ZZ/ d\ZZ)^\times$.
Note that the above expression \eqref{eq:qExpFFm} contains the full  $e^{2\pi i \Re\tau}$  Fourier mode sector, i.e.\ it contains both the $q^1 \bar{q}^0$ term as well as the infinite tower of exponentially suppressed corrections $q (q\bar{q})^n$ for $n>0$. 
Since $\Hpm_{\Delta_{2s}}$ does not have any $(q\bar{q})^n$ terms, one can restrict to $\cFFpm{s}{m}{k}$, and the only sources of $(q\bar{q})^n$ are the depth-two $\bsv$, for which the full $q^{>0}\bar{q}^{>0}$ terms were given in section 7.1 of Part~I. All of them are accompanied by Laurent polynomials with rational coefficients.  

\subsection{Even example}

The same kind of analysis can be performed for the $q,\bar q$-expansion (\ref{qbarqexp}) of the even
modular functions $\FFp{s}{m}{k}$. As an example we can focus on $\FFp{6}{2}{6}$, where the $q^1 \bar q^0$ coefficient is given by
\begin{align}
\FFp{6}{2}{6}\big|_{q^1\bar{q}^0} &= - \frac{ 691 y^5}{3192564375}  - \frac{  y}{10800}  - \frac{ 5123}{4914000}    
- \frac{  7061}{1310400 y} - \frac{  6151}{374400 y^2} - \frac{  23239}{ 748800 y^3}
 \notag \\
&\quad - \frac{  89}{2600 y^4} - \frac{  89}{5200 y^5}  + \zeta_3 \bigg(  \frac{1}{ 480 y^2}  + \frac{7 }{480 y^3}  
+ \frac{7 }{160 y^4}  + \frac{21}{ 320 y^5} + \frac{21 }{512 y^6}
 \bigg) 
\label{evenex.1} \\
&\quad + \frac{ 21 \zeta_{11}}{512 y^6} +  \frac{\Lambda(\Delta_{12},13)}{ 17275 \Lambda(\Delta_{12},11)} 
\bigg( 1 + \frac{ 15}{2 y} + \frac{ 105}{4 y^2} + \frac{ 105}{2 y^3} + \frac{  945}{16 y^4}  + \frac{  945}{32 y^5}   \bigg)
\notag
\end{align}
and strongly resembles its odd counterpart (\ref{eq:1InstFFm}).
These two examples \eqref{eq:1InstFFm} and \eqref{evenex.1} once more illustrate the fact that the ratio of L-values $\frac{\Lambda(\Delta_{2s},t_1)}{ \Lambda(\Delta_{2s},t_2)}$, appearing in the perturbative expansion of the non-zero Fourier modes, has $t_1,t_2$ odd for $\FFp{s}{m}{k}$ as compared to $t_1, t_2$ even for $\FFm{s}{m}{k}$.

While the seed functions of the odd $\FFm{s}{m}{k}$ are series of 
$q^n{-}\bar q^n$ as in (\ref{eq:PoincFFm}), the seed functions of the even $\FFp{s}{m}{k}$ contain $q^n{+}\bar q^n$ and furthermore have zero modes 
involving $\mathbb Q$ multiples of $y^{m+k}$ and $\zeta_{2m-1} y^{k-m+1}$ such as
\begin{align}
\FFp{6}{2}{6} &=  \PS \left[ \frac{ 691 y^8}{373530031875}  - \frac{   691 y^5 \zeta_3}{3192564375} + \frac{  1382 y^5 }{1064188125} \Re {\cal E}_0(4, 0^2) \right]_\gamma \label{evenex.2} \\
&=  \PS \left[ \frac{ 691 y^8}{373530031875}  
- \frac{   691 y^5 }{3192564375} \bigg( \zeta_3 + 
 \sum_{n=1}^{\infty} n^{-3} \sigma_3(n) (q^n + \bar q^n) \bigg)
 \right]_\gamma\, ,
\notag
\end{align}
see section 3.3 of Part~I for a construction at generic $s,m,k$, and \eqref{eq:genseed} for a quick recap. 
We can again employ the relations between the Fourier
expansion of seed functions and Poincar\'e series reviewed in appendix A of Part~I to generate examples
of how the ratios of L-values as in (\ref{eq:1InstFFm}) and (\ref{evenex.1}) arise from Kloosterman sums and integrals as in (\ref{eq:qExpFFm}).

%%%%%
\subsection{General comments}
%%%%%

As we discussed in Part~I, the seed functions for the various $\FFpm{s}{m}{k}$ can all be written as finite, rational linear combinations of building blocks whose Fourier coefficients of $e^{2\pi i \ell \Re \tau}$
take the simple form
\begin{align}
c_\ell(\Im\tau) \sim \left\lbrace \begin{matrix}\phantom{\pm}  |\ell|^{-r} \sigma_{2m-1}(|\ell|) (\pi \Im\tau)^{m+k-r} e^{-2\pi |\ell| \Im\tau} &:\ \ell>0\,,\\[2mm]
\pm  |\ell|^{-r} \sigma_{2m-1}(|\ell|) (\pi \Im\tau)^{m+k-r} e^{-2\pi |\ell| \Im\tau}&:\ \ell<0\,,
\end{matrix}\right. \label{eq:GenSeed}
\end{align}
for integers $r$ in the range $m{+}1\leq r \leq 2m{-}1$ and where the sign $\pm$ for the negative Fourier modes $\ell<0$ is adapted to the modular function $\FFpm{s}{m}{k}$ considered.

In \cite{Dorigoni:2019yoq} it was explained how to extract the asymptotic expansion for $\Im \tau \to \infty$ for the Poincar\'e sum of a seed of the form \eqref{eq:GenSeed} and how to derive its Laurent polynomial $d_{0,0}(y)$ for the zero Fourier mode, discussed in full detail in Part~I. It is furthermore possible to exploit the asymptotic nature of such an expansion to obtain, via resurgence analysis, the exponentially suppressed terms in the same Fourier mode sector, i.e.\ the terms $\sim (q\bar{q})^n$ with $n>0$.

In a similar spirit, we believe that it should be possible, starting from the Poincar\'e sum of the general seed \eqref{eq:GenSeed}, to extract its asymptotic expansion for $\Im \tau \to \infty$ in any Fourier mode sector. Unlike for the zero-mode sector, no such general expression is at the present time known for \eqref{eq:GenSeed}. 
For example, it would be extremely interesting to start from the expression \eqref{eq:qExpFFm} for the first Fourier mode $e^{2\pi i \Re\tau}$ of  $\FFm{6}{2}{5}(\tau) $, or the analogous expression for $\FFp{6}{2}{6}$ starting from~\eqref{evenex.2}, and to derive their asymptotic expansions for $\Im \tau \to \infty$.
Similar to what was done in~\cite{Dorigoni:2019yoq} the integral in~\eqref{eq:qExpFFm} could be done term-wise after expanding the $\tau$-dependent part of the exponential in an absolutely convergent series. This yields multiple, partly divergent, infinite sums over Kloosterman sums. The analytic continuation of these sums is left for future work.

Irrespective of the explicit result, we can still make some predictions.  Firstly, due to the presence of this novel, and extremely non-trivial, Kloosterman sum $S(\ell,1;d)$ in the first Fourier mode (\ref{eq:qExpFFm}) of $\FFm{6}{2}{5}$ (or (\ref{Kloosum}) for the generic Fourier mode), we should find that this asymptotic expansion truncates after finitely many terms. More importantly, this asymptotic expansion has either rational numbers, single odd zetas or L-values in its coefficients, reproducing all the $q^1 \bar{q}^0$ terms given in \eqref{eq:1InstFFm} and similarly \eqref{evenex.1} for the even example $\FFp{6}{2}{6}$. 
Secondly, the asymptotic nature of such an expansion should also hide and encode the presence of an infinite tower of exponentially suppressed corrections, i.e.\ the $q (q\bar{q})^n$ for $n>0$, each one of them accompanied by a Laurent polynomial in $y$ with rational coefficients.
%

%%%%%%%%%%%%%%%%%%%
%%%%%%%%%%%%%%%%%%%
%%%%%%%%%%%%%%%%%%%

\medskip
\subsection*{Acknowledgements}
We are grateful to Francis Brown for his generous help in understanding his results on multiple 
modular values, providing us with explicit examples beyond the ones published in~\cite{Brown2019} 
and valuable comments on an earlier version of this work. 
We would like to thank Eric D'Hoker, Nikolaos Diamantis, Mehregan Doroudiani, Herbert Gangl, Jan Gerken, Martijn Hidding, Kim Klinger-Logan, Nils Matthes, Stephen D. Miller, Bram Verbeek, and Federico Zerbini
for valuable discussions and collaboration on related topics. DD and OS thank the Albert Einstein Institute Potsdam for kind 
hospitality during several stages of this project and creating a stimulating atmosphere. 
OS is supported by the European Research Council under ERC-STG-804286 UNISCAMP.

\appendix

%%%%%%%%%%%%%%%%%%%
%%%%%%%%%%%%%%%%%%%
%%%%%%%%%%%%%%%%%%%

\section{Recap of seed functions}
\label{app:seeds}
For the convenience of the reader, in this appendix we summarise the final results of Part~I for the seed functions $\seedpm{s}{m}{k} $ associated with the modular invariant solutions \eqref{eq:PSsol} to the Laplace systems \eqref{eq:Laps}.

In the even case we have the seeds 
\begin{align}
\label{eq:genseed}
\seedp{s}{m}{k}(\tau) &=\notag
(-1)^{k+m} \frac{ {\rm B}_{2k} {\rm B}_{2m}  (4y)^{k+m} }{(2k)! (2m)! (\mu_{k+m}-\mu_s) }
-(-1)^{k} \frac{4 {\rm B}_{2k} (2m{-}3)! \zeta_{2m-1}  (4y)^{k+1-m} }{(2k)!(m{-}2)!(m{-}1)! (\mu_{k-m+1}-\mu_s)}\\
&- (-1)^{k} \frac{ 2  {\rm B}_{2k} \Gamma(2m)
}{(2k)! \Gamma(m)} \sum_{\ell=k-m+1}^{k-1} g^+_{m,k,\ell,s} (4y)^\ell \Re \mathcal{E}_0 (2m,0^{k+m-\ell-1})\,,
\end{align}
with $\mu_s=s(s{-}1)$, iterated Eisenstein integrals $\mathcal{E}_0 (2m,0^{p})$ defined in \eqref{eq:E0depth1} and rational coefficients $g^+_{m,k,\ell,s}$ given by
\begin{align}
\label{eq:fkm}
g^+_{m,k,\ell,s} = \frac{\Gamma(\ell)}{\Gamma(\ell{+}s)} \sum_{i=\ell}^{k-1}  \frac{ (\ell{+}1{-}s)_{i-\ell}\Gamma(s{+}i)\Gamma(m{+}k{-}i{-}1)}{\Gamma(k{-}i)\Gamma(i{+}1)\Gamma(m{-}k{+}i{+}1)}\,.
\end{align}

Similarly, in the odd case we have the seeds
\begin{align}
\label{eq:seedm}
\seedm{s}{m}{k} =
%-
i (-1)^{k} \frac{  {\rm B}_{2k} (2m{-}1)! }{2\Gamma(2k) \Gamma(m)} \sum_{\ell=k-m+1}^{k} g^{-}_{m,k,\ell,s} (4y)^\ell \Im \mathcal{E}_0 (2m,0^{k+m-\ell-1})
\end{align}
with rational coefficients $g^{-}_{m,k,\ell,s}$ given by
\begin{align}
g^{-}_{m,k,\ell,s} =  \frac{\Gamma(\ell)}{\Gamma(\ell{+}s)} \sum_{i=\ell}^{k}  \frac{ (\ell{+}1{-}s)_{i-\ell}\Gamma(s{+}i)\Gamma(m{+}k{-}i)}{\Gamma(k{-}i{+}1)\Gamma(i{+}1)\Gamma(m{-}k{+}i)}\, ,
\label{eq:seedm.B}
\end{align}
see sections 3.3 and 5.2 of Part I for the derivation of (\ref{eq:genseed}) and (\ref{eq:seedm}), respectively.

\section{Examples of cocycles}
\label{app:morecocyc}

In this appendix, we spell out further examples of the cocycles (\ref{eq:Hnoninv}) of the solutions
$\Hpm_{\Delta_{2s}}$ to homogeneous Laplace equations.
In the ancillary file, all cocycles up to $2s=26$ are given along with the ratios of the completed L-values inside the critical strip.

\subsection{Weight 16}
\label{subappendixA}

For the cusp form of weight 16 in the Hecke normalisation $\Delta_{16} = q + O(q^2)$, 
the ratios of L-values
\begin{align}
\Lambda(\Delta_{16},8): 
\Lambda(\Delta_{16},10):
\Lambda(\Delta_{16}, 12):
\Lambda(\Delta_{16},14)&=
\frac{35}{468}: \frac{49}{468}:\frac{ 245}{936}: 1   \,,
\label{ratios16} \\
\Lambda(\Delta_{16}, 9):
\Lambda(\Delta_{16},    11):
\Lambda(\Delta_{16}, 13): 
\Lambda(\Delta_{16}, 15) 
&=
\frac{3617}{98280}: \frac{3617}{51480}
: \frac{3617}{16380}: 1   \,,
\notag 
\end{align}
and similar ratios $\Lambda(\Delta_{16},t) :\Lambda(\Delta_{16}, 14) $ and
$\Lambda(\Delta_{16},t) :\Lambda(\Delta_{16}, 15) $ with $t<8$
following from the reflection formula~\eqref{eq:Lrefl}
lead to the cocycles
{\allowdisplaybreaks
\begin{subequations}
\begin{align}
\Hp_{\Delta_{16}}(\tau) - \Hp_{\Delta_{16}}(-\tfrac1{\tau})  &=  \frac{ \pi^{15} \Lambda(\Delta_{16},15) (1{-}\tau \bar\tau)}{ 2520 y^7}
\bigg\{ 1 + \tau^6 \bar\tau^6 - 
 \frac{3617 }{780} (\tau^2 {+} \bar\tau^2) (1 {+} \tau^4 \bar\tau^4)  \notag \\*
 &\ \ + 
  \frac{25319 }{10296} (1 {+} \tau^2 \bar\tau^2) (\tau^4 {+} \bar\tau^4) - 
  \frac{3617 }{14040} (\tau^6 {+} \bar\tau^6) - 
  \frac{22979}{2340} \tau \bar\tau (1 {+} \tau^4 \bar\tau^4)  \notag \\*
 &\ \ + 
  \frac{647443 }{51480} \tau \bar\tau (\tau^2 {+} \bar\tau^2) (1 {+} \tau^2 \bar\tau^2) - 
  \frac{25319 }{8580} \tau \bar\tau (\tau^4 {+} \bar\tau^4) - 
  \frac{372551 }{25740} \tau^2 \bar\tau^2 (\tau^2 {+} \bar\tau^2)  \notag \\*
 &\ \ + 
  \frac{1089559 }{51480} \tau^2 \bar\tau^2 (1 {+} \tau^2 \bar\tau^2) - 
  \frac{461756}{19305} \tau^3 \bar\tau^3 \bigg\}\,,
\\
\Hm_{\Delta_{16}}(\tau) - \Hm_{\Delta_{16}}(-\tfrac1{\tau})  &=   \frac{i \pi^{15} \Lambda(\Delta_{16},14) (\tau{+} \bar\tau)}{ 360 y^7}
\bigg\{  1 + \tau^6 \bar\tau^6 - 
  \frac{1225 }{936} (\tau^2 {+} \bar\tau^2) (1 {+} \tau^4 \bar\tau^4)  \notag \\*
 &\ \ + 
  \frac{49}{156} (1 {+} \tau^2 \bar\tau^2) (\tau^4 {+} \bar\tau^4) - 
  \frac{5}{468} (\tau^6 {+} \bar\tau^6) - \frac{490}{117} \tau \bar\tau (1 {+} \tau^4 \bar\tau^4)  \notag \\*
 &\ \ + 
  \frac{392}{117} \tau \bar\tau (\tau^2 {+} \bar\tau^2) (1 {+} \tau^2 \bar\tau^2) - 
  \frac{20}{39} (\tau^5 \bar\tau {+} \tau \bar\tau^5) + 
  \frac{3577}{468} \tau^2 \bar\tau^2 (1 {+} \tau^2 \bar\tau^2) \notag \\*
 &\ \  - 
  \frac{655}{156} \tau^2 \bar\tau^2 (\tau^2 {+} \bar\tau^2) - \frac{80}{9} \tau^3 \bar\tau^3  \bigg\} \, .
\end{align}
\end{subequations}}%

\subsection{Weight 18}
\label{subappendixB}

For the cusp form of weight 18 in the Hecke normalisation $\Delta_{18} = q +O(q^2)$, 
the ratios of L-values
\begin{align}
\Lambda(\Delta_{18},10): 
\Lambda(\Delta_{18},12):
\Lambda(\Delta_{18}, 14):
\Lambda(\Delta_{18},16)&=
\frac{1}{120}: \frac{1}{24}: \frac{11}{60}: 1  \,,
\label{ratios18} \\
\Lambda(\Delta_{18}, 11):
\Lambda(\Delta_{18},  13):
\Lambda(\Delta_{18}, 15): 
\Lambda(\Delta_{18}, 17) 
&= \frac{43867}{5544000}: \frac{43867}{1310400}:\frac{ 43867}{270000}: 1\,,
\notag 
\end{align}
along with the remaining values, including $\Lambda(\Delta_{18}, 9)=0$, due to (\ref{eq:Lrefl}) lead to the cocycles
\begin{subequations}
\begin{align}
\Hp_{\Delta_{18}}(\tau) - \Hp_{\Delta_{18}}(-\tfrac1{\tau})  &=  \frac{ \pi^{17} \Lambda(\Delta_{18},17) (1{-}\tau^2 \bar\tau^2)}{ 20160 y^8}
\bigg\{ 1 + \tau^6 \bar\tau^6 - 
  \frac{307069}{67500} (\tau^2 {+} \bar\tau^2) (1 {+} \tau^4 \bar\tau^4)  \notag \\
 &\ \ + 
  \frac{43867}{18720} (1 {+} \tau^2 \bar\tau^2) (\tau^4 {+} \bar\tau^4) - 
  \frac{43867}{198000} (\tau^6 {+} \bar\tau^6) - 
  \frac{175468}{16875} \tau \bar\tau (1 {+} \tau^4 \bar\tau^4) \notag \\
 &\ \  + 
  \frac{43867}{2925} \tau \bar\tau (\tau^2 {+} \bar\tau^2) (1 {+} \tau^2 \bar\tau^2) - 
  \frac{43867 }{12375} \tau \bar\tau (\tau^4 {+} \bar\tau^4) + 
  \frac{318769}{11700} \tau^2 \bar\tau^2 (1 {+} \tau^2 \bar\tau^2)  \notag \\
 &\ \  - 
  \frac{29785693}{1485000} \tau^2 \bar\tau^2 (\tau^2 {+} \bar\tau^2) - 
  \frac{6536183}{185625} \tau^3 \bar\tau^3 \bigg\}\,,
\\
\Hm_{\Delta_{18}}(\tau) - \Hm_{\Delta_{18}}(-\tfrac1{\tau})  &=   \frac{i \pi^{17} \Lambda(\Delta_{18},16) (\tau{+} \bar\tau)(1{+} \tau \bar \tau)}{ 2520 y^8}
\bigg\{  1 + \tau^6 \bar\tau^6 -  \frac{77 }{60} (\tau^2 {+} \bar\tau^2) (1 {+} \tau^4 \bar\tau^4)
 \notag \\
 &\ \  +   \frac{7}{24} (1 {+} \tau^2 \bar\tau^2) (\tau^4 {+} \bar\tau^4) - 
  \frac{97}{20} \tau \bar\tau (1 {+} \tau^4 \bar\tau^4) + 
  \frac{469}{120} \tau \bar\tau (\tau^2 {+} \bar\tau^2) (1 {+} \tau^2 \bar\tau^2)  \notag \\
 &\ \  - 
  \frac{31}{60} \tau \bar\tau (\tau^4 {+} \bar\tau^4) + 
  \frac{1247}{120} \tau^2 \bar\tau^2 (1 {+} \tau^2 \bar\tau^2) - 
  \frac{319}{60} \tau^2 \bar\tau^2 (\tau^2 {+} \bar\tau^2)  \notag \\
 &\ \  - 
  \frac{1}{120} (\tau^6 {+} \bar\tau^6) - \frac{196 }{15} \tau^3 \bar\tau^3 \bigg\} \, .
\end{align}
\end{subequations}

%%%%%%%%%%%%%%%%%%%%%%%%%%%%%%%%%%%%%%%%%%
%%%%%%%%%%%%%%%%%%%%%%%%%%%%%%%%%%%%%%%%%%
%%%%%%%%%%%%%%%%%%%%%%%%%%%%%%%%%%%%%%%%%%

\bibliography{cites}
\bibliographystyle{utphys}

\end{document}